\documentclass[a4paper,11pt]{article}
\usepackage{jcappub}

\usepackage[dvipsnames]{xcolor}
\usepackage{hyperref}
\usepackage{natbib}
\usepackage{amsmath}
\usepackage{amssymb}
\usepackage{empheq}
\usepackage{graphicx}
\usepackage{parskip}
\usepackage{hyperref}
\usepackage{appendix}
\usepackage[utf8]{inputenc}
\hypersetup{
    colorlinks=true,
    linkcolor=blue,
    filecolor=magenta,      
    urlcolor=cyan,
}
\usepackage{mathtools}

\usepackage[object=vectorian]{pgfornament} 
\usepackage{tikz}
\newcommand{\sectionline}{%
  \noindent
  \begin{center}
  {
    \resizebox{0.5\linewidth}{1ex}
    {{%
    {\begin{tikzpicture}
    \node  (C) at (0,0) {};
    \node (D) at (9,0) {};
    \path (C) to [ornament=85] (D);
    \end{tikzpicture}}}}}%
    \end{center}
  }

\usepackage{enumitem}

\newcommand{\phienics}{$\varphi$\texttt{enics}}
\newcommand{\fenics}{\texttt{FEniCS}}
\newcommand{\meshExp}{\texttt{ArcTanExp}}
\newcommand{\meshPow}{\texttt{ArcTanPowerLaw}}

\newcommand{\figref}[1]{Figure~\ref{#1}}  
\newcommand{\Figref}[1]{Figure~\ref{#1}}  
\newcommand{\eqnref}[1]{Eq.~\eqref{#1}}
\newcommand{\eqnsref}[1]{Eqs.~\eqref{#1}}
\newcommand{\secref}[1]{Sec.~\ref{#1}}
\newcommand{\Secref}[1]{Sec.~\ref{#1}}

\newcommand{\Appref}[1]{Appendix~\ref{#1}}
\newcommand{\ie}{i.e.\ }
\newcommand{\eg}{e.g.\ }

\newcommand{\ud}{\ensuremath{\mathrm{d}}}
\newcommand{\udd}[1]{\ensuremath{\mathrm{d}#1\,}}

\newcommand{\lapRad}{\ensuremath{\nabla_r}}

\newcommand{\norm}[1]{\ensuremath{\left\lVert #1\right\rVert}}
\newcommand{\inner}[2]{\ensuremath{\left\langle #1,#2\right\rangle}}

\newcommand{\rSrc}{\ensuremath{r_{\rm s}}}
\newcommand{\mSrc}{\ensuremath{M_{\rm S}}}
\newcommand{\stepWid}{\ensuremath{w}}

\newcommand{\fldOne}{\ensuremath{\pi}}
\newcommand{\fldLt}{\ensuremath{\phi}}
\newcommand{\fldHvy}{\ensuremath{H}}
\newcommand{\mOne}{\ensuremath{m_{\fldOne}}}
\newcommand{\mLt}{\ensuremath{m_{\fldLt}}}
\newcommand{\mHvy}{\ensuremath{m_\fldHvy}}
\newcommand{\paramNL}{\ensuremath{\epsilon}}
\newcommand{\kinmix}{\ensuremath{\alpha}}
\newcommand{\UVScl}{\ensuremath{\Lambda}}
\newcommand{\higgsNL}{\ensuremath{\lambda}}

\newcommand{\opCorr}[1]{\ensuremath{O_{#1}}}

\newcommand{\itp}[1]{\ensuremath{#1^{(k+1)}}}

\newcommand{\nEl}{\ensuremath{N}}

\newcommand{\nounits}[1]{\ensuremath{\bar{#1}}}
\newcommand{\mpl}{\ensuremath{M_{\rm P}}}
\newcommand{\fldscl}{\ensuremath{\mu}}
\newcommand{\rscl}{\ensuremath{\kappa}}

\newcommand{\rSrcS}[1][]{\ensuremath{\nounits{t}^{#1}\rSrc^{#1}}}

\title{\phienics: Vainshtein screening with the finite element method}
\date{}

\author[a]{Jonathan Braden,}
\author[b]{Clare Burrage,}
\author[b,c]{Benjamin Elder,}
\author[b,d]{and Daniela Saadeh}

\affiliation[a]{Canadian Institute for Theoretical Astrophysics, University of Toronto, 60 St. George Street, Toronto, Canada}
\affiliation[b]{School of Physics and Astronomy, University of Nottingham, University Park, Nottingham, U.K.}
\affiliation[c]{Department of Physics and Astronomy, University of Hawai'i, 2505 Correa Road, Honolulu, USA}
\affiliation[d]{Institute of Cosmology and Gravitation, University of Portsmouth, Portsmouth, U.K.}

\emailAdd{daniela.saadeh@port.ac.uk}

\abstract{
Within the landscape of modified theories of gravity, progress in understanding the behaviour of, and developing tests for,  screening mechanisms has been hindered by the complexity of the field equations involved, which are nonlinear in nature and characterised by a large hierarchy of scales. This is especially true of Vainshtein screening, where the fifth force is suppressed by high-order derivative terms which dominate within a radius much larger than the size of the source, known as the Vainshtein radius.

In this work, we present the numerical code \phienics, building on the \texttt{FEniCS} library, to solve the full equations of motion from two theories of interest for  screening: a model containing high-order derivative operators in the equation of motion and one characterised by nonlinear self-interactions in two coupled scalar fields. We also include functionalities that allow the computation of  higher-order operators of the scalar fields in post-processing, enabling us to check that the profiles we find are consistent solutions within the effective field theory. These two examples illustrate the different challenges experienced when trying to simulate such theories numerically, and we show how these are addressed within this code. The examples in this paper assume spherical symmetry, but the techniques may be straightforwardly generalised to asymmetric configurations. This article therefore also provides a worked example of how the finite element method  can be employed to solve the screened equations of motion. \phienics{} is publicly available and can be adapted to solve other theories of screening.
}

\begin{document}

\maketitle
\flushbottom

\section{Introduction}
The theory of general relativity is an extremely successful description of gravity on Earth and within the solar system. On cosmological scales the standard cosmological model, $\Lambda$CDM, relies on general relativity as the theory of gravity, but  is in agreement with observations only after the introduction of a new, dark, matter component, and either fine tuning of the cosmological constant or the introduction of a second new component, called dark energy \cite{Aghanim:2018eyx}.  Even then, tensions may be starting to arise between differing measurements of the Hubble constant \cite{Bernal:2016gxb,Freedman:2017yms,Riess:2019cxk,Wong:2019kwg}. 

It therefore behoves us to consider the possibility that general relativity may be modified, in particular on large cosmological scales.  However, such modifications must still be in agreement with local tests of gravity, which see no deviation from general relativity \cite{Will:2014bqa}.  This leads to the development of theories with screening mechanisms, where nonlinearities mean that modifications can be large on cosmological scales, but small locally \cite{Joyce:2014kja,Ishak:2018his}.

The simplest and most common modification of gravity is the introduction of an additional scalar degree of freedom in the gravitational sector. The additional scalar mediates a fifth force, which must be suppressed within the solar system via a screening mechanism to be consistent with observations. Such suppression can be achieved if the scalar field displays non-trivial self-interactions, which take the form of nonlinear terms in the Lagrangian. 
The presence of the nonlinear terms needed for screening should not be surprising, as both general relativity, still our current best theory of gravity, and the Higgs model, the only fundamental scalar to have been detected, require non-trivial self interactions of their fields.  Nonlinear terms for the scalar mode occur naturally within, for example, $f(R)$ theories \cite{Sotiriou:2008rp}, massive gravity \cite{Hinterbichler:2011tt,deRham:2010kj} and Horndeski theories \cite{Horndeski:1974wa,Deffayet:2009mn}. In Refs.~\cite{Desmond:2018sdy,Desmond:2018kdn} it has been suggested  that a fifth force with screening may be present on galactic scales.  Theories of dark energy and modified gravity, including those theories that possess screening, are reviewed in Refs.~\cite{Joyce:2014kja,Bull:2015stt,Koyama:2015vza,Ishak:2018his,Slosar:2019flp}. 

Of particular interest for this work are Galileon scalar-tensor theories.  Galileon theories were introduced in Ref.~\cite{Nicolis:2008in}, as higher-order derivative scalar theories that, nevertheless, have second-order equations of motion.  In flat space-time, the theories were also required to be invariant, up to total derivatives, under shifts of the vacuum expectation value of the scalar field and shifts in its gradient.  As these theories were only required to respect the symmetry up to total derivatives, we will refer to these theories as Wess-Zumino Galileons following Ref.~\cite{Hinterbichler:2011tt}. The presence of the symmetry helps protect the particular form of these theories from quantum corrections.

Wess-Zumino Galileon theories can be generalised in many different ways.  We can require that the symmetry be completely respected at the level of the action (not just up to total derivatives), but relax the requirement that the equations of motion be second order. The resulting terms in the equation of motion may be higher order in derivatives, but the resulting Ostrogradski ghost instability appears only at the cut-off of the effective field theory.  The discussion in Ref.~\cite{Hinterbichler:2011tt} indicates that the UV behaviour of these theories may differ from Wess-Zumino Galileons.  If we do not insist on the symmetry, but do require the equations of motion to remain second order in derivatives, then the Wess-Zumino Galileons generalise to the Horndeski scalar-tensor theory \cite{Horndeski:1974wa,Deffayet:2009mn}, where the equations of motion remain second order even around a curved background. Indeed the Horndeski model can be generalised even further to beyond-Horndeski \cite{Gleyzes:2014dya,Gleyzes:2014qga} and Degenerate Higher-Order Scalar-Tensor (DHOST) theories \cite{Langlois:2015cwa,Langlois:2015skt,Achour:2016rkg,Crisostomi:2016czh}.  

A final Galileon model of interest here is the UV complete Galileon of Ref.~\cite{deRham:2017imi}. This is a two scalar field model which has quadratic kinetic terms for the two scalars, but has a nonlinear potential and kinetic mixing between the two scalars. Considered just as a description of the two scalar fields, this theory is complete in the UV. If one field is heavier than the other, and is `integrated out' of the theory by substituting its equation of motion back into the action, then the resulting theory for the light scalar is a Galileon theory, with higher order derivatives in the equations of motion.  

 \sectionline
 
In order to test theories with screening against observations, we need an accurate description of how the screening mechanism operates. As the theory is, by definition, nonlinear, and may contain non-canonical kinetic terms, it is challenging to solve analytically except in very simplified circumstances. 
The nonlinear structure of  the theory can also pose a challenge to numerical simulation. Nevertheless, previous  simulations of Vainshtein screening have successfully shed light on the phenomenology of the model. Ref.~\cite{Ogawa:2018srw} showed that the Galileon force can be enhanced inside a hole in a planar object. A violation of the equivalence principle has been shown in finite-differencing simulations of a Galileon model for two extended bodies  \cite{Hiramatsu:2012xj}, and for two point sources in finite-element simulations of $P(X)$ theories \cite{Kuntz2019}. Ref. \cite{White2020} solved for the cubic Galileon field in two- and three-body systems on solar system scales, using finite-differencing techniques. Systems with time evolution have also been studied.  In particular, a four-dimensional numerical code to study scalar gravitational radiation emitted from binary systems and probe the Vainshtein mechanism in situations that break the assumption of staticity and spherical symmetry was presented in Ref.~\cite{Dar:2018dra}.

Considerable effort has also been devoted to the numerical characterisation of chameleon-like screening.
The time evolution of a chameleon and symmetron field around spherically symmetric black holes was studied in Ref.~ \cite{Frolov_2017} using a combination of pseudo-spectral and Gauss-Legendre methods.
In the context of laboratory tests of gravity, the behaviour of the chameleon and symmetron inside an experimental chamber has been studied using finite-differencing and finite-element techniques.  This lead to experimental bounds on the chameleon \cite{Upadhye_2006, Elder_2016} and symmetron models \cite{Jaffe_2017, Brax:2018zfb}. Ref.~\cite{Elder_2020} studied the symmetron force in Casimir experiments, forecasting the sensitivity of future experiments that are realisable with the current state of the art technology. In Ref.~\cite{Burrage:2017shh}, the finite element method was used to study the dependence of chameleon screening on the source shape, assuming cylindrical symmetry. All these studies solved the full equation of motion, without restricting the behaviour to specific approximating regimes.

Much effort has gone into studying the cosmological evolution of these theories numerically.  It was shown in Ref.~\cite{Barreira:2015xvp} that N-body simulations in theories with Vainshtein screening could be sped up by refining the mesh only in regions where the fifth force is not suppressed.  An alternative approach to speeding up such calculations by introducing a screening factor was introduced in Ref.~\cite{Winther:2014cia}, and further attempts to speed up calculations of chameleon screening mechanisms are detailed in Ref.~\cite{Bose:2016wms}. \texttt{ISIS}, an  N-body cosmological code with scalar fields based on \texttt{RAMSES} \cite{Llinares:2013jza}, uses a nonlinear multi-grid solver that can treat a large class of scalar-tensor theories of modified gravity that possess screening mechanisms. This was extended to include disformal couplings to matter in \cite{Hagala:2015paa}. The results of a code comparison project for these N-body modified gravity codes were reported in Ref.~\cite{Winther:2015wla}. In Ref.~\cite{Barreira:2013eea}, the \texttt{ECOSMOG} code was used to simulate linear and nonlinear growth of the large-scale structure in Cubic Galileon gravity. Subsequently, it was extended to use adaptive mesh refinement in Ref.~\cite{Li:2013nua}, and was applied to quartic Galileons in Ref.~\cite{Li:2013tda}. Various Boltzmann codes have also been written to simulate cosmological evolution in theories of modified gravity, including \texttt{ISiTGR} \cite{Dossett:2011tn}, \texttt{MGCAMB} \cite{Hojjati:2011ix,Zucca:2019xhg}, \texttt{EFTCAMB} \cite{Hu:2013twa,Hu:2014oga} and   \texttt{hi}$\_$\texttt{class} \cite{Zumalacarregui:2016pph,Bellini:2019syt}.  A comparison of these codes can be found in Ref.~\cite{Bellini:2017avd}. Simulations of $f(R)$ theories of modified gravity, including the hydro-dynamical code \texttt{SHYBONE}, are reported in Ref.~\cite{Arnold:2019zup}.  A recent review of simulation techniques for modified gravity theories, in particular those with screening, can be found in Ref.~\cite{Llinares:2018maz}. These works have focused on simulating the cosmological solutions to the Galileon equations of motion, and the resulting consequences for observational probes.  This has necessarily required making approximations to, or truncations of, the Galileon equations of motion.

In this work we focus on solving the full nonlinear field equations for screened scalar fields around isolated compact objects. To properly test screening mechanisms and obtain robust constraints, one must obtain solutions to the full field equations.  In doing so, one immediately encounters a large hierarchy of scales.  The simulation grid must be sufficiently large (of order several Compton wavelengths $m^{-1}$) to accurately capture the Yukawa suppression, while simultaneously having a small enough spacing to capture any features in the source-vacuum transition.  At minimum, the source will have a characteristic size $\rSrc$, possibly with additional substructure of much smaller width $w$.  This leads to potentially large hierarchies $m\rSrc$ and $mw$.  Standard numerical approaches using uniformly spaced grids are ill-suited to such problems, as the number of grid-points must exceed the largest of these hierarchies.  Incorporating this hierarchy of scales accurately is extremely computationally burdensome even in cases of spherical symmetry, where the equations reduce to a single spatial dimension, and is practically intractable when symmetry assumptions are relaxed and additional spatial dimensions are required.  To overcome this challenge, nonuniform meshes are required, which cluster lattice points near narrow features (\ie boundary layers), while sparsely sampling regions where the solutions are smooth.  One particularly natural way to do this, which we explore in this paper, is to use finite-element methods. 

In this work, we develop a numerical code -- \phienics{} -- applying the finite element method to two case studies  of interest for Vainshtein screening: a theory with high-order derivative operators in the equation of motion and a second theory with nonlinear potential self-interactions combined with kinetic mixing.  The latter of these theories is (a slightly modified form of) the UV-complete Galileon, and the former is the resulting low-energy theory obtained when the heavy field is integrated out. This relationship is explored in more detail in Ref.~\cite{physics_paper}. However, for the purposes of this paper, these models provide two illustrations of different types of theory that exhibit screening, and so serve as good test beds to demonstrate the versatility of the \phienics{} code and the finite element method.

\phienics{} is developed for the specific purpose of providing prospective users with an easy-to-use framework to test their theories of screening.
The finite element method relies on an integral form of the field equations that is better equipped for high-derivative operators, as it helps to lower their order. Additionally, Neumann boundary conditions, typically placed at the origin, are natural in the formulation, without the need to impose them `by hand'.  These advantages are not unique to the finite element method, however, whose key advantage is its flexibility  to deal with complex geometries.  As such, the finite element method is employed in a number of fields in physics and engineering. Although, in this work, we restrict to spherical geometries where the solutions found by the code can be compared with analytic results under specific assumptions, the flexibility of the finite element method will allow the code to be straightforwardly extended to deal with less symmetric systems.  

This paper is structured as follows. In Sec.~\ref{Sec:Theories}, we briefly review the Galileon models we use to demonstrate the \phienics{} code, and the density profiles for the massive source adopted for this purpose. Section~\ref{Sec:numerical_method} describes the numerical method used to solve the equation(s) of motion for the scalar field(s), including a description of the meshes used. Section~\ref{Sec:Examples} gives examples of the successful working of the code, and of the types of results that can be obtained from it. We conclude in Sec.~\ref{sec_conclusion}. Further technical details of the code -- in particular, tests of numerical convergence -- are included as appendices.

Throughout, we work in units with $c=\hbar=1$, and use the reduced Planck mass $\mpl=1/\sqrt{8\pi G}$. Our metric convention is $(-,+,+,+)$. 

\phienics{} is available from \url{https://github.com/scaramouche-00/phienics}. It is developed in \texttt{Python}, with both \texttt{Python 2} and \texttt{3} currently supported\footnote{However, please note \texttt{Python 2} support will be discontinued.}. The code documentation is hosted at \url{https://phienics.readthedocs.io}.

\section{Massive Galileons and Vainshtein screening} \label{Sec:Theories}
In this section we briefly present our two case-study theories. They are chosen to demonstrate the ability of \phienics{} to deal with higher-derivative operators and the presence of multiple fields. Furthermore, these theories are related in the manner discussed in Ref.~\cite{deRham:2017imi,physics_paper}. The equations of motion and boundary conditions required to specify solutions are given in \secref{Sec:eqn-of-motion-and-bc}, and the source profiles around which we solve for the behaviour of the scalar fields are detailed in \secref{Sec:source-profiles}.

Most field theories must be viewed as low-energy effective field theories (EFTs), and scalar field theories used to explore screening are no exception.
To ensure the validity of any solution we have found, it is necessary to compute higher-order operators omitted in the low-energy description and ensure they are smaller than the terms included when defining the equations of motion.
The ability to compute these operators accurately is a key ability of \phienics{}. We introduce this aspect in more detail in \secref{Sec:theory-screening}.

\subsection{Equations of motion and boundary conditions} \label{Sec:eqn-of-motion-and-bc}
We solve for the behaviour of the Galileon fields(s) around a static, spherical compact source in Minkowski space-time, neglecting any space-time curvature coming from the presence of the source, or back-reaction due to the energy density stored in the scalar field profile.
At minimum, the compact source is characterised by an overall size $\rSrc$ and mass $\mSrc = 4\pi\int_0^{\infty} \udd{r} r^2\rho(r)$, where $\rho$ is the source's energy density.

We consider two models:

\begin{enumerate}
\item A theory of a massive scalar field characterised by higher-derivative operators with equation of motion
\begin{equation}
  \Box\fldOne - \mOne^2\fldOne - \paramNL \frac{\Box( (\Box\fldOne)^n )}{\UVScl^{3n-1}} = \frac{\rho}{\mpl} \, .
  \label{Eq:IR}
\end{equation}
Here $\fldOne$ is a massive scalar field (of mass $\mOne$), $\UVScl$ is a cut-off scale with units of mass, and 
$\rho$ is the energy density of the source object.
The dimensionless parameter $\paramNL$ regulates the strength of the nonlinear term, which is expected to give rise to Vainshtein screening. In the following, this theory will be referred to as the `{\bf single-field theory}'.

\item A theory of two coupled massive scalar fields with nonlinear self-interactions, originally proposed in \cite{deRham:2017imi} in the context of massive Galileons, with equations of motion

\begin{subequations}\label{Eq:UV}
\begin{align}
  \Box\fldLt - \mLt^2\fldLt - \kinmix\Box\fldHvy &= \frac{\rho}{\mpl}~,  \label{Eq:UVa} \\
  \Box\fldHvy - \mHvy^2\fldHvy - \kinmix\Box\fldLt - \frac{\higgsNL}{3!} \fldHvy^3 &= 0 \label{Eq:UVb} \, .
\end{align}
\end{subequations}
Here, $\fldLt$ is a light cosmological scalar field of mass $\mLt$, $\fldHvy$ is a self-interacting more massive scalar field of mass $\mHvy$, $\rho$ is the energy density of the compact source,
$\kinmix$ is the dimensionless coupling between the two scalar fields, and $\higgsNL$ is the dimensionless parameter controlling the strength of the self-coupling of $\fldHvy$.
Typically, $\mLt \ll \mHvy \ll \mSrc$. Additionally, $\mLt \rSrc \ll 1$, \ie the range of the fifth force is much greater than the characteristic size of the source. The Compton wavelength $\mHvy^{-1}$ of the heavy field $\fldHvy$ can be either greater or smaller than the source radius. In the following, this theory will be referred to as the `{\bf two-field theory}'.
\end{enumerate}

In Ref.~\cite{deRham:2017imi} the two-field model was shown to be UV complete, in the absence of the coupling to matter fields.
Including additional couplings to matter necessarily introduces higher order operators, and so may change the UV-complete nature of the theory. Also in Ref.~\cite{deRham:2017imi}, it was shown that integrating-out the heavy field $H$ from the two-field model leaves a single-field higher-derivative theory of the form of~\eqnref{Eq:IR} with $n=3$.
The single-field higher derivative model of~\eqnref{Eq:IR} is only valid up to the cut-off $\UVScl$. 
      
As the field configurations we consider are static, the d'Alembertian operator, $\Box$, becomes the Laplacian, $\nabla^2$. Further, for spherically symmetric systems considered in this work, the Laplacian becomes 
$\lapRad^2 \equiv \frac{1}{r^2}\frac{\partial}{\partial r}\left(r^2 \frac{\partial}{\partial r}\right)$. 
Under these symmetry assumptions,~\eqnref{Eq:IR} becomes:
\begin{equation}
\lapRad^2\fldOne - \mOne^2\fldOne - \paramNL\frac{\lapRad^2( (\lapRad^2\fldOne)^n )}{\UVScl^{3n-1}} = \frac{\rho}{\mpl} \, ,
\label{IR_EoM}
\end{equation}
and similarly~\eqnref{Eq:UV} becomes
\begin{subequations}
\begin{align}
   \lapRad^2\fldLt - \mLt^2\fldLt - \kinmix\lapRad^2\fldHvy &= \frac{\rho}{\mpl}~, \notag \\
   \lapRad^2\fldHvy - \mHvy^2\fldHvy - \kinmix\lapRad^2\fldLt - \frac{\higgsNL}{3!}\fldHvy^3 &= 0 \, .
\end{align}
\label{UV_EoM}
\end{subequations}

These equations must be supplemented by appropriate boundary conditions.  Since we are interested in static solutions, far from the source the fields must take values that minimise their potentials:
\begin{equation}\label{eqn:bc_infinity}
  \fldOne(\infty) = \fldLt(\infty) = \fldHvy(\infty) = 0 \, .
\end{equation}
It is standard to impose a regularity condition at the origin: 
\begin{equation}\label{eqn:bc_origin}
 \frac{\partial}{\partial r}\fldOne(0) = \frac{\partial}{\partial r}\fldLt(0) = \frac{\partial}{\partial r}\fldHvy(0) = 0 \, .
\end{equation}
Even though other choices are, in principle, possible, all known physical theories display this behaviour for extended sources.
This choice of Neumann boundary conditions avoids cusps or singularities at the origin, which would cause the energy density of the scalar field(s) to be infinite. Although diverging potentials are present around point sources in Newtonian gravity and electrostatics, these are always regulated when extended sources are considered.

Finally, the single-field theory is order four, and thus requires two additional boundary conditions.
As shown in~\secref{Sec:high_order_ops} and~\secref{Sec:IR_theory_weak_form}, it is convenient to think of $\lapRad^2\fldOne$ as an independent dynamical degree of freedom.
Far from the source, the fields $\fldOne,\fldLt$, and $\fldHvy$ will be close to their vacuum values.
Due to their nonzero masses, the fields decay as $\sim e^{-m_{\star} r}/r$ far from the source.  
This also constrains the behaviour of the field derivatives at infinity.
In the single-field theory, for example, $\frac{\partial}{\partial r}\pi, \lapRad^2\pi, \left[\left(\lapRad^2\fldOne\right)^n\right]$, and $\frac{\partial}{\partial r} \left[\left(\lapRad^2\fldOne\right)^n\right]$ all decay as $e^{-mr}/r$ or $e^{-3mr}/r^3$ for sufficiently large $r$. 
We therefore use
\begin{equation}
    \left[\left(\lapRad^2\fldOne\right)^n\right](\infty)=0~,
\end{equation}
as a second Dirichlet boundary condition for the single-field theory.

For the final boundary condition, we note the quantities $\lapRad^2\pi(0)$, $\left[\left(\lapRad^2\pi\right)^n\right](0)$, and $\frac{\partial}{\partial r} \left[\left(\lapRad^2\pi\right)^n\right](0)$ are all expected to be finite but nonzero \cite{physics_paper}.
We therefore demand $\frac{\partial}{\partial r} [\left(\lapRad^2\pi\right)^n](0) < \infty$ as a fourth boundary condition for the single-field theory. These boundary conditions are consistent with the requirements of the UV completion of the theory.

Summarising, the complete set of boundary conditions for the single-field theory is:
\begin{equation} \label{Eq:full_BC_IR}
\left\lbrace \pi(\infty)=0; \frac{\partial}{\partial r}\pi(0)=0;  \left[\left(\lapRad^2\pi\right)^n\right](\infty) = 0; \frac{\partial}{\partial r} \left[\left(\lapRad^2\pi\right)^n\right](0)<\infty \right\rbrace~,
\end{equation}
and 
\begin{equation} \label{Eq:full_BC_UV}
\left\lbrace \phi(\infty)=0; H(\infty)=0; \frac{\partial}{\partial r}\phi(0)=0; \frac{\partial}{\partial r} H(0)=0 \right\rbrace~,
\end{equation}
for the two-field theory.

\subsection{Source profiles} \label{Sec:source-profiles}
As stated above, we want to find static scalar field profiles around a static, spherically symmetric source of mass $\mSrc$ and characteristic radius $\rSrc$. 
For a radial source density profile given by a step-function $\rho(r)= (3\mSrc/4 \pi r_{\rm step}^3)\Theta(r_{\rm step}-r)$, analytic approximations allow us to estimate the form of the field profile. 
These analytic results provide a useful test of the code. 
For density profiles that vary with $r$, an analytic approximation to the form of the solution is not generally known. 

For illustration, we will consider three source profiles: a smoothed top-hat, a truncated cosine, and a `wedding cake' profile made up of three stacked Gaussians.
Each of these profiles has a natural intrinsic length scale: the location of the boundary layer for the smoothed top-hat, the period of the cosine, or the half-width of the widest Gaussian.  For the step profile mentioned above, this intrinsic scale is $r_{\rm step}$.
However, the dynamics we are interested in are sensitive to the total mass of the object, so we define $\rSrc$ as smallest radius enclosing $95$\% of the mass, rather than these intrinsic scales.

To relate the natural theory scales (denoted here by $r_{\rm th}$) to our definition of $\rSrc$, we introduce a rescaling parameter $\bar{t}$ so that $r_{\rm th} = \bar{t}\rSrc$.  Denoting a profile characterised by theory scale $r_{\rm th}$ as $\rho(r|r_{\rm th})$, we obtain $\nounits{t}$ by solving $4\pi \int_0^{\rSrc}\udd{r} r^2 \rho(r|\bar{t}\rSrc) = 0.95 \, \mSrc$.\footnote{We solve this using \texttt{Scipy}'s inbuilt Broyden method, with $\nounits{t}^{(0)}=1$ as an initial guess.}

\begin{figure}
    \centering
    \includegraphics[width=\textwidth]{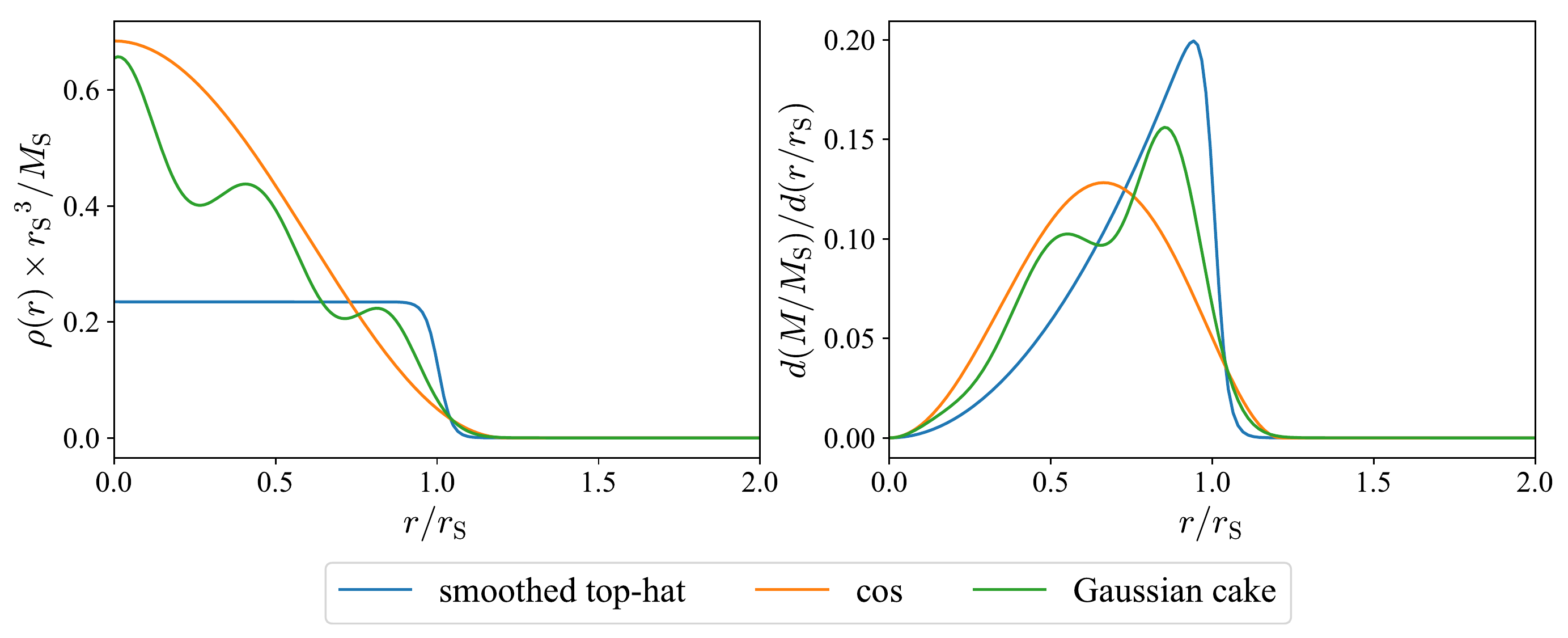}
    \caption{\textit{Left}: The dimensionless source profiles in Eqs.~\eqref{Eq:top_hat_source}, \eqref{Eq:cos_source}, and \eqref{Eq:GCake_source}. \textit{Right}: The dimensionless radial mass density $\left(\frac{\rSrc}{\mSrc}\right)\frac{dM}{dr} = 4\pi\rho \left(\frac{r}{\rSrc}\right)^2 \frac{\rSrc^3}{\mSrc}$ (\ie the mass of a spherical shell of radius $r$) for the same source profiles.  In both cases the mass and radial distance are normalised to the source mass and radius, respectively.}
    \label{Fig:source_profiles}
\end{figure}

The expressions for the three example density profiles, shown in~\figref{Fig:source_profiles}, are:
\begin{enumerate}
\item A smoothed top-hat profile:
\begin{align}
  \rho(r) &=  \frac{\mSrc}{ 4 \pi(-2\stepWid^3)\textrm{Li}_3(-e^{\nounits{t}\rSrc/\stepWid}) } \frac{1}{\exp{\frac{r - \nounits{t}\rSrc}{ \stepWid }} + 1 }  
\label{Eq:top_hat_source}
\end{align}
where $\textrm{Li}_3(x)$ is the polylogarithm function of order 3. In the limit $\frac{\stepWid}{\rSrc}\rightarrow 0$ the density profile in~\eqnref{Eq:top_hat_source} becomes a step function. In this limit, we find $\nounits{t}^{-3} = 0.95$, so $\rho(r)= \frac{3 \mSrc}{4 \pi (\nounits{t} \rSrc)^3}\Theta\left(\nounits{t}\,\rSrc-r\right) = 0.95\frac{3 \mSrc}{4 \pi \rSrc^3}\Theta\left(\nounits{t}\,\rSrc-r\right)$.\footnote{With our definition of source radius $\rSrc$ (\ie the radius enclosing 95\% of the source mass) the transition of the limiting top hat profile does not occur at $\rSrc$.}
Choosing a profile that is smoother than a step function avoids shocks in the derivatives of the true solution that are difficult to resolve. For this source profile, analytic solutions may be found in the limit of a step function and specific sub-cases.

\item A truncated cosine profile:
\begin{equation}
\rho(r) =  
\begin{dcases}
\frac{3\pi \mSrc}{4 (\nounits{t}\rSrc)^3(\pi^2-6)} \left[\cos\left(\pi\frac{r}{\nounits{t}\rSrc}\right)+1\right] & \text{if } r \leq \nounits{t} \rSrc \\
0 & \text{otherwise}
\end{dcases}
\label{Eq:cos_source}
\end{equation}
describing a more slowly varying source profile. Note that even with this simple source distribution, there is no known analytic solution for the field profile.

\item A `Gaussian wedding cake' profile, \ie a linear combination of three Gaussians making a three-layered shape:
\begin{equation}
    \rho(r) = \frac{\mSrc}{4 \pi (\nounits{t}\rSrc)^3 X}\left[A_1 e^{-\frac{1}{2}\left(\frac{r}{\sigma_1}\right)^2} + A_2 e^{-\frac{1}{2}\left(\frac{r-\mu_2}{\sigma_2}\right)^2} + A_3 e^{-\frac{1}{2}\left(\frac{r-\mu_3}{\sigma_3}\right)^2}\right] \, ,
    \label{Eq:GCake_source}
\end{equation}
with the parameters $\mu_i$, $\sigma_i$, and $A_i$ controlling to locations, widths, and relative heights of the peaks.
The normalisation factor $X$ is set by the constraint ${4\pi\int_0^\infty \ud r r^2\rho(r) = \mSrc}$.
For calculations in this paper, we take $\mu_2=\nounits{t}\rSrc/3$, $\mu_3=2 \, \nounits{t}\rSrc/3$, $\sigma_1=\nounits{t}\rSrc/9$, $\sigma_2=\nounits{t}\rSrc/7$, and $\sigma_3=\nounits{t}\rSrc/12$.  The dimensionless heights $A_{1,2,3}$ are chosen so $\rho(0)=3/2 \times \rho(\mu_2)=3 \times \rho(\mu_3)$ and $A_3=1$. 
The resulting normalisation factor is $X\approx 0.4186008$.
The form of this profile is chosen to illustrate the flexibility of the finite element method in solving for non-monotonic source profiles with internal structure, rather than to represent a physical system of interest.

\end{enumerate}

In~\eqnsref{Eq:top_hat_source},~\eqref{Eq:cos_source}, and \eqref{Eq:GCake_source}, the symbol $\pi=3.14159...$ is a normalisation parameter and does not represent the field $\pi$ in~\eqnref{Eq:IR}. 

Numerically, the smoothed-top hat profile is the most difficult to solve for, as it has a boundary layer, introducing an additional hierarchy. The truncated cosine has a discontinuous second derivative, which can cause numerical challenges. The `Gaussian wedding cake' profile is the easiest to obtain numerically, as it does not present discontinuities or internal structures on very different scales than its overall radius.  However, it is the most challenging profile to understand analytically.

The user can easily implement their favourite source profile by adding a subclass in the \texttt{source} module using the existing examples as a template.  
More details are given in \Appref{Sec:Code_structure} and the \phienics{} documentation.

\subsection{Screening and higher-order operators} \label{Sec:theory-screening}

To compare the relative strengths of the scalar and gravitational forces -- which in turn allows one to determine whether the theory displays Vainshtein screening -- \phienics{} computes the Newtonian potential from the Poisson equation
\begin{equation}
  \lapRad^2\Phi_{\rm N} = \frac{\rho}{2 \mpl^2} \, .
\end{equation}
To make a fair numerical comparison, we compute $\Phi_{\rm N}$ using the same mesh (see~\secref{Sec:FEM} for a definition) and source density profile used to compute the scalar field profile in the theory we are comparing to.
Note that when the scalar field couples to matter with a coupling strength $M=\mpl$, the ratio of the scalar fifth force to the Newtonian gravitational force around a compact object is two.

When considering a theory as a low-energy effective field theory, as we are here for the single-field model, one should check whether the assumption that higher-order operators can be neglected when computing field profiles is indeed valid.  
This is an important test of the validity of any solution. The specific scheme used in \phienics{} is especially optimised for the computation of such operators: 
\begin{equation}
    Q_n^{\mathrm{(single-field)}} \propto \frac{\paramNL^n}{\UVScl^{3n-1}} \lapRad^2( ( \lapRad^2\fldOne )^n )
    \label{Eq:Qn_IR}
\end{equation}
and
\begin{equation}
    Q_n^{\mathrm{(two-field)}} \propto \frac{\alpha^n}{\mHvy^{3n-1}} \lapRad^2( ( \lapRad^2\fldLt )^n )
    \label{Eq:Qn_UV}
\end{equation}

In~\eqnref{Eq:Qn_IR}, the dimensionless parameter $\paramNL$ controls the strength of the nonlinear terms in $\fldOne$. In Eq.~\eqref{Eq:Qn_UV}, the dimensionless parameter $\kinmix$ controls the effect that the heavy field $\fldHvy$ and its associated nonlinearities has on the light field $\fldLt$. 
Computing the operators in Eq.~\eqref{Eq:Qn_IR} and \eqref{Eq:Qn_UV} across the radial domain accurately requires considerably more care than the field profiles or gradients alone. We discuss this aspect in Sec.~\ref{Sec:high_order_ops}, where we describe how these operators are computed within the \phienics{} code.

\section{Numerical method}\label{Sec:numerical_method}
In this section, we describe our numerical approach to solve the equations of motion~\eqnref{IR_EoM} and~\eqnref{UV_EoM}, with boundary conditions~\eqnref{Eq:full_BC_IR} and~\eqnref{Eq:full_BC_UV}, respectively.
As analytic solutions are not generally known, numerical methods are essential.
A number of difficulties arise that any numerical method must address:
\begin{enumerate}
\item we are solving a nonlinear boundary value problem, requiring a high dimensional nonlinear solver;
\item there are large hierarchies in the problem, in particular:
\begin{itemize}
\item[--] the ratio of the overall source size to the Compton wavelength of the fields; and 
\item[--] the ratio of the overall source size to the size to the scales of its internal structure;
\end{itemize}
\item for the single-field model (\eqnref{IR_EoM}) and in the computation of the $Q_n$ operators (\eqnref{Eq:Qn_IR} and~\eqnref{Eq:Qn_UV}), we must deal with numerically ill-behaved high-order derivative operators.
\end{enumerate}

In the one-dimensional case considered here, a large hierarchy between the overall source radius and typical scales of its internal structure may lead to a similarly large hierarchy of scales in the structure of the sourced solutions.   For example, if the source term is constant except in a narrow transition region, we expect the field solutions to possess a boundary layer.
This is a manifestation of complex solution geometry derived from complexity in the source itself.
For the spherically symmetric case considered here, the possible geometric complexity is limited by the lack of angular degrees of freedom.
Absent these simplifying symmetries, solutions can be substantially more complex.
In these cases, the numerical grid used to compute the solution must be sufficiently flexible to capture the full structure of the field solutions.
A natural approach, which we will use here and outline below, is the finite element method.
For the simple spherically symmetric case studied here, more powerful techniques are available, such as globally based pseudospectral methods~\cite{boyd01,Bond:2015zfa}.
However, unlike finite element methods, these global methods can be difficult to generalise to more complex source geometries.
Therefore, in this paper we pursue a finite element approach, providing a simple test case in which to illustrate the utility of the methods, and providing a foundation on which to base further studies where the assumption of spherical symmetry is relaxed.

In the remainder of this section, we outline our methods to resolve each of these difficulties. First, in \secref{subsec:Newton-method} we describe the Newton method for solving the equations of motion, which addresses their nonlinear nature. Next, in \secref{Sec:FEM} we review our spatial discretisation scheme---the finite element method, briefly mentioning key features that allow us to address the issues of both large hierarchies and high-order derivatives. We continue by introducing the weak form of the differential equations in \secref{Sec:weak_form}. 
In the final two subsection, we outline our approach to deal with large hierarchies and high-order derivatives.
To tackle the large hierarchies, we apply nonlinear transformations of the integration domain, described in \secref{Sec:Meshes}.
The finite element method provides a natural way to implement such transformations.
Finally, we introduce a modified form of the equations of motion which is better suited to dealing with high derivative operators, presented in \secref{Sec:high_order_ops}.

\subsection{Nonlinear Solver: Newton Method} \label{subsec:Newton-method}
In this subsection, we discuss how we approach the nonlinearity of the equations of motion~\eqnref{IR_EoM} and~\eqnref{UV_EoM}.
Of course, we cannot solve the continuum problem on a finite computer, so we must solve a discrete version of the equations.
We use an (appropriately) discretised version of the Newton method applied to differential equations, which we briefly outline here.  
We restrict to a single differential equation, since the generalisation to multiple coupled differential equations is straightforward.
Conceptually, we want to solve for the equation of motion at each lattice site. However, in practice we will seek to find a solution of the equation of motion projected into a convenient discrete basis to be outlined in \secref{Sec:FEM}. Our nonlinear solver is thus a high dimensional root finding problem, seeking the coefficients of the expansion of the solution in our discrete basis.

We begin by expressing our differential equation in the form $\mathcal{F}[u] = 0$. We now assume we have an approximate solution $u^{(k)}$ and wish to find the true solution $u_{\rm true}$.
If our guess is sufficiently close, we are justified to expand in the difference between the current guess and the true solution
\begin{equation}\label{eqn:newton-frechet}
  0 = \mathcal{F}[(u_{\rm true}-u^{(k)})+u^{(k)}] \approx \mathcal{F}[u^{(k)}] + \int\ud{\bf x} \frac{\delta \mathcal{F}}{\delta u({\bf x})}[u^{(k)}]\left(u_{\rm true}-u^{(k)}\right)  \, .
\end{equation}
This gives us a linear equation for the difference $\delta u^{(k)} \equiv u_{\rm true}-u^{(k)}$, which is sourced by the violation of the equation of motion $-\mathcal{F}[u^{(k)}]$.
We then construct the next iterative approximation
\begin{equation}
  u^{(k+1)} = u^{(k)} + \delta u^{(k)} \, ,
\end{equation}
where $\delta u^{(k)}$ is the solution to~\eqnref{eqn:newton-frechet}.
Finally, integrating against $\frac{\delta\mathcal{F}}{\delta u({\bf x})}[u^{(k)}]$ and substituting~\eqref{eqn:newton-frechet}, we obtain an equation for $u^{(k+1)}$ directly in terms of $u^{(k)}$
\begin{equation}\label{eqn:newton-direct}
  \int\ud{\bf x}\frac{\delta\mathcal{F}}{\delta u({\bf x})}[u^{(k)}]u^{(k+1)}  = - \mathcal{F}[u^{(k)}]+ \int\ud{\bf x}\frac{\delta\mathcal{F}}{\delta u({\bf x})}[u^{(k)}]u^{(k)} \, .
\end{equation}
If the initial guess is sufficiently close to the correct solution, repeated iterations of this procedure will converge to a solution to the equations of motion.
Our discretisation procedure consists of first expressing these Newton iterations in weak form, and then discretising them using the finite element method. These aspects are discussed in \secref{Sec:FEM} and \secref{Sec:weak_form}.

Since we want a convergent solution to the equations of motion, there are two natural stopping criteria for our iterations:
\begin{enumerate}
\item Residual: Since a true solution to the equations of motion satisfies $\mathcal{F}[u_{\rm true}] = 0$, we require that the \textit{residual} $\mathcal{F}[u^{(k)}]$ satisfies 
\begin{equation}
 \norm{\mathcal{F}[u^{(k)}]} \leq \epsilon_{\rm rel}^{\rm (R)} \norm{\mathcal{F}[u^{(0)}]} + \epsilon_{\rm abs}^{\rm (R)},
 \label{Eq:Residual_criterion}
\end{equation}  
\item Change in the solution: As the sequence of solutions ${u^{(k)}}$ convergences, the change in the solution $\norm{u^{(k)} - u^{(k-1)}}$ tends to $0$. A convergence criterion can then be written as:
\begin{equation}  
  \norm{u^{(k)}-u^{(k-1)}} \leq \epsilon_{\rm rel}^{\rm (S)} \norm{u^{(0)}} + \epsilon_{\rm abs}^{\rm (S)}.
\label{Eq:Change_criterion}
\end{equation}
\end{enumerate}
In the above, $\epsilon_{\rm rel}^{\rm (\cdot)}$ and $\epsilon_{\rm abs}^{\rm (\cdot)}$ are user set relative and absolute tolerances, $u^{(0)}$ is the initial guess, and $\norm{\cdot}$ is some norm on the solution space.
Note that the tolerances $\epsilon_{\rm rel}^{\rm (\cdot)}$ and $\epsilon_{\rm abs}^{\rm (\cdot)}$ need not be the same for both criteria.
Relative criteria have the advantage that they do not depend on the characteristic size of $u$ (for the change in solution criterion) or that of the terms in the equations of motion (for the residual criterion). However, if the initial guess is very close to the actual solution, they may enforce an unreasonable number of iterations on the solver, potentially beyond the accuracy the scheme is actually capable of achieving. In this case, absolute criteria become helpful because they stop the iterative process once an absolute tolerance has been met. In the formulae shown, relative and absolute criteria have been combined to retain both advantages.  The specific norm used in \phienics{} will be explained in Sec.~\ref{Sec:Examples_weak_form}.

As is clear from the above derivation, the key ingredient for success of this method is to start with a sufficiently accurate initial guess.  Often, this is the most difficult step in the numerical solution of a nonlinear boundary value problem.  
Our approach for the screening problems studied in this paper is discussed in Sec.~\ref{Sec:IR_theory_weak_form}-\ref{Sec:UV_theory_weak_form}.
For the problems considered here, we find these guesses to be sufficient to guarantee convergence.
However, for other problems, additional tricks may be needed.
This may include expanding the basin of convergence of the nonlinear solver by combining a gradient descent with Newton iterations or allowing variable Newton steps.  It may also include improved initial guesses by considering a family of theories with an adjustable parameter that allows us to deform from some theory we can solve (either numerically or analytically) to the theory of interest.  Starting from the solvable theory, we can adjust the parameter in a series of steps, using the solution at each intermediate step as the initial guess of the next step.  Since these improvements were not needed for the problems studied here, we leave their implementation to future work.

\subsection{Spatial Discretisation: The Finite Element Method}
\label{Sec:FEM}

In order to obtain a discrete approximation to the Newton iterations in~\eqnref{eqn:newton-frechet}, we use the finite element method, which we briefly outline here.
For an accessible reference on the finite element method, see Ref.~\cite{langtangen2019introduction}.
As will be clear from the presentation, many implementation choices are possible, and we make use of the publicly available \fenics{} library\footnote{\url{https://fenicsproject.org/}} \cite{FEniCS_citations, Old_FEniCS_book} for the required functionality\footnote{The \fenics{} library was also used to study the phenomenology of screening in Refs.~\cite{Burrage:2017shh,Kuntz2019}.}.  To avoid unnecessary technical complications, here we restrict the discussion to implementation in one spatial dimension.

Our first step is to replace the infinite radial domain $[0,\infty)$ by a large but finite domain $[0,r_{\rm max}]$. 
In doing so, we must ensure that imposing the Dirichlet boundary condition at a finite radius (instead of infinity) does not distort the numerical solution.
In other words, we must ensure that the systematic error made by applying the Dirichlet boundary condition at a finite radius is small.\footnote{Another option is to expand in basis functions that are themselves defined on the semi-infinite interval, in which case the Dirichlet boundary condition can be imposed at infinity.} 
For the problems of interest in this paper, the Compton wavelength is the largest scale and therefore controls the fields' behaviour at infinity.
This means $r_{\rm max}$ must be orders of magnitude larger than the Compton wavelength of all the fields involved. 
To see this explicitly, we can approximate the profile $\psi$ of a field of mass $m$ far away from the source as $\psi(\rSrc)\left(\frac{\rSrc}{r}\right)e^{-m(r-\rSrc)}$, where $\psi(\rSrc)$ is the value at the source radius. In order for the true value of the field at $r_{\rm max}$, \ie $\psi_{\rm true}(r_{\rm max})$,
to be approximated by $0$, \ie the boundary value, this difference must be small. A good starting point is requiring $\psi_{\rm true}(r_{\rm max})/\psi(\rSrc) \ll 1$, from which it follows:
\begin{equation}
  \frac{r_{\rm max}}{\rSrc} + \frac{1}{m\rSrc}\ln\left(\frac{r_{\rm max}}{\rSrc}\right) \gtrsim 1 + \frac{1}{m\rSrc}\ln\left|\frac{\psi(\rSrc)}{\psi_{\rm true}(r_{\rm max})}\right| \, .
\end{equation}
This estimate can be improved by applying a similar condition to the derivatives appearing on the equation of motion. 
In practice, one needs to check that changing the specific value of $r_{\rm max}$ does not change the numerical solution. 
For the values of $r_{\rm max}$ used in production runs, we verified that the results were insensitive to the precise numerical value.

The next step, from which the finite element scheme derives its name, is to divide this large (but finite) interval into a collection of $N$ ``elements'', each storing a local representation of the functions defined on the interval.  
These elements are typically of varying sizes, and in two or more dimensions can even have varying geometries. 
These local function spaces are then stitched together to give a function defined on the whole domain.  
This is particularly powerful in more than one spatial dimension, where the cells can be joined together in highly nontrivial ways, allowing for an accurate representation of extremely complex geometries.  
In the one-dimensional case considered here, the adaptive nature of these elements provides a natural way to resolve the various hierarchies mentioned above.  To fully specify our discretisation, we are now left with a number of choices:
\begin{itemize}
  \item How do we choose the sizes of each element and distribute them through the domain?
  \item How do we represent a function (and its derivatives) within each element?
  \item How do we deal with the boundaries joining the elements together?
\end{itemize}

Our first step is to size our individual elements and position them in the domain.
For the one-dimensional case considered here, this amounts to deciding the location of the left ($r_{l,{\rm min}}$) and right ($r_{l,{\rm max}}$) endpoints of the $l$th element.
Equivalently, we may specify the length of each element and the position of its center.
A convenient approach, which we use here, is to use a monotonically increasing mapping function $T(x)$ from an underlying variable $x$ to the (dimensionless) radial variable $r$ appearing in our equations.
In other words, the radial element vertices $r_i$ are related to the element vertices $x_i$ in the original space by
\begin{equation}
 r_i = \rSrc T(x) \, .
\end{equation}
If we uniformly size the elements in the $x$ space, then the size of the element $\Delta r$ at radial position $r$ will be approximately
\begin{equation}
  \Delta r(r) \approx \Delta x \frac{\partial T}{\partial x}(T^{-1}(r)) \, .
\end{equation}
More details of the specific mapping functions are provided in~\secref{Sec:Meshes}.

\begin{figure}\label{fig:grid-figure}
\centering
  \includegraphics[width=0.7\textwidth]{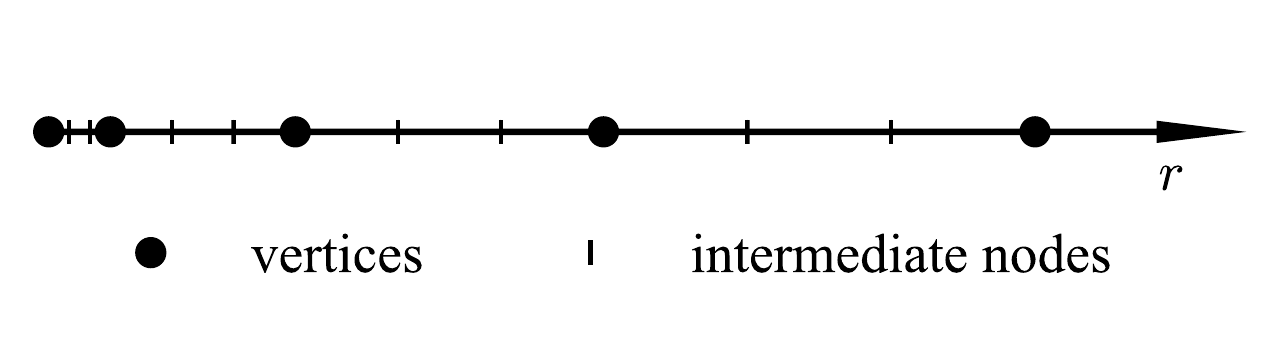}
  \caption{An illustration of a finite element discretisation with $4$ elements. In this example, the degree of interpolating polynomials is $3$.}
\end{figure}

Next we choose our local representation for functions within each element.
A plethora of choices are available, and we follow the prescription used in \fenics, which we outline below.
This gives us access to a powerful library for the functionalities of the finite element method.
Within each element, we choose a set of basis functions $B_i$ and locally expand the function
\begin{equation}
  f_l(r) = \sum_{i=0}^{O}c_iB_i(r)\, , \qquad r \in [r_{l,{\rm min}},r_{l,{\rm max}}]
\end{equation}
where the index $l$ labels the element, and we have expanded in a basis of order $O$.
$f_l(r)$ indicates the restriction of the function $f$ to the element $l$.
From this expansion, numerical derivatives are defined by
\begin{equation} \label{Eq:FEM_derivatives}
  \frac{\ud^n}{\ud r^n}f_l(r) = \sum_{i=0}^{O}c_i\frac{\ud^n}{\ud r^n}B_i(r) \, .
\end{equation}
Note, however, that if the expansion is truncated at low-order, higher-order numerical derivatives may poorly approximate the true derivative of the function.
To choose a set of basis functions $B_i$ and interpolation order $O$, we use Lagrange polynomials, which are specified by choosing a set of $O+1$ collocation points $r_i$ for $i=0,\dots O$.
In the \fenics{} library, two points are given by the cell extrema, whereas the other $O-1$ points are taken to be uniformly spaced within each element.
A simple collocation grid is illustrated in~\figref{fig:grid-figure}.

To obtain a global approximation for a function we must join neighbouring elements together.  For a connected domain in one-dimension, we take 
\begin{equation}
  r_{l+1,{\rm min}} = r_{l,{\rm max}}~,
\end{equation}
for $l=1,\dots \nEl-1$ where $\nEl$ is the number of elements, and
\begin{equation}
  r_{1,{\rm min}} = 0 ,\quad r_{N,{\rm max}} = r_{\rm max} \, ,
\end{equation}
for the boundary elements.
At this point in the construction, the functions are piecewise continuous, but discontinuities can arise at the element boundaries.  If we wish to work in a continuous function space, we make use of the fact that the element boundaries are collocation points, and impose an additional global constraint that function values at the cell boundaries are equal in neighbouring cells.  
For a discretisation using $\nEl$ elements, each with an order $p$ spectral approximation, there are then $\nEl (p+1)$ independent degrees of freedom for the discontinuous function space, and $\nEl p + 1$ independent degrees of freedom for the piece-wise continuous function space.
As explained in Sec.~\ref{Sec:cont_vs_disc}, at various points we will find it advantageous to either impose global continuity or to work within a piece-wise continuous function space.
  
\subsection{Weak Form of Differential Equations}
\label{Sec:weak_form}
The finite element method naturally lends itself to the integral (weak) form of the field equations.
Consider a differential operator $\mathcal{D}[u]$ acting on a function $u$, and the corresponding equation $\mathcal{D}[u] = 0$. In our case, this equation is a Newton iteration as in Eq.~\eqref{eqn:newton-frechet}. We consider the inner product $\left\langle\mathcal{D},v\right\rangle$ between $\mathcal{D}$ and a test function $v$ selected from some test function space.
A weak solution to the differential equation is one for which this inner product vanishes for \emph{any} function in the test space.
This is, of course, a weaker condition than demanding that the equation $\mathcal{D}[u(x)] = 0$ holds at every point $x$ (\ie the strong form of the equation).
As a result, weak solutions can accommodate the presence of shocks and point-wise discontinuities in solutions. As test space, we use the Sobolev space $H^1$ of square-integrable functions in $\mathbb{R}^3$ whose gradient is also square-integrable in $\mathbb{R}^3$.
Numerically, we approximate this (infinite dimensional) test space by the same basis of finite elements used to expand our solution.
Systems of equations can be expressed as components of a single vector equation, with the scalar product taken with respect to a vector of test functions.

Since we work with spherically symmetric solutions in three spatial dimensions, it is reasonable to use the inner product
\begin{equation}
\langle f,g\rangle = \int_0^{\infty} \ud r r^2 fg  \,  .
\end{equation}
Note that by introducing the $r^2$ weight into our inner product, we are effectively looking for weak solutions to $r^2\mathcal{D}$.
Without loss of generality, we can assume our test functions vanish at any points where we have specified Dirichlet boundary conditions.
Inhomogeneous boundary conditions can be converted into homogeneous boundary conditions by writing $u = u_{\rm bc} + \Delta u$, where $u_{\rm bc}$ is some function satisfying the inhomogeneous boundary conditions.  We then solve for $\Delta u$ instead, which lives in a function space satisfying homogeneous boundary conditions.
The integral projections appearing in the weak formulation must be estimated numerically.
These are approximated internally by \fenics{} using the \texttt{assemble} method, which implements accurate quadrature or tensor methods to numerical integration. 

We can also use the weak form of the differential equation to reduce the order of differential operators.
Consider, for example, the radial Laplacian of a function
\begin{equation} \label{Eq:int_by_part}
  \inner{v}{\lapRad^2 f} = \int_0^\infty \ud r\, r^2 v\frac{1}{r^2}\frac{\partial}{\partial r}\left(r^2\frac{\partial f}{\partial r}\right) = -\inner{\lapRad v}{\lapRad f} + \left[r^2v\frac{\partial f}{\partial r}\right]_0^\infty \, .
\end{equation} 
Therefore, up to a boundary term, we can swap projections of $\lapRad^2f$ onto the test function $v$ with projections of $\lapRad f$ with the gradient $\lapRad v$. For our set of boundary conditions~\eqnref{Eq:full_BC_IR} and Eq.~\eqref{Eq:full_BC_UV}, the boundary term vanishes. We will make use of this several times in~\secref{Sec:Examples_weak_form}.

\subsection{Mesh Construction: Distributing and Sizing the Elements}
\label{Sec:Meshes}

To accurately capture the large hierarchies present in theories of screening, we take advantage of the flexibility of the finite element method to specify non-uniform spatial discretisations.
As explained above, for an $\nEl$ element discretisation on a connected one-dimensional domain, this is accomplished by specifying a collection of $\nEl+1$ element vertices.
In what follows, we denote this collection of vertices (which define the lengths and locations of the elements) a mesh.
To generate a radial mesh, we begin with a uniform mesh on the interval $[x_{\rm min},x_{\rm max}]\subset\mathbb{R}$.
We then introduce a nonlinear map from this uniform mesh into our radial coordinates
\begin{equation}
  r_i = T(x_i) \ \mathrm{for} \ x_i = x_{\rm min} + i\left(\frac{x_{\rm max}-x_{\rm min}}{\nEl}\right) \, ,\ i=0,\dots, \nEl \, .
\end{equation}
Finally, the left and right boundaries of element $l$ are identified as $r_{l,{\rm min}}=r_{l-1}$ and $r_{l,{\rm max}} = r_{l}$, where $l=1,\dots, \nEl$ labels the elements.
A subtle point (discussed in \secref{sec:expMesh}, \secref{sec:powMesh}, and \secref{sec:decluster} below), is that the boundaries of the preimage uniform mesh can depend on the map $T$ and the outer radial boundary of the final mesh $r_{\rm max}$.

Our goal is to create a mesh that is finest around the source-vacuum transition (where the sharpest variation in the field profiles are expected) and very coarse far from the source (where the solutions slowly decay with a characteristic scale given by the Compton wavelength).
This allows us to represent a large simulation box and multiple large hierarchies with a reasonable number of points.

The mapping functions we introduce below have a non-monotonic radial vertex density, with the maximal point density (and thus the greatest spatial resolution) occurring at some radius $r_{\rm mpd}$ in the interior of the interval.  It is often convenient to fix the radial position of this maximal vertex density to the location where either the source or solution is most rapidly varying.  A common example occurs when the source profile includes sharp features, such as the top hat, which manifest as corresponding sharp features in the field profiles.
Because our inner radial vertex is zero, one strategy to reposition $r_{\rm mpd}$ is to simply rescale all of the radial points using a new transform $T_{\rm scl}(x) = T(x)/c$.
The scale $c$ is obtained by solving $T''(x_{\rm mpd}) = 0$ for $x_{\rm mpd}$ and then setting $c = T(x_{\rm mpd})/r_{\rm mpd}$.
The extrema of the original $x$ mesh are then calculated to ensure the size of the radial domain is unchanged.
For convenience, this procedure is included in \phienics\footnote{Our implementation in \phienics{} assumes that $r_{\rm mpd} = \rSrc$.  We find this is useful for the smoothed top-hat function, but not the Gaussian wedding cake or truncated cosine profiles.  However, in more general problems it may be convenient to shift the location of maximal resolution, so we have introduced $r_{\rm mpd}$ as a free parameter in the above discussion.}.

Additionally, we want to avoid clustering too many points near the origin, which can lead to either additional numerical noise or slower convergence as the number of elements is increased.
We find enforcing linear scaling of the map near the origin is sufficient to avoid these numerical issues.

After building a nonlinear mesh, it is sometimes necessary to either increase or decrease the density of mesh vertices in some regions.  \phienics{} includes built-in functionality for each of these tasks.  In the former case, targeted refinement of the mesh can be achieved by splitting elements in half within specified radial intervals, thus increasing the number of elements.
For the former case, \phienics{} provides functionality to repel (or decluster) element vertices from a specific radius.   Unlike the refinement algorithm, the declustering algorithm does not alter the number of mesh vertices.  This algorithm is described in sub-section \ref{sec:decluster}.
After producing the mesh, we always check that mesh vertices are sufficiently spaced to be resolved well within machine precision (\ie $10^4$ times the \texttt{FEniCS} machine precision of $3\times 10^{-16}$).

We now describe the two parametrised nonlinear transformations $T(x)$ used in this paper, as well as additional modifications that may be useful in specific problems. 
Note that parameter choices that lead to small residuals when solving the equations of motion may lead to noisy calculations of the operators $Q_n$ in post-processing.
This is likely because when more nodes are placed than are necessary to capture the relevant modes in a function, the only effect is to enhance round-off error by summing additional modes that are unresolved by machine-precision. The interested reader can find more details in \Appref{Sec:Tests}.

Users can implement their own choice of transformation, which may make use of \phienics{}'s linear refinement and point removal features. This is done by adding a custom mesh subclass, following the examples of the \meshExp{} and \meshPow{} subclasses included with \phienics.
These two transformations are outlined in this section. Further details are given in \Appref{Sec:Code_structure}.

\subsubsection{\meshExp{} mesh}\label{sec:expMesh}
\begin{figure}[h!]
\centering
\includegraphics[width=0.9\textwidth]{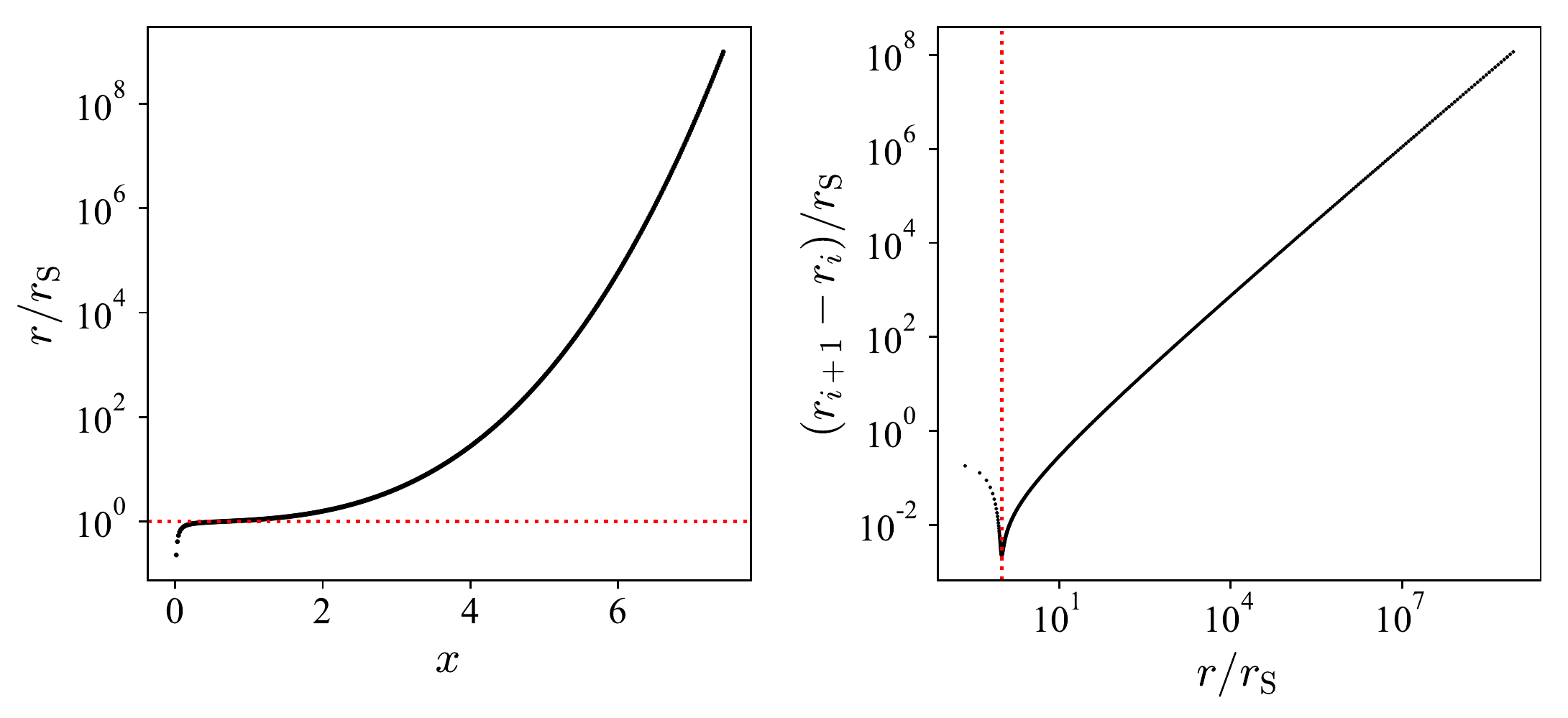}
\caption{\emph{Left:} A sample distribution of element vertices $r_i$ for the \meshExp{} mesh (\eqnref{Eq:ATExpMesh}) on a truncated domain with $r_{\rm max}=10^9\rSrc$ and $500$ elements.  The mapping parameters are $k=25$, $a=5\times10^{-2}$, and $b=3\times10^{-2}$. \emph{Right}: The corresponding lengths of each element. In both panels, the red dotted line indicates the source radius.}
\label{Fig:ATExpMesh}
\end{figure}

This mesh is generated by the nonlinear transformation:
\begin{equation}
  r = T(x) = \rSrc\frac{2}{\pi}\arctan{(kx)} \exp{\left(a x^3 + b x\right)}~,
  \label{Eq:ATExpMesh}
\end{equation}
where $k>0, a,b\geq0$, and $a$ and $b$ are not simultaneously zero.
\Figref{Fig:ATExpMesh} gives an example of the resulting element vertices.

The $\arctan$ factor regulates vertex density around the origin while maintaining linear scaling. Meanwhile, the cubic exponential factor clusters vertices around the source radius $\rSrc$, while spreading them out at large radii.  The specific point density at $\rSrc$ can be tuned using the parameters $a$ and $b$. The flatter the transformation at $T^{-1}(\rSrc)$, the denser the number of points at $\rSrc$, so that higher values of $b/a$ make it less dense.
Note that $b$ and $a$ also control the spacing at large radii, which can lead to a poor sampling of the Yukawa suppressed regime for extreme parameter values.

The extrema of the starting linear mesh $x_{\rm min}=T^{-1}(r_{\rm min})$ and $x_{\rm max}=T^{-1}(r_{\rm max})$ are obtained by solving $T(x_*)-r_*=0$ numerically, with $*={\rm min,max}$, using \texttt{Scipy}'s implementation of the Newton method. To obtain an initial guess for the algorithm, we approximate the inverse transformation $T^{-1}$ as linear around the origin, and take
\begin{equation}
  x_{\rm min}^{(0)} = \frac{r_{\rm min}}{T^{\prime}(0)} \, . 
\end{equation}
For the outer boundary, we approximate $T(x) \approx \rSrc e^{ax+bx^3}$.
If $a\neq 0$, we obtain a guess for $x_{\rm max}^{(0)}$ using Cardano's formula for cubic equations.
We define $p\equiv b/a$,
$q(r) = \frac{1}{a}\log{\left(\frac{\rSrc}{r}\right)}$
, and $D(r) = ( q(r)/2 )^2 + (p/3)^3$.
Our initial guess for $x_{\rm max}$ is then:
\begin{equation}
  x_{\rm max}^{(0)} =
  \sqrt[3]{ -\frac{q(r_{\rm max})}{2} + \sqrt{D(r_{\rm max})}} + \sqrt[3]{ -\frac{q(r_{\rm max})}{2} - \sqrt{D(r_{\rm max}) } } \, .
\end{equation}
Cardano's formula is justified because the exponential term in the transformation is monotonic, so there is only one real root to $T(x)-r=0$.
When $b\neq0$ and $a=0$, we instead use $x_{\rm max}^{(0)} = \frac{1}{b}\log\left(\frac{r_{\rm max}}{\rSrc}\right)$.

\subsubsection{\meshPow{} mesh}\label{sec:powMesh}
\begin{figure}[h!]
\centering
\includegraphics[width=0.9\textwidth]{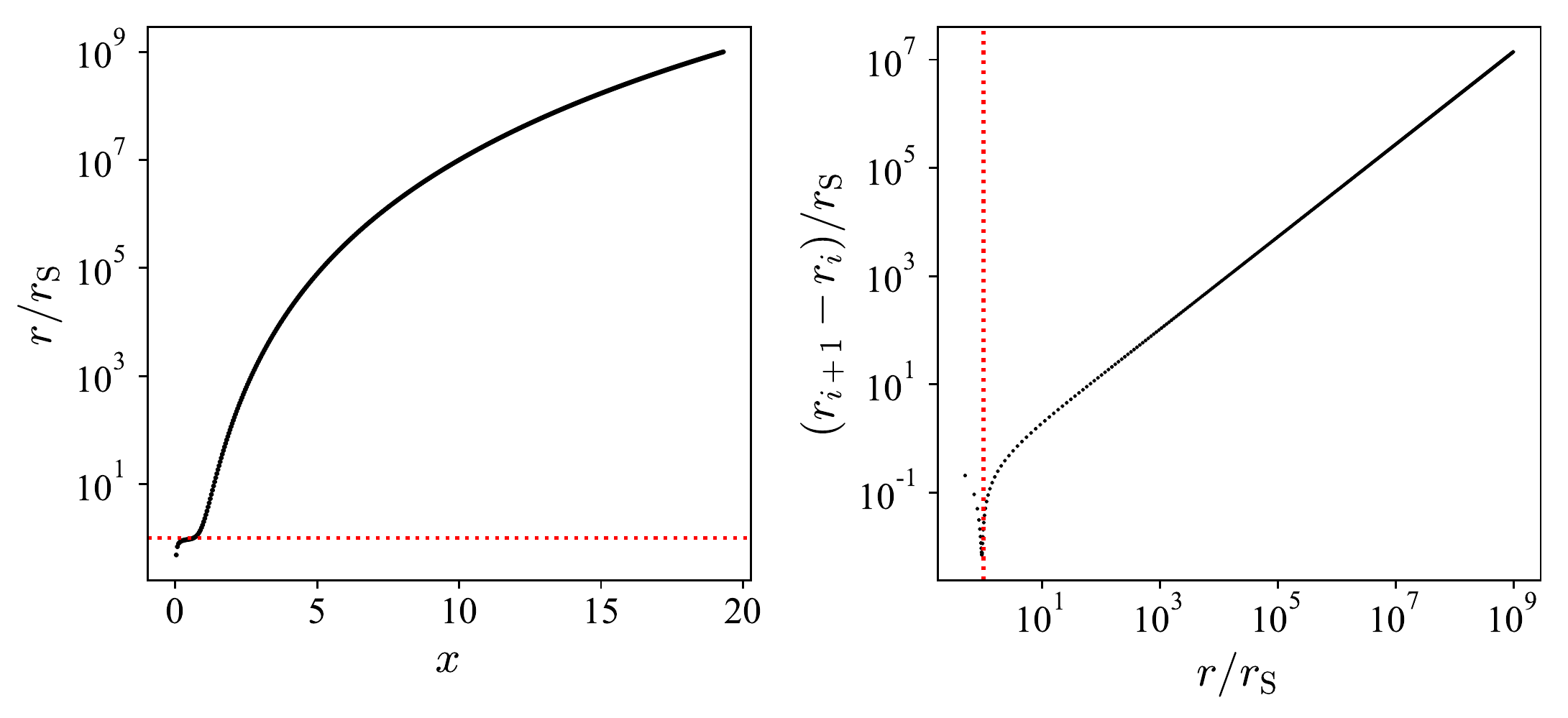}
\caption{\emph{Left}: A sample distribution of element vertices $r_i$ for the \meshPow{} mesh (\eqnref{Eq:ATPLMesh}), with mapping parameters $k=25, \gamma=7$, and $r_{\rm max}=10^9 \rSrc$. The finite element decomposition consists of $500$ elements. \emph{Right}: The corresponding lengths of each element.  In both panels, the red dotted line indicates the source radius.}
\label{Fig:ATPLMesh}
\end{figure}

This mesh is generated by the transformation:
\begin{equation}
  T(x) = \rSrc \left(\frac{2}{\pi} \arctan{ (kx) } + x^{\gamma}\right)~,
\label{Eq:ATPLMesh}
\end{equation}
where $k>0$ and $\gamma \geq 1$.
This transformation is visualised in \figref{Fig:ATPLMesh}.

As in \eqnref{Eq:ATExpMesh}, the $\arctan$ term allows us to enforce linear scaling at the origin, which avoids  placing an unnecessary number of points at $r=0$. Compared to the previous mesh, this transformation distributes points more uniformly across the box, therefore placing a larger number of points at larger radii.

As in the previous case, we obtain the extrema of the starting mesh by using the Newton's method. At small radii, it is still a good approximation to describe the inverse transformation as linear around the origin, whereas at large $r$, where the power-law term dominates, $x\approx (r/\rSrc-1)^{1/\gamma}$. Therefore we take
\begin{equation}
  x_{\rm min}^{(0)}=\frac{r_{\rm min}}{T^{\prime}(0)}~,
\end{equation}
and
\begin{equation}
  x_{\rm max}^{(0)}=\left(\frac{r_{\rm max}}{\rSrc}-1\right)^{1/\gamma}~,
\end{equation}
as initial guesses for the algorithm.

\subsubsection{Linear Refinement and Vertex Declustering}
\label{sec:decluster}
For all choices of mesh, \phienics{} allows the mesh to be linearly refined in a specified radial domain, which allows the user to further resolve specific features. This is obtained by using \texttt{FEniCS}'s inbuilt \texttt{refine} method, which halves cells in a specified interval $[r_{\rm start}, r_{\rm stop}]$. This procedure is repeated a user-set number of times.

Occasionally, numerical noise in specific parts of the domain, particularly at the Vainshtein radius where the nonlinear operators briefly go to $0$, may propagate into surrounding regions due to the nonlinear nature of the equations of motion.
This leakage results in numerical inaccuracies in the computed solution.
To alleviate this problem, \phienics{} includes an option to decluster points at some radius of interest $r_{\rm rm}$.

The point declustering is implemented as follows.\footnote{We found that this algorithm gave better results than fusing cells together.}
The user provides a radial declustering point $r_{\rm rm}$, declustering parameter $k_{\rm rm}$, and radial extent $A_{\rm rm}$.
Given a nonlinear transformation $T$, such as \eqnref{Eq:ATExpMesh} or \eqnref{Eq:ATPLMesh}, we define a new transformation $T_*$:
\begin{equation}
    T_*(x) \equiv \frac{T(x)}{c} \, T_{\rm rm}(x) \, ,
    \label{Eq:Tstar}
\end{equation}
where
\begin{equation}
    T_{\rm rm}(x) = \frac{A_{\rm rm}}{\pi} \arctan{\left(k_{\rm rm}\left(x-x_{\rm rm}\right)\right)} + 1 + \frac{A_{\rm rm}}{2}~.
\end{equation}
In this expression, $x_{\rm rm}$ is a derived parameter determined from the user-supplied declustering radius via $r_{\rm rm} = T_{*}(x_{\rm rm})$.

The new function $T_{\rm rm}$ adjusts the slope of the transformation $T_*$ at $r_{\rm rm}$. For the \meshExp{} transformation  \eqnref{Eq:ATExpMesh}, choosing $k_{\rm rm} \gtrsim k_c \equiv 3\sqrt[3]{a} \log{(r_{\rm rm}/\rSrc)}^{2/3} + b$ decreases the slope, whereas $k_{\rm rm} \lesssim - k_c$ increases it; for the \meshPow{} transformation \eqnref{Eq:ATPLMesh}, $k_{\rm rm} \gtrsim k_c \equiv \left(\frac{1 + A_{\rm rm}/2}{r_{\rm rm}/\rSrc} \right)^{1/\gamma} \sqrt{ \gamma^2 - 1}$ decreases the slope, whereas  $k_{\rm rm} \lesssim - k_c$ increases it.\footnote{Please note these are rules of thumb only. Also, for $|k| \lesssim k_c$, the slope is changed, but not enough to place a new local maximum or minimum in the vertex distance.} In the former case, points are repelled (\ie declustered) from the location $x_{\rm rm}$ leading to decreased resolution, while in the latter they are attracted (\ie clustered) leading to increased resolution.  The user-supplied parameter $A_{\rm rm}$ sets the radial extent of the (de)clustering effect. Although clustering and declustering are both possible with this algorithm, we note that it was designed with declustering at the Vainshtein radius in mind, which is its main use.

Introducing the factor $T_{\rm rm}$ generically shifts the radial location with the densest clustering of points. We counteract this unwanted effect by introducing the scaling parameter $c$ in Eq.~\eqref{Eq:Tstar}, as discussed earlier in this section. 
We want to ensure that the maximal vertex density occurs at the source radius while simultaneously keeping the declustering/clustering at $r_{\rm rm}$.  This gives us the following three conditions: $T_*^{\prime\prime}(x_{\rm s})=0$ at $x_{\rm s}=T_*^{-1}(\rSrc)$, and $T_*(x_{\rm rm})=r_{\rm rm}$ for the unknown variables $\{x_{\rm s},x_{\rm rm},c\}$. We can solve explicitly for $c=T(x_{\rm rm})\left( 1+\frac{A_{\rm rm}}{2} \right)/r_{\rm rm}$, leaving two equations:
\begin{subequations}
 \label{Eq:system_declustering}
\begin{align}
   T_*(x_{\rm s}) & = \rSrc~, \\ 
   T_*^{\prime\prime}(x_{\rm s}) &= 0 \, ,
\end{align}
\end{subequations}
in two unknowns $\{x_{\rm s}, x_{\rm rm}\}$. We solve Eqns.~\eqref{Eq:system_declustering} by using \texttt{Scipy}'s built-in \texttt{root} method, which applies a modification of the Powell hybrid method \cite{MINPACK}.

If the user implements clustering at $r_{\rm rm}$ (\ie negative $k_{\rm rm}$), it is important to check that $T^\prime_*$ remains positive around $x_{\rm rm}$, so the transformation $T_*$ is monotonic.
We automate this check in \phienics{} by resampling $T^\prime_{*}$ in the vicinity of $x_{\rm rm}$ and explicitly checking that it remains positive.\footnote{To resample, we first define new coordinates $y$ via $x(y) = \sqrt{1 + A_{\rm rm}/2} \tan{(y/10)} / | k_{\rm rm}| + x_{\rm rm}$.  We then select eleven resampling points $y_i$ uniformly distributed in the interval $y \in[-10 \arctan{(1)},-10 \arctan{(1)}]$.}

The remaining task is to update the extrema of the starting mesh, which will generally shift. After updating the derivative of the transformation as $T_*^{\prime}(x) = T^{\prime}(x) T_{\rm rm}(x) /c + T(x) {T_{\rm rm}}^{\prime}(x) / c$, we repeat the procedure of solving $T_*(x_{\rm min,max}) = r_{\rm min,max}$ using the Newton method.

\subsection{Treatment of High-Order Derivative Operators}
\label{Sec:high_order_ops}
A key challenge in the single-field theory (Eq.~\eqref{Eq:IR}) is to compute the highly nonlinear term $\lapRad^2( (\lapRad^2\pi)^n )$ accurately. A similar challenge is encountered in the computation of the $Q_n$ operators \eqnref{Eq:Qn_IR} and \eqnref{Eq:Qn_UV}, which encode corrections to the effective IR theory. 

To address the former challenge, we write the equations of motion as a system of three lower-order equations
\begin{align}
  \left(\begin{array}{c} Y-\mOne^2\fldOne -\epsilon\lapRad^2 W \\ Y \\ W \end{array}\right)
  =
  \left(\begin{array}{c} \frac{\rho}{\mpl} \\\lapRad^2\fldOne  \\ Y^n \end{array} \right)~,
  \label{Eq:IR_strong_system}
\end{align}
through the introduction of the auxiliary variables $W$ and $Y$.

Similarly, we re-express the system of equations of the two-field theory
\begin{align}
  \left(\begin{array}{c} Y - \mLt^2\fldLt - \kinmix Z \\ Z-\mHvy\fldHvy - \kinmix Y - \frac{\higgsNL}{6}\fldHvy^3 \\ Y \\ Z \end{array}\right)
  =
  \left(\begin{array}{c} \frac{\rho}{\mpl} \\ 0 \\ \lapRad^2\fldLt \\ \lapRad^2\fldHvy \end{array}\right) \, ,
\label{Eq:UV_strong_system}
\end{align}
through the introduction of the auxiliary variables $Y$ and $Z$.
For the two-field theory, this formulation is not needed to compute the field profiles.
However, we find that introducing $Y$ as an independent variable improves the calculation of the operators $Q_n$ in post-processing.
This is most likely because we have introduced additional degrees of freedom to resolve structure in the Laplacian, which are subsequently constrained by the equations of motion.
The dimensionless units used in our numerics are presented in \Appref{Rescalings}. 
However, in this section we express the equations in their native form for readability.

In the single-field theory, the introduction of the the auxiliary variable  is sufficient to accurately compute the nonlinear term in the equations of motion.
Meanwhile, the $Q_n$ operators obtained in post-processing require some more work, as they are not constrained by the equations of motion, with the exception of the $Q_3^{\rm (single-field)} \propto \lapRad^2W$ term in the single-field theory.

We begin by expanding
\begin{equation}
  Q_n \propto \lapRad^2(Y^n) = nY^{n-1}\lapRad^2Y + n(n-1)Y^{n-2}\lapRad Y\cdot\lapRad Y \, ,
\label{Eq:Qn_expanded}
\end{equation}
and we use the right-hand side of this expression for numerical evaluations.
Furthermore, we project this expression onto the same function space as $Y$.
Using some simple test profiles for $Y$, we verified that this yielded more accurate results than directly evaluating the left-hand side.
We also did heuristic checks of accuracy by comparing numerical calculations of the left- and right-sides of \eqnref{Eq:Qn_expanded}.
The most likely reason for the improved accuracy is that the Laplacian upweights the short-wavelength (\ie high-order polynomial) modes, which can amplify both short-distance numerical noise and aliased short-wavelength power generated by the high degree of nonlinearity.

High orders of nonlinearity introduce one additional subtlety.  The most natural set of units to evaluate the equations of motion do not necessarily lead to dimensionless $Y$'s that are order one.  As a result, high-order powers may either underflow or overflow floating point precision.  Moreover, numerical aliasing may become more pronounced as $Y$ increases due to the nonlinearity.
Therefore, for practical computations it is convenient to first compute the rescaled operator
  \begin{equation}
    Q_n^K = \lapRad^2\left(\left(K\lapRad^2\phi\right)^n\right) \, ,
  \end{equation}
for some $K$. From this we obtain
\begin{equation}
    Q_n^{\rm num} = Q_n^{K}/K^n \, .
\end{equation}
Note that this is equivalent to working with a different set of of dimensionless units.

\section{Example Solutions} \label{Sec:Examples}

In this section, we present results from \phienics{} for three representative models --- one for the single-field theory and two for the two-field theory.  First, we derive the weak form of the Newton iterations in \secref{Sec:Examples_weak_form}.  This serves as a worked example for future \phienics{} users.  We also outline some additional numerical details specific to each example.  \Secref{Sec:Results} presents the numerical solutions computed by \phienics.  Finally, we compare our numerical results to theoretical predictions in \secref{Sec:Theo_tests}.  For the interested reader, numerical convergence tests are presented in \Appref{Sec:Tests}.  These tests provide an important test of our numerics, and are meant as a basic template for future \phienics{} users to ensure accuracy of their results.

In our single-field example, we consider the field behaviour around a source of large mass similar to the star Antares.  This example allows us to explore the behaviour of a nonlinear, high-derivative theory in the presence of a realistic source.  We choose $\mOne$ so that the Compton wavelength is much longer than the Vainshtein radius, which is typically the case of physical interest.

For the two-field theory, we consider two cases.  In the first, the Compton wavelengths of both scalar fields are comparable to the size of the source.  We expect features to be excited by the source-vacuum transition, with a characteristic size dependent on the width of the transition region. 
Characterising these features analytically, or even numerically, is challenging, so this model is a good testing ground for our numerical method. 

In the second two-field model, we instead consider a different hierarchy of scales: $\mLt\rSrc \ll \mHvy\rSrc \ll 1$. This allows us to test the regime in which the Compton wavelength of the heavy field is larger than the source (which results in different numerical and phenomenological behaviour). This hierarchy also allows us to demonstrate the behaviour of extremely light fields $\phi$, arguably the regime of greatest interest for these theories.

The parameters specifying our three models are:
\begin{enumerate}[label={(M\arabic*)},leftmargin=1.25cm]
    \item Single-field theory; the model parameters are ${\mOne=10^{-50}\mpl}$, ${\paramNL=3\times 10^{-3}}$, $\UVScl=10^{-39}\mpl$, and ${n=3}$.   The source has mass ${\mSrc=5\times 10^{39}\mpl}$ and radius $\rSrc=7\times 10^{45}\mpl^{-1}$. \label{Ex:Antares}
    \item Two-field theory; the model parameters are $\mLt=10^{-48} \mpl$, $\mHvy=10^{-46} \mpl$, ${\alpha=0.4}$, and ${\lambda=0.7}$. The source has mass ${\mSrc=10^{10} \mpl}$ and radius ${\rSrc=10^{47}\mpl^{-1}}$. \label{Ex:Wiggle_model}
    \item Two-field theory; the model parameters are $\mLt=10^{-51} \mpl$, $\mHvy=10^{-48} \mpl$, $\alpha=0.4$; we consider both $\lambda=0$ and $\lambda=0.7$. The source has mass $\mSrc=10^{10} \mpl$ and radius $\rSrc=10^{47}\mpl^{-1}$. \label{Ex:Test_model}
\end{enumerate}

Our results for models \ref{Ex:Antares} and \ref{Ex:Wiggle_model} are given in~\secref{Sec:Results}, where we present numerically computed field profiles and derived quantities for the source profiles defined in Eqs.~\eqref{Eq:top_hat_source}-\eqref{Eq:GCake_source}. We will instead use Model \ref{Ex:Test_model} to illustrate numerical convergence tests in \Appref{Sec:Tests}, as well as to compare against theoretical predictions for the special linear subcase $\lambda=0$ in the UV theory (for which expressions were given in \cite{Tobys_thesis}).
Unless otherwise stated, for the smoothed top-hat profile Eq.~\eqref{Eq:top_hat_source}, we use a width $w=0.02 \, \rSrc$.

\subsection{Weak form of the Newton iterations} \label{Sec:Examples_weak_form}
In this section, we derive the weak form of the Newton iterations as discussed in \secref{Sec:weak_form}.
The dimensionless units used in \phienics{} are presented in \Appref{Rescalings}.  However,  in this section we will use dimensionful variables for readability.
We also provide some additional technical details about the choice of continuous or discontinuous function spaces, and the stopping criteria for our Newton iterations.
Throughout, we denote our current estimate of variable $u$ by $u^{(k)}$, with the subsequent iteration denoted $u^{(k+1)}$.

\subsubsection{Single-field, higher-derivative theory}
\label{Sec:IR_theory_weak_form}

In \eqnref{Eq:IR_strong_system}, we expressed the original equation of motion of the single-field theory as a system of equations that allows us to lower the order of the nonlinear, high-derivative operator. In order to be solved iteratively using the Newton method, \eqnref{Eq:IR_strong_system} must be linearised as shown in \eqnref{eqn:newton-frechet}. For the single-field theory, the linearised system of equations is:

\begin{align}
  \left(\begin{array}{ccc}
    -\mOne^2 & 1 & -\frac{\paramNL}{\UVScl^{3n-1}}\lapRad^2 \\
    -\lapRad^2 & 1 & 0 \\
    0 & -n(Y^{(k)})^{n-1} & 1
  \end{array}\right)
  \left(\begin{array}{c}
    \itp{\fldOne} \\ \itp{Y} \\ \itp{W} 
  \end{array}\right) 
  &= \notag \\
  \left(\begin{array}{c}
    \itp{Y} - \mOne^2\itp{\fldOne} - \frac{\paramNL}{\UVScl^{3n-1}} \lapRad^2\itp{W} \\
    \itp{Y} - \lapRad^2\itp{\fldOne}  \\
    \itp{W} - n (Y^{(k)})^{n-1} \itp{Y} 
\end{array}\right)
&=
\left(\begin{array}{c} \frac{\rho}{\mpl} \\ 0 \\ (1-n)(Y^{(k)})^n  \end{array}\right) \, ,
\end{align}
with boundary conditions
\begin{equation}
  \left\lbrace \itp{\fldOne}(r_{\rm max}) = 0; \lapRad\itp{\fldOne}(0) = 0; \itp{W}(r_{\rm max}) = 0; \lapRad \itp{W}(0) = 0 \right\rbrace \, .
\end{equation}
We interpret this system of equations as a single vector equation and integrate each side against a vector test function $(v_1, v_2, v_3)$ 
\begin{align}
  \left\langle \left(
  \begin{array}{c}
    \itp{Y} - \mOne^2\itp{\fldOne} - \paramNL \frac{\lapRad^2 \itp{W}}{\UVScl^{3n-1}} \\
    \lapRad^2 \itp{\fldOne} -  \itp{Y} \\
    \itp{W} - n (Y^{(k)})^{n-1} \itp{Y}
  \end{array} \right),
  \left( \begin{array}{c} v_1 \\ v_2 \\ v_3 \end{array} \right) \right\rangle =
  \left\langle \left( \begin{array}{c} \frac{\rho}{\mpl} \\ 0 \\ (1-n) (Y^{(k)})^n  \end{array} \right),
  \left( \begin{array}{c} v_1 \\ v_2 \\ v_3 \end{array} \right) \right\rangle \, .
\end{align}

The weak form is therefore:
\begin{align}
  \begin{split}
    &\int \itp{Y} v_1 r^2 \udd{r} - \int \mOne^2\itp{\fldOne} v_1 r^2 \udd{r} - \frac{\paramNL}{\UVScl^{3n-1}} \int \lapRad^2\itp{W} v_1 r^2 \udd{r}  \\
    + &\int \lapRad^2\itp{\fldOne} v_2 r^2 \udd{r}  - \int \itp{Y} v_2 r^2 \udd{r}  \\
    + &\int \itp{W} v_3 r^2 \udd{r} - n \int (Y^{(k)})^{n-1}\itp{Y} v_3 r^2 \udd{r}  \\
    = & \int \frac{\rho}{\mpl} v_1 r^2 \udd{r} + (1-n) \int (Y^{(k)})^n v_3 r^2 \udd{r}  \, .
  \end{split}
\end{align}

Following \eqnref{Eq:int_by_part}, we can express all the terms that contain a Laplacian as products of gradients by integrating by parts.
For example,
\begin{align}
  \begin{split}
    \int \lapRad^2 \itp{\fldOne} v_2 r^2 \udd{r} &= \left[ \frac{\partial\itp{\fldOne}}{\partial r} v_2 r^2 \right]_0^{\infty} - \int_0^{\infty} \lapRad\itp{\fldOne} \cdot \lapRad v_2 r^2 \udd{r}~, \\
    &= - \int_0^{\infty} \lapRad\itp{\fldOne} \cdot \lapRad v_2 r^2 \udd{r}~,
  \end{split}
\end{align}
since the boundary term is zero for our choice of boundary conditions \eqnref{Eq:full_BC_IR}.
This gives us the weak form of the equations used in \phienics
\begin{align}
  \begin{split}
    &\int \udd{r} r^2 \left(\itp{Y}-\mOne^2\itp{\fldOne} + \frac{\paramNL}{\UVScl^{3n-1}}\lapRad \itp{W}\lapRad\right)v_1  \\
    - &\int \udd{r} r^2\left(\lapRad\itp{\fldOne}\lapRad + \itp{Y}\right) v_2 \\
    + &\int \udd{r} r^2 \left(\itp{W} -n(Y^{(k)})^{n-1}\itp{Y}\right)v_3 \\
    = &\int \udd{r} r^2\frac{\rho}{\mpl} v_1 + (1-n) \int \udd{r} r^2 (Y^{(k)})^n v_3  ~.
  \end{split}
  \label{IR_NL_weak_form}
\end{align}

As a final step, we require an initial guess as the starting point for Newton's method. This is provided by the limit in which the nonlinear term is dominant. In this case, we solve the Poisson equation
\begin{equation}
 -\lapRad^2 W^{(0)} = \frac{\Lambda^{3n-1}}{\epsilon} \frac{\rho}{\mpl}~,
 \label{Eq:NL_initial_guess}
\end{equation}
for $W^{(0)}$. From this, we compute $Y^{(0)}=\sqrt[n]{W^{(0)}}$, and finally we obtain $\pi^{(0)}$ from the Poisson equation $\lapRad^2\fldOne^{(0)}=Y^{(0)}$. 
In principle, another initial guess could be found by setting $\paramNL = 0$ to remove the nonlinear term.
However, this almost never leads to convergence, except for extremely small values of the nonlinearity parameter $\paramNL$, hinting that the nonlinearities are indeed very important, as expected.\footnote{Setting $\paramNL=0$ gives a good guess far from the source, where the linear theory is recovered, but nonlinearities generally dominate closer to the source.}

\subsubsection{Two-field, second-derivative theory}
\label{Sec:UV_theory_weak_form}
The system of equations for the UV theory in Eq.~\eqref{Eq:UV_strong_system} is linearised as:
\begin{align}
\left(\begin{array}{c}
   \itp{Y} - \mLt^2 \itp{\fldLt} - \kinmix \itp{Z} \\
   \itp{Z} - \mHvy^2 \itp{\fldHvy} - \kinmix \itp{Y} - \frac{\higgsNL}{2} (\fldHvy^{(k)})^2 \itp{\fldHvy} \\
   \itp{Y} - \lapRad^2\itp{\fldLt} \\
   \itp{Z} - \lapRad^2\itp{\fldHvy}
\end{array}\right)
=
\left(\begin{array}{c} \frac{\rho}{\mpl} \\ -\frac{\higgsNL}{3}(\fldHvy^{(k)})^3 \\ 0 \\ 0 \end{array}\right) \, ,
\end{align}
which is the strong form of the Newton iterations.
These are supplemented by the boundary conditions
\begin{equation}
  \left\lbrace \itp{\fldLt}(r_{\rm max})=0; \itp{\fldHvy}(r_{\rm max})=0; \lapRad\itp{\fldLt}(0)=0; \lapRad \itp{\fldHvy}(0)=0 \right\rbrace \, .
\end{equation}
As with the single-field theory, we interpret the system of equations as a single vector equation and take the scalar product of both sides with a vector test function $(v_1, v_2, v_3, v_4)$.
We then integrate by parts to turn integrals of Laplacians into products of gradients, as in Eq.~\eqref{Eq:int_by_part}.  This gives us the final form of our equations
\begin{align}
  \begin{split}
    &\int \udd{r} r^2 \left(\itp{Y}-\mLt^2\itp{\fldLt}-\kinmix\itp{Z}\right)v_1  \\
     + &\int \udd{r} r^2 \left(\itp{Z}-\mHvy^2\itp{\fldHvy}-\kinmix\itp{Y} - \frac{\higgsNL}{2}(\fldHvy^{(k)})^2\itp{\fldHvy} \right)v_2  \\
     + &\int \udd{r} r^2\left(\lapRad\itp{\fldLt} \lapRad + \itp{Y} \right) v_3 - \int \udd{r} r^2\left(\lapRad \itp{\fldHvy} \lapRad + \itp{Z} \right)v_4   \\
     = &\int \udd{r} r^2 \frac{\rho}{\mpl} v_1 + \int \udd{r} r^2 \frac{\higgsNL}{3} (\fldHvy^{(k)})^3 v_2
  \end{split}
  \label{Eq:UV_Newton_iteration}
\end{align}
For our initial guess, we use the solution for $\higgsNL = 0$ as a ``linear guess''. Unlike the single-field theory, in all cases examined this choice led to convergence. This suggests very different behaviour in the UV-complete two-field theory compared to the one-field effective IR theory.

\subsubsection{Numerical Implementation} \label{Sec:cont_vs_disc}
In the following, we will use piecewise continuous functions to characterise quantities we integrate directly when solving the equations of motion, specifically the field profiles and the auxiliary variables $Y$, $Z$, and $W$. 
In fact, when integrating the equations of motion, we found that using the same mathematical representation for all physical terms gave the lowest residuals to the strong form of the equations. 
Intuitively, in this case the auxiliary variables are both constrained to be Laplacians of some function and to satisfy the equations of motion. This double information can be used to stitch together the values of the functions at the vertices correctly.

The situation is different when derivatives are computed in post-processing, without supplemental information from the physics. This is useful, for instance, when computing the fifth force arising from a scalar field, which is proportional to its gradient, or in the computation of the $Q_n$ operators. In this case, the gradient is computed as in \eqnref{Eq:FEM_derivatives}. To preserve the most mathematical information, we use a discontinuous function space in this case. We have found in the tests described in Appendix~\ref{Sec:Tests} that this choice gives the most accurate results.
This is because the approximating polynomials are finite order, so there will be discontinuities in some derivatives of the function. For instance, if linear (\ie degree 1) Lagrange polynomials are chosen to approximate a field profile, its gradient will be piece-wise constant (and discontinuous). 

As the stopping criterion for the Newton iterations, we employ a residual criterion as in Eq.~\eqref{Eq:Residual_criterion}, together with the norm
\begin{equation}\label{eqn:projection_norm}
  \norm{\mathcal{F} \cdot B_{l,o}}_{\infty}\equiv \mathrm{max}\left\lbrace\lvert\langle \mathcal{F},B_{l,o}\rangle\rvert : l = 1,\dots,\nEl, o = 0,\dots,p\right\rbrace \, , 
\end{equation}
where the $B_{l,o}$'s are the $o$-th basis function in the $l$-th element as described in~\secref{Sec:FEM}. Here $N$ is the number of elements, and $p$ is the order of the interpolating polynomials. This norm can be thought of as the infinity norm $\norm{V}_{\infty} \equiv \underset{i=1,\cdots,Np+1}{\max} \{ | V_i | \}$ (\ie maximum absolute component) of a vector whose $(Np+1)$ entries are the scalar products\footnote{The quantity $\mathcal{F}=\int \mathcal{F} v r^2 \udd{r}$ is called the \textit{weak residual}.} $\langle \mathcal{F}, B_{l,o} \rangle = \int \mathcal{F} B_{l,o} r^2\udd{r}.$ 

When presenting results, we will also find it helpful to present them in terms of the infinity norm of a (discretised) function $f$, which is defined to be:
\begin{equation}\label{eqn:collocation_norm}
 \lVert f \rVert_{\infty} \equiv \underset{i=1,\cdots,Np+1}{\max} \{ | f(r_i) | \}~, 
\end{equation}
\ie infinity norm of the vector made up by the function values at mesh nodes.

 \subsection{Numerical solutions} \label{Sec:Results}
\begin{figure}[h!]
  \centering
  \includegraphics[width=\textwidth]{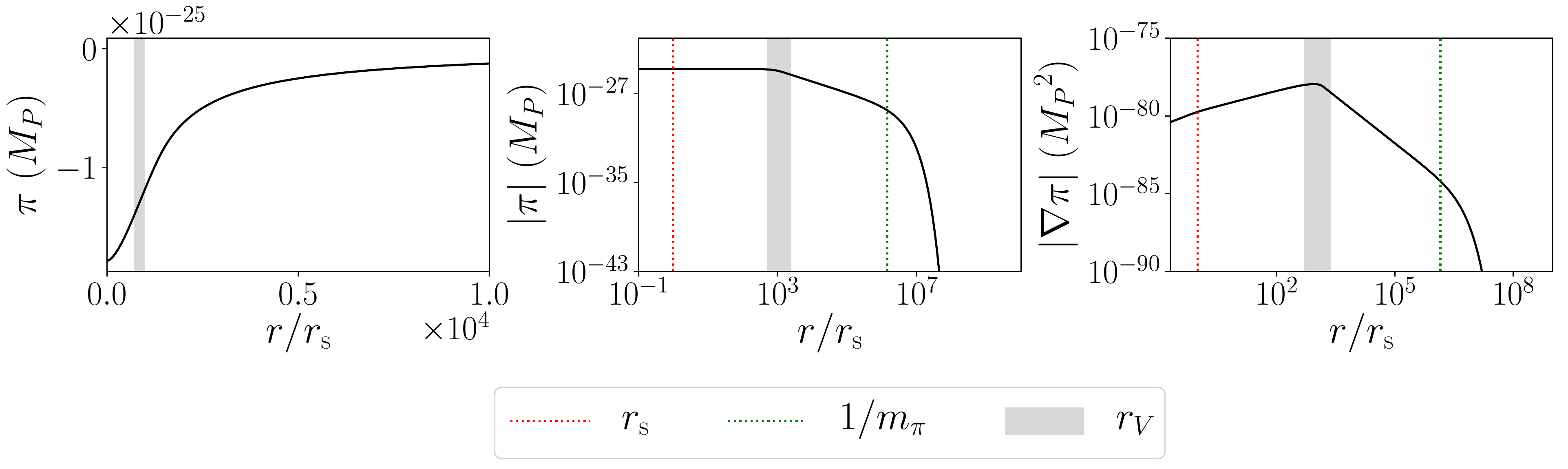}
  \caption{\phienics{}'s output for the single-field theory, example model \ref{Ex:Antares}.  \emph{Left and middle}: The field profile around the source in linear scale (\emph{left}) and across the box in log-log scale (\emph{middle}).  \emph{Right}: The gradient of the field $\pi$, which is proportional to the fifth force. 
The numerical profile was computed using an \meshPow{} mesh with $2300$ elements, $k=14$, $\gamma=8$, and $r_{\rm max}=10^{13}\, \rSrc$.  The interpolating polynomials were degree $7$, and the overall field scale was taken to be $\fldscl_{\fldOne}=10^{-35} \mpl$ (see \Appref{Rescalings} for the definition).   The source profile was a smoothed top-hat.  For this choice of mass/Compton wavelength $1/\mOne$, the details of the field profile and gradient are similar to those obtained from the other two source profiles. }
  \label{Fig:Antares_profile}
\end{figure}

 In Fig.~\ref{Fig:Antares_profile}, we show \phienics{}'s output for model \ref{Ex:Antares}. The first two panels show the $\fldOne$ profile near the source (left panel) and throughout the simulation volume (middle panel).  The right panel shows the gradient of the scalar field $\lapRad\pi$, which is proportional to the fifth force.  The grey band in Fig.~\ref{Fig:Antares_profile} indicates the scale at which the behaviour of the numerical solutions changes, and therefore represents our numerical estimate of the Vainshtein radius of this system.
 The source radius $\rSrc$ and Compton wavelength $\mOne^{-1}$ are indicated by a dashed red and dashed green line, respectively. As expected, the field reaches its minimum (\ie maximum excursion) in the middle of the source, then damps slowly within the Vainshtein radius. Outside of the Vainshtein radius, we recover the expected $\sim e^{-\mOne r}/r$ decay of a massive scalar field. The large Compton wavelength in this model (relative to both the Vainshtein and source radii) mean structures within the source are not detected.
 Therefore, the field profiles and gradients are very similar for all three source profiles in this case. 
As shown in~\Appref{Rescalings}, these results are applicable for other sources which keep the effective parameters $\nounits{m}_\fldOne\equiv\mOne\rSrc$ and $\nounits{\paramNL}\equiv\paramNL\left(\frac{\mSrc}{\mpl}\right)^{n-1}\left(\rSrc\UVScl\right)^{1-3n}$ unchanged.

\begin{figure}
  \centering
  \includegraphics[width=\textwidth]{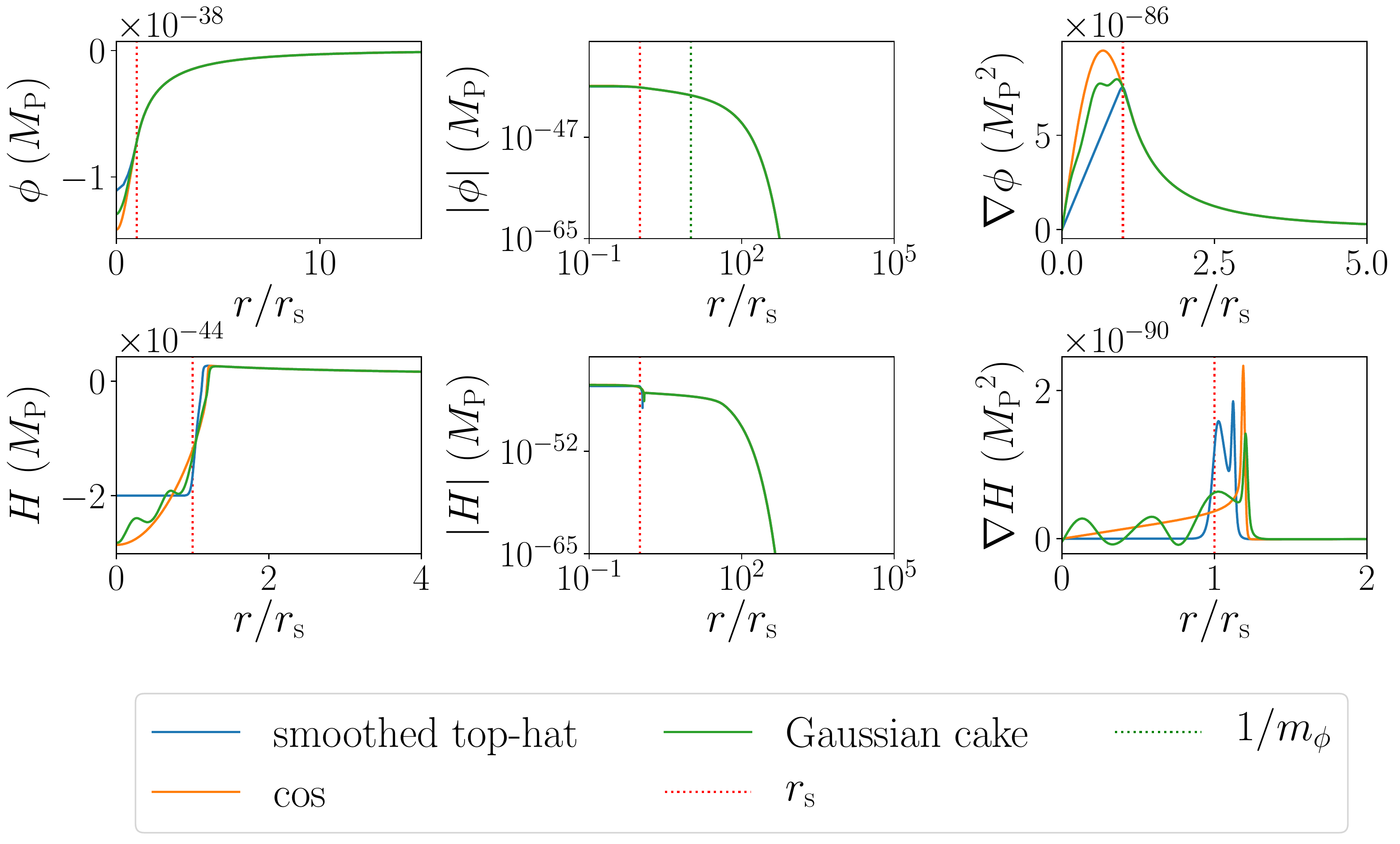}
  \caption{\phienics{}'s output for the two-field theory example model \ref{Ex:Wiggle_model}, in the presence of a smoothed top-hat, cosine or Gaussian cake source profile. Top panels: $\fldLt$'s field profile around the source (left panel) and across the simulation box (middle panel), $\fldOne$'s gradient around the source. Bottom panels: Same for field $\fldHvy$. These plots have been obtained with the following settings: for all source profiles, \meshExp{} mesh with $250$ elements before refinement, $a=5\times10^{-2}$, $b=10^{-2}$, $r_{\rm max}=10^9\, \rSrc$; degree of interpolating polynomials: $5$; $\fldscl_{\fldLt}=10^{13} \mpl$, $\fldscl_{\fldHvy}=10^{12} \mpl$ (see \Appref{Rescalings} for the definition of the numerical parameters $\fldscl_{\fldLt},\fldscl_{\fldHvy}$). For the smoothed top-hat profile: $k=20$, additionally twice linearly refined between $r/\rSrc=1.05$ and $1.2$; for the cos profile: $k=1$, additionally thrice linearly refined between $r/\rSrc=1.1$ and $1.25$; for the Gaussian cake profile: $k=1$,  additionally thrice linearly refined between $r/\rSrc=1.1$ and $1.3$.}
	\label{Fig:Wiggle_model_profiles}
\end{figure}

In Fig.~\ref{Fig:Wiggle_model_profiles}, we show the output for model \ref{Ex:Wiggle_model}. As expected, in this case the impact of the source structure is dramatic. In particular, the heavy field profile and its gradient develop complicated features thanks to the smaller Compton wavelength.
The oscillations in the field profile near the surface of the source arise because H and $\phi$ are not eigenmodes for linear fluctuations about the vacuum.
This mode mixing, combined with the nonlinearity of the system, gives rise to the oscillations observed near the edge of the source.

\begin{figure}
	\centering
	\includegraphics[width=0.7\textwidth]{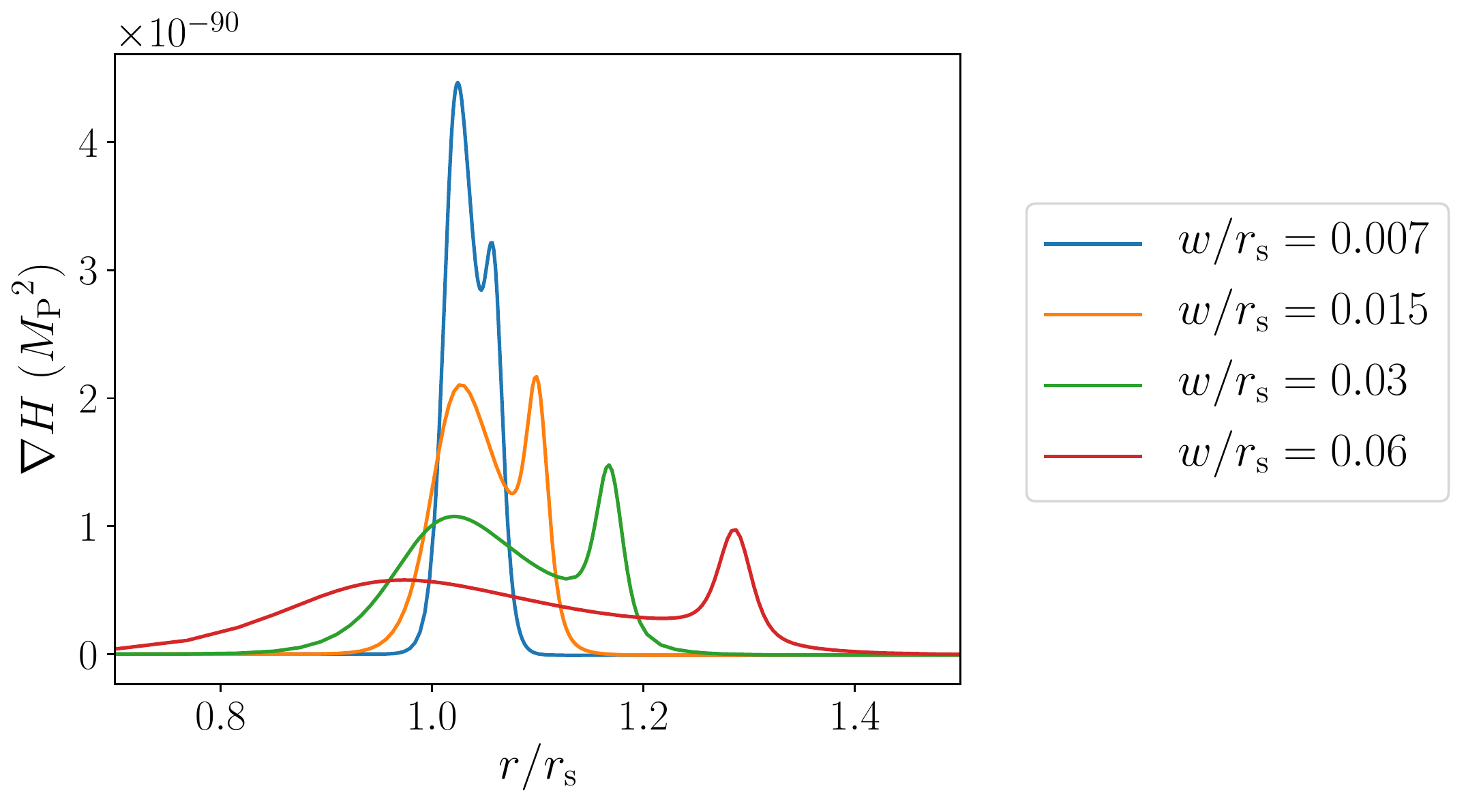}
	\caption{Behaviour of the heavy field gradient $\lapRad\fldHvy$ with increasing source steepness, for a smoothed top-hat profile of varying width $w$ in model \ref{Ex:Wiggle_model}. The gradient exhibits features that are a consequence of $\fldHvy$ not being an eigenstate of the linear operator.
	This plot has been obtained with the following settings: \meshExp{} mesh with $250$ elements, $k=20$, $a=5\times10^{-2}$, $b=10^{-2}$, $r_{\rm max}=10^9\, \rSrc$; degree of interpolating polynomials: $5$; $\fldscl_{\fldLt}=10^{13} \mpl$, $\fldscl_{\fldHvy}=10^{12} \mpl$ for all models. For $w=0.007 \, \rSrc$: additionally thrice linearly refined between $r/\rSrc=1$ and $1.1$. For $w=0.015 \, \rSrc$: additionally twice linearly refined between $r/\rSrc=1.05$ and $1.15$. For $w=0.03 \, \rSrc$: additionally twice linearly refined between $r/\rSrc=1.13$ and $1.2$. For $w=0.06 \, \rSrc$: additionally twice linearly refined between $r/\rSrc=1.2$ and $1.35$. 
	}
	\label{Fig:Width_of_wiggles}
\end{figure}

To better understand these features, we study them using a smoothed top-hat source profile of varying width --- the results are in Fig.~\ref{Fig:Width_of_wiggles}. The features we see in the field profile at distances close to the source radius are induced by the sharp change in density at the surface of the source mass.  As seen in Fig.~\ref{Fig:Width_of_wiggles}, the amplitude of the oscillatory features decreases as the steepness of the density profile is decreased.  The characteristic structure of these features evident in Fig.~\ref{Fig:Width_of_wiggles}, with two peaks in $\lapRad H$ close to the source radius, is a result of the field $H$ not being an eigenmode of the linear system,
 and instead a mix of the light and heavy modes.  Both modes are excited by the change in density at the source-vacuum transition, and so two fluctuations with different characteristic scales are seen in $H$. 
Although not detailed here, we verified that \phienics{} resolves these complicated features accurately using methods analogous to those in \Appref{Sec:Tests}.

\begin{figure}
	\centering
	\includegraphics[width=\textwidth]{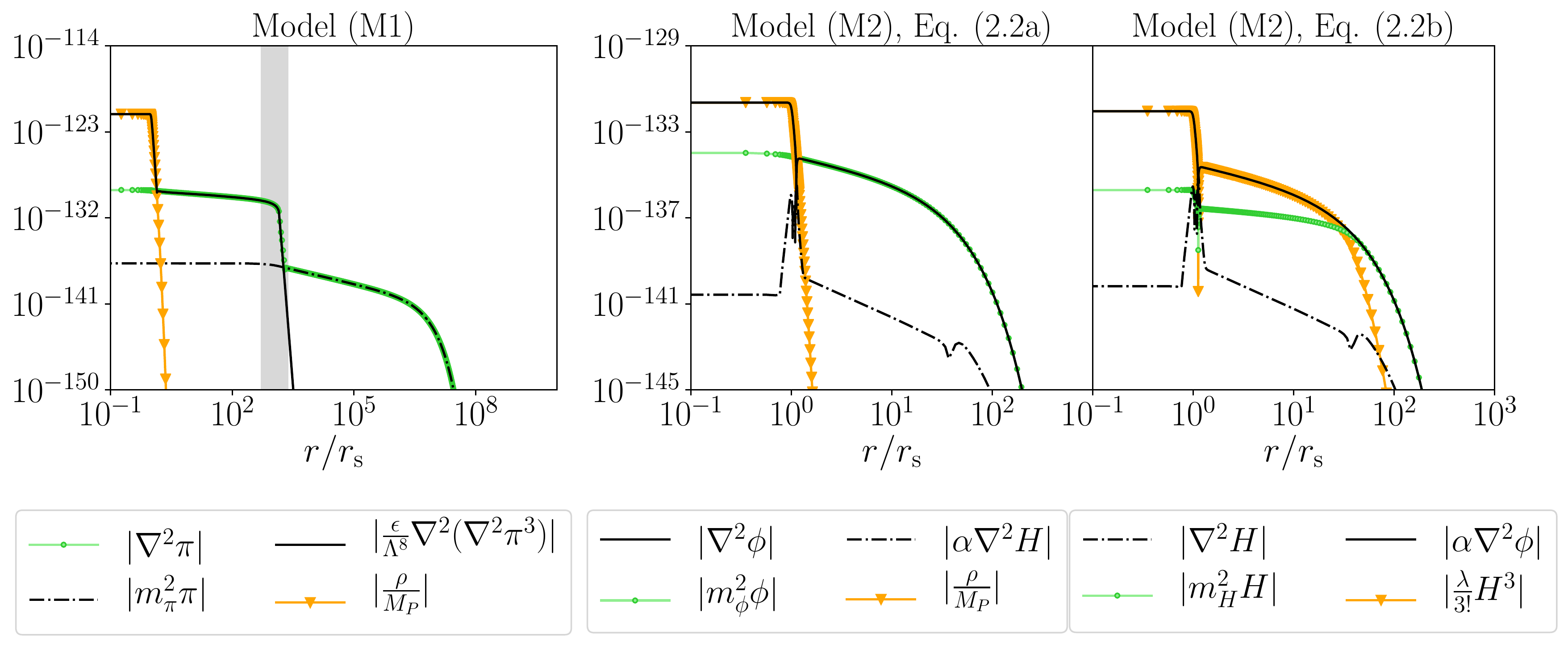}
	\caption{Left panel: terms in the equation of motion Eq.~\eqref{Eq:IR} for the single-field theory, model \ref{Ex:Antares}. The grey shaded area indicates the Vainshtein radius. Middle and right panels: terms in the equations of motion Eq.~\eqref{Eq:UV} for the two-field theory, model \ref{Ex:Wiggle_model}. For all panels, the vertical axis is in units $\mpl^3$. This plot has been obtained with the same settings as those in Figure~\ref{Fig:Antares_profile} and \ref{Fig:Wiggle_model_profiles} for a smoothed top-hat source of width $w=0.02 \, \rSrc$.}
	\label{Fig:Eqn_terms_across_box}
\end{figure}

For both theories, the behaviour of the terms in the equations of motion Eqs.~\eqref{Eq:IR}-\eqref{Eq:UV} across the radial domain can be seen in \figref{Fig:Eqn_terms_across_box}. For readability, we only plot them for a top-hat source of width $w=0.02 \, \rSrc$.  
The qualitative behaviour of the terms outside of the source does not change for the other source profiles.

\subsection{Comparison with theoretical expectations} \label{Sec:Theo_tests}
We now compare the results obtained with \phienics{} against  theoretical expectations for the behaviour of these theories. As discussed earlier the analytic solutions to these theories' equations of motion are not known in general, but in certain limits and for certain parameter values a useful comparison is possible.

\subsubsection{Comparison of profiles and equations of motion}

When considering the equations of motion for the theories of interest, it is easiest to understand the resulting behaviour of the field(s) if certain terms in the equation of motion dominate over others.  For example, in the single-field theory (\eqnref{Eq:IR}) if the dominant terms are the Laplacian of $\pi$ and the mass term $m_{\pi}^2\pi$, then we can expect that the field $\pi$ has either exponential growth or decay. Our analytic intuition for the behaviour of the solutions to the equations of motion is also informed by what we know about the boundary conditions for the theory.  In particular, the field evolves to a constant value, its vacuum expectation value, far away from the compact source. From this we can conclude that there is a region of space sufficiently far away from the source, where `sufficiently' remains to be determined, within which the field value is close to its vacuum expectation value and gradients are small.  In this regime, in the single-field theory, the dominant terms in the equation of motion are the Laplacian and mass terms, and because we know the boundary condition at infinity we can conclude that far from the source we have 
\begin{equation}
    \pi \sim \frac{1}{r}e^{-m_{\pi}r} \, .
    \label{eq:yukawa}
 \end{equation} %
We can further anticipate the existence of other regimes with different behaviour.  As the gradients of the field grow as we approach the source,  the term in the equation of motion containing higher powers of derivatives could dominate over the Laplacian of the field $\pi$.  Inside the source, the term proportional to the density of the source will be significant. 

For the single-field theory and model \ref{Ex:Antares}, Fig.~\ref{Fig:Eqn_terms_across_box} shows the different terms in the equation of motion across the radial domain.
We observe three regimes, which are separated by the source radius $\rSrc$ and the Vainshtein radius $r_V \sim 8.6 \times10^2 \, \rSrc$.
Inside the source, the nonlinear (solid black line) and source (orange line with triangles) terms dominate, tracking each other. Outside of the source but well inside the Vainshtein radius, the source term decays rapidly and the kinetic terms become dominant, as can be seen by the nonlinear (solid black line) and Laplacian (green line with circles) terms. Outside of the Vainshtein radius, the Laplacian (green line with circles) tracks the mass term (black dot-dashed line), and the field's behaviour is described by Eq.~(\ref{eq:yukawa}). As a result a further change in the behaviour of the solution is seen at the Compton wavelength $\mOne^{-1}$.  For $r<\mOne^{-1}$ we see a power-law decay of the field, and for $r>\mOne^{-1}$ we see exponential decay.

This behaviour can be understood as follows. Inside the Vainshtein radius, we can neglect the mass term $-\mOne^2\pi$, so that the equation of motion Eq.~\eqref{IR_EoM} can be approximated as $\lapRad^2\pi - \frac{\epsilon}{\Lambda^{3n-1}}\lapRad^2(\lapRad^2\pi^n) \sim \frac{\rho}{M_P}$ and integrated once. Within the Vainshtein radius
we also expect that the nonlinear term will be the dominant term driving the evolution of the field.
Inside the source, we therefore neglect the Laplacian of $\pi$, which reduces the equation of motion to the Poisson-like equation $- \frac{\epsilon}{\Lambda^{3n-1}}\lapRad^2(\lapRad^2\pi^n) \sim \frac{\rho}{M_P}$ for $r \lesssim r_V$. This is in agreement with Fig.~\ref{Fig:Eqn_terms_across_box}, where the nonlinear and source terms overlap within the source. In analogy with gravitation (except for a sign), we expect $(\lapRad^2\pi)^3$ to reach its maximum at the origin and then decay as $(\lapRad^2\pi)^3 \sim 1/r$ outside of the source, as long as we remain inside the Vainshtein radius. We indeed see in Fig.~\ref{Fig:Eqn_terms_across_box} that the Laplacian scales as $\lapRad^2\fldOne \sim 1/\sqrt[3]{r}$ in the region $r_S<r<r_V$. 
Finally, outside the Vainshtein radius, the equation of motion takes the linear 
form, and we expect the field to decay as $\sim e^{-\mOne r}/r$.

For the two-field theory with a step-function source, an exact solution for the linear (\ie $\lambda=0$) case was given in Ref.~\cite{Tobys_thesis} (see Eqns.~(4.23)-(4.26) within). This provides a useful test case for our numerical solution obtained with \phienics{}.
Although this comparison does not test the nonlinear solver --- as it requires only a single iteration of a linear solver --- we can make consistency checks. Importantly, it allows us to check that our grid can capture sharp features.
In particular, numerical solutions obtained for a smoothed top-hat distribution \eqnref{Eq:top_hat_source} should converge to those corresponding to a step-function source as the width of the source-vacuum transition relative to $\rSrc$ is taken to zero.\footnote{Note that this test would not allow checking the accuracy of the features displayed in \figref{Fig:Width_of_wiggles} as they are not present for $\lambda=0$.
}

For this comparison we will use model~\ref{Ex:Test_model} with $\lambda=0$. 
For every choice of $r_{\rm max}$, there is a radius after which the solution will be sensitive to the finiteness of the box. For the purposes of this test we use $r_{\rm max}=10^9 \, \rSrc$ and compare the value of our numerical solutions for radii $r < 10^5 \, \rSrc$.

\begin{figure}
  \centering
  \includegraphics[width=\textwidth]{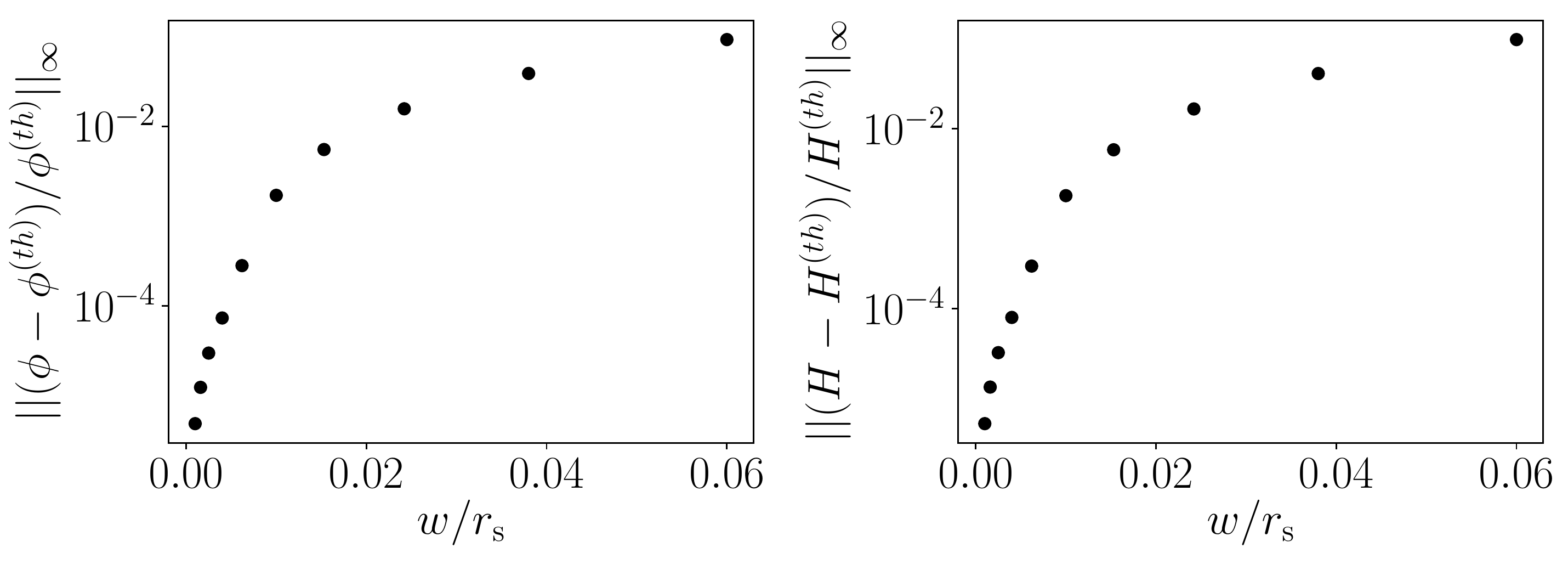}
  \caption{Infinity norm (\ie maximum absolute value at mesh nodes, whether mesh vertices or intermediate collocation points) of the relative error between the solution obtained with \phienics{} for the model \ref{Ex:Test_model} ($\lambda=0$), and the analytical result of Ref.~\cite{Tobys_thesis}, Eqns.~(4.23)-(4.26). The left and right panels show, respectively, the results for $\fldLt$ and $\fldHvy$. For all solutions, we used the following settings: \meshExp{} mesh with $800$ elements, $k=7$, $a=10^{-4}$, $b=0$, $r_{\rm max}=10^9\, \rSrc$; degree of interpolating polynomials: $5$. }
  \label{Fig:UV_linear_test}
\end{figure}

\Figref{Fig:UV_linear_test} shows the maximum relative difference between \phienics{}'s solution and the analytical result, for both fields. We have used the sequence of source widths:
\begin{equation}
 w\rSrc^{-1}=\{ 1,  1.6,  2.5,  4,  6.2, 10,  15.3,  24.2,  38,  60 \} \times 10^{-3}~,
\end{equation}
which were chosen to be approximately equispaced logarithmically.
As expected, we see the numerical solution approach the analytic solution (with $w=0$) as we decrease the width of the boundary layer connecting the source to the vacuum.
For all models, we observe that the maximum error is localised at the origin, where the field reaches its minimum, and sharply decreases outside of the source.

\subsubsection{Operators $Q_n$}

Above we demonstrated the existence of various dynamical regimes, each characterised by a different combination of dominant terms in the equations of motion.  In the single-field case, in particular, the nonlinear term was important within the Vainshtein radius $r_{\rm V}$.  To verify that the low-energy effective theory is valid in this nonlinear regime, we must evaluate further higher-order corrections on these solutions, which have the form of the $Q_n$'s in Eq.~\eqref{Eq:Qn_IR}-\eqref{Eq:Qn_UV}. If the size of these terms is comparable to those we have kept in the equation of motion, then our effective field theory breaks down.\footnote{It should be noted that the leading $Q_n$ operators being small, for $n$ finite, is a necessary though not sufficient condition for the operator series to converge, and therefore for the theory to be predictive.} For the case $n=3$ considered here, only $Q_3^{\rm (single-field)}$ is directly evaluated by the integrator.  The others are obtained in post-processing.  Given the highly nonlinear nature of these operators, and the shift of power to smaller scales induced by the gradients, it is important to verify that this post-processing produces accurate results.
Here we will verify that our post-processed $Q_n$'s agree with approximate analytic calculations motivated from the dynamical behaviour observed above.

In particular, we will consider the following operators: 
\begin{equation}
 {\opCorr{p}}^{\mathrm{(single-field)}} \equiv (-1)^{p+1} \binom{3p}{p} \frac{\paramNL^p}{2p+1}  \lapRad^2\left((\lapRad^2 \pi)^{2p+1}\right) \UVScl^{-6p-2}~,
 \label{Eq:On_IR}
\end{equation}
for the single-field theory and
\begin{equation}
 {\opCorr{p}}^{\mathrm{(two-field)}} \equiv (-1)^{p+1}  \binom{3p}{p} \frac{\kinmix^{2p+2}}{2p+1} \left(\frac{\higgsNL}{3!}\right)^p \lapRad^2\left((\lapRad^2 \fldLt)^{2p+1}\right) \mHvy^{-6p-2}~,
 \label{Eq:On_UV}
\end{equation}
for the two-field theory. These two types of operator have the same functional form, and they differ only in the dependence of the constant prefactor on the parameters of the two theories.
These operators are motivated by the connection between the single-field theory with $n=3$ and the low-energy ($E \ll \mHvy$) limit of the two-field theory, where higher-order operators of this type may be expected to arise \cite{physics_paper}.
Note, in particular, the correction $\opCorr{p}$ is related to our original operators $Q$ via $\opCorr{p} \propto Q_{2p+1}$.

\paragraph{Single-Field Theory:}

\begin{figure}
	\centering
	\includegraphics[width=\textwidth]{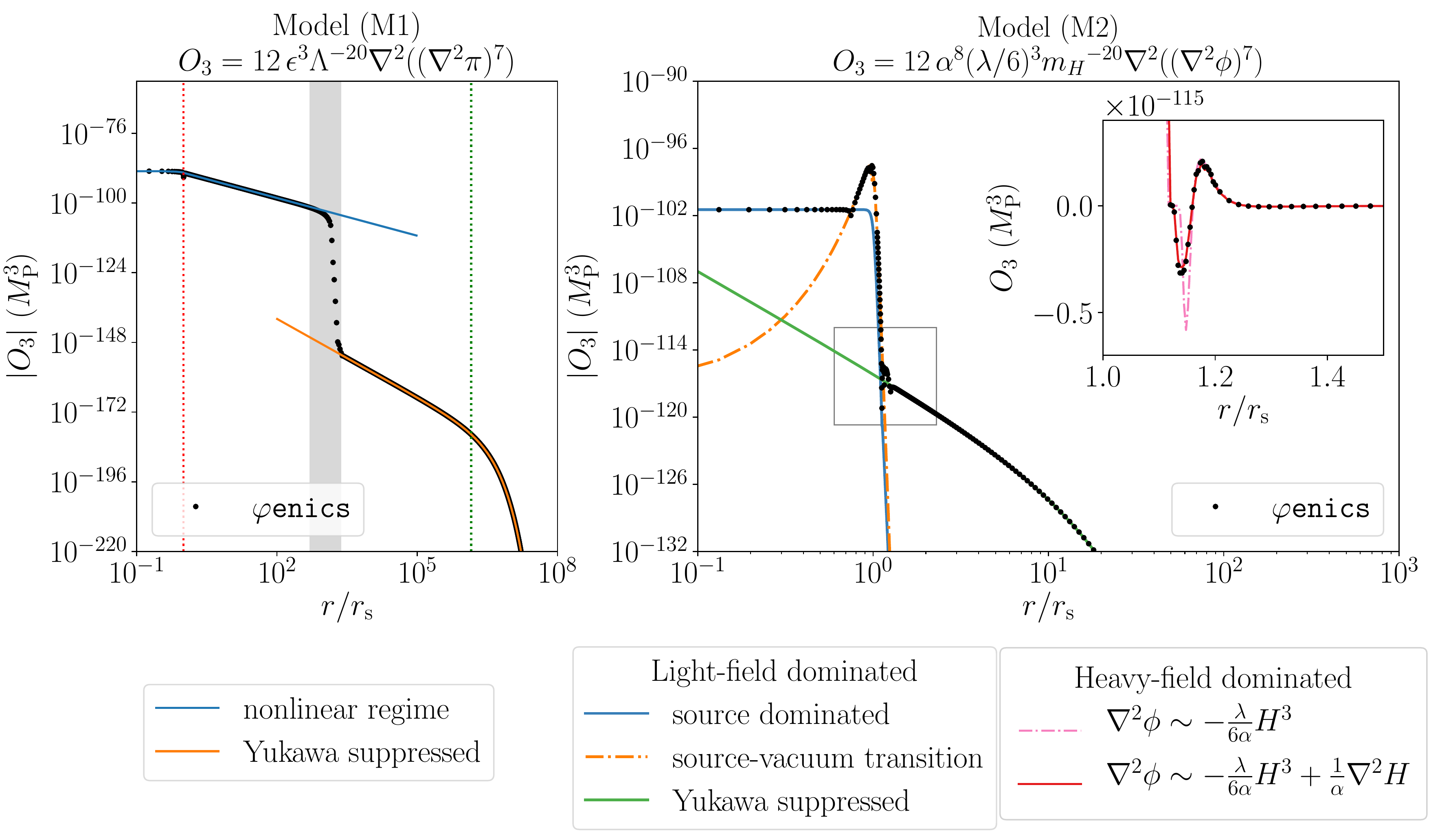}
	\caption{Comparison of \phienics{}'s prediction for the operator $O_3$ defined in Eq.~\eqref{Eq:On_IR}-\eqref{Eq:On_UV} for the single-field (left) and two-field (right) theory example models \ref{Ex:Antares} and \ref{Ex:Wiggle_model}. In the single-field theory, we recover the nonlinear regime and the Yukawa suppression, matching between them. The red and green vertical lines indicate, respectively, the source radius and the Compton wavelength of the field $\fldOne$, whereas the grey band highlights the Vainshtein radius. The rich phenomenology of the two-field theory is reflected in the many regimes that characterise it.  The light field $\fldLt$ dominates across most of the radial domain, where we can compare against three regimes (defined in the text): source domination, source-vacuum transition and Yukawa suppression. In a small region just outside the source radius, indicated with the grey box, the heavy field dominates.  We show this (in linear scale) in the inset, where we display the leading term contributing to the operator (\ie the nonlinear term $\higgsNL H^3/(6\kinmix)$) as well as the next-order correction, \ie the Laplacian $\kinmix^{-1}\lapRad^2H$. This plot was obtained with the following settings: \meshExp{} mesh with $250$ elements before refinement, $k=3, a=5\times10^{-2}, b=10^{-2}, r_{\rm max}=10^9 \, \rSrc$, twice linearly refined between $r/\rSrc=1.05$ and $r/\rSrc=1.2$; degree of interpolating polynomials: 5; $\fldscl_{\fldLt}=10^{13} \mpl$, $\fldscl_{\fldHvy}=10^{12} \mpl$. Also for the single-field theory, the normalisation constant $a_{\rm piv}$ is computed at a pivot radius $r_{\rm piv} = 10^4 \, \rSrc$, and for the two-field theory at $r_{\rm piv}=10 \, \rSrc$.}
	\label{Fig:Operators}
\end{figure}

We first derive predictions for the single-field theory based on the dynamical behaviour outlined above.
For the single-field theory, we distinguish between the nonlinear regime (`NL' in the following) at scales of $0 < r < r_V$ (combining the nonlinear behaviour of the field inside and outside the source) and Yukawa suppression (`YS') for $r > r_V$. As discussed above, the term nonlinear in $\pi$ dominates the evolution of the field inside the Vainshtein radius,
which yields the following result for $W\equiv(\lapRad^2\pi)^n$: 
\begin{align}
  W_{\rm NL}=\frac{1}{4\pi\paramNL}\frac{\UVScl^{3n-1}\mSrc}{\rSrcS\mpl}
  \begin{dcases}
    \frac{1}{2}\left(3-\frac{r^2}{\rSrcS[2]}\right), &\quad \text{for } 0 < r\leq \rSrcS \\
    \frac{\rSrcS}{r}, &\quad \text{for } \rSrcS \leq r \leq r_{\rm V}
    \end{dcases} \, .
  \label{Eq:IR_On_NL}
\end{align}
From this result, we obtain the Laplacian $(\lapRad^2\pi)_{NL}\equiv {W_{NL}}^{1/n}$, which is subsequently used to evaluate $\opCorr{p}$:
\begin{align}
  \opCorr{p, \mathrm{NL}} =\frac{\UVScl^{1-\beta}}{\rSrcS[2]}2c_p
  \paramNL^{p-\beta} \beta
  \left(\frac{\mSrc}{8\pi\rSrcS\mpl}\right)^{\beta}
  \begin{dcases}
    \left(3-\frac{r^2}{\rSrcS[2]}\right)^{\beta-2} \left( (1+2\beta) \frac{r^2}{\rSrcS[2]}-9 \right) & \text{for } 0 < r\leq \rSrcS~, \\
     2^{\beta-1}(\beta-1) \left(\frac{\rSrcS}{r}\right)^{\beta+2} & \text{for } \rSrcS \leq r\leq r_V~,
  \end{dcases}
\end{align}
where we have defined $\beta\equiv(2p+1)/n$ and $c_p\equiv (-1)^{p+1} \binom{3p}{p} \frac{1}{2p+1}$. 
Outside of the Vainshtein radius, the field decays as $\pi \approx \pi(r_{\rm piv})\frac{r_{\rm piv}}{r}e^{-\mOne (r-r_{\rm piv})}$, where $r_{\rm piv} > r_{\rm V}$ is some pivot scale.
The undetermined normalisation is calibrated to simulations by computing  $a_{\rm piv}\equiv \fldOne(r_{\rm piv})r_{\rm piv}e^{\mOne r_{\rm piv}}$, which is nearly independent of the choice of $r_{\rm piv}$, provided $r_{\rm piv} > r_{\rm V}$.
We obtain:
\begin{equation}
  \opCorr{p,\mathrm{YS}} = \UVScl \mOne^2 c_p (2p+1) \paramNL^p \left(a_{\rm piv}\frac{\mOne^3}{\UVScl^3}\frac{e^{-\mOne r}}{\mOne r} \right)^{2p+1} \left( 2p+1+ \frac{4p}{\mOne r} + \frac{2p}{\mOne^2r^2}  \right) \, .
  \label{Eq:IR_On_YS}
\end{equation}
We compare these analytic single-field predictions to the post-processing calculations of \phienics{} in the left panel of Fig.~\ref{Fig:Operators}.  We see excellent agreement with the nonlinear solution for $r < r_{\rm v}$, and with the Yukawa suppressed solution for $r > r_{\rm V}$.  
The Vainshtein radius $r_{\rm V}$ provides the radius at which the two solutions match onto each other, which is a second parameter that must be obtained from simulations.
The apparent discrepancy at $r=\rSrc$ is due to our source profile being a smoothed top-hat profile instead of a step function.


\paragraph{Two-Field Theory:}
For the two-field theory, similarly detailed predictions are not available in all regimes.
Instead, we will make consistency checks by comparing the numerical value of the post-processed operators with analytic approximations based on the dynamical field behaviour observed in Fig.~\ref{Fig:Eqn_terms_across_box}.
The middle panel of that figure shows that the Laplacian $\lapRad^2\phi$ tracks either the source (for $r\lesssim \rSrc$) or mass term (for $r\gtrsim \rSrc$) of the light field almost everywhere in the simulation volume.
The exception is a small region just outside of the source radius ($r \sim \rSrc$), where the heavy-field term $\alpha\lapRad^2 H$ briefly dominates. Based on this observation, we expect the operator $\opCorr{3}$ to also demonstrate these different regimes of behaviour. 
To fix terminology, we refer to these two regimes as `light-field dominated' and `heavy-field dominated', respectively.

In the light-field dominated regime, we distinguish three cases:
\begin{itemize}
\item \emph{source dominated}: Well inside the source, the Laplacian $\lapRad^2\phi$ tracks the source term, which is approximately constant for our smoothed top-hat profile. In the same region, the heavy-field term $\alpha\lapRad^2 H$ is negligible. The mass term $m_{\phi}^2\phi$ is small compared to the source term, but varies more rapidly, so it cannot be neglected. We can then write $\lapRad^2(\lapRad^2\fldLt)^{2n+1} \sim \lapRad^2( (\frac{\rho}{\mpl}+\mLt^2\fldLt)^{2n+1} )$. Because the source density, $\rho\approx \rho_0 \equiv \rho(r=0)$, is approximately constant around the origin, we can write: 
\begin{equation} \label{Eq:On_source_dom_step}
	\lapRad^2 \left( \left(\frac{\rho}{\mpl}+\mLt^2\fldLt \right) \right)^{2n+1} \approx \sum_{q=0}^{2n+1} \binom{2n+1}{q} \left( \frac{\rho_0}{\mpl} \right)^{2n+1-q} \mLt^{2q} \, \lapRad^2 (\fldLt^q) \, .
\end{equation}
For the term $\lapRad^2( \fldLt^q )$ in this expression, we obtain $q \phi^{q-2} ( \phi\lapRad^2\phi + (q-1) (\lapRad\phi)^2 )$. Knowing a priori which of the last terms dominates in $\lapRad^2( \phi^q )$ is challenging.
We evaluate them with \phienics{} using the dimensionless units of \Appref{Rescalings}, which are better suited to this kind of comparison.
Due to the Neumann boundary condition, we expect $\lvert \phi \lapRad^2\phi \rvert \gg (\lapRad\phi)^2$ around the origin and numerically we see that this is true. We can therefore rewrite the right-hand side of Eq.~\eqref{Eq:On_source_dom_step} as $\left(\frac{\rho}{\mpl}\right)^{2n+1} \frac{\nabla^2\phi}{\phi} \sum q \binom{2n+1}{q} \left( \frac{\mLt^2\phi}{\rho/\mpl} \right)^q$. We can see from Fig.~\ref{Fig:Eqn_terms_across_box} that, for model \ref{Ex:Wiggle_model} and around the origin, the mass term is much smaller than the source term, i.e. $\mLt^2\phi \ll \rho/\mpl$. Therefore, we expect the $q=1$ term to dominate, and we have: 
\begin{equation}
  {\opCorr{p}} \approx c_p(2p+1)\kinmix^{2p+2}\left(\frac{\higgsNL}{6}\right)^p\left(\frac{\rho_0}{\mpl\mHvy^3}\right)^{2p}\left(\frac{\mLt}{\mHvy}\right)^2\lapRad^2\phi~.
\end{equation}

\item \emph{source-vacuum transition}: At the source-vacuum transition $\rSrc$, the operator $\opCorr{p}$ is instead dominated by the rapid variation of the source profile.
This contribution can be computed exactly from \eqnref{Eq:top_hat_source}, leading to the approximation:
\begin{equation}
  \begin{split}
    \opCorr{p} \approx & \frac{\mHvy}{w^2} c_p(2p+1)\kinmix^{2p+2} \left(\frac{\higgsNL}{6}\right)^p\left(\frac{\rho(r)}{\mpl\mHvy^3}\right)^{2p+1}\frac{e^{-\frac{r-\rSrcS}{w}}}{\left(1+e^{-\frac{r-\rSrcS}{w}}\right)^2} \times \\
    & \times \left( e^{-\frac{r-\rSrcS}{w}}\left(1 + 2p -\frac{2 w}{r}\right) -1 - \frac{2 w}{r} \right)~.
  \end{split}
\end{equation}

\item \emph{Yukawa suppressed}: Similarly to the single-field theory, by evaluating the normalisation $a_{\rm piv}\equiv \fldLt( r_{\rm piv} ) r_{\rm piv}e^{\mLt r_{\rm piv}}$ at some pivot scale $r_{\rm piv}$ in the regime where $\fldLt$ obeys the linearised equation of motion, we obtain: 
\begin{equation}
  \begin{split}
    {\opCorr{p}} =& c_p(2p+1) \left(\frac{ \lambda}{6}\right)^p \alpha^{2p+2}\mHvy \mLt^2 \left(a_{\rm piv} \frac{\mLt^3}{\mHvy^3} \frac{e^{-\mLt r}}{\mLt r} \right)^{2p+1} \\
      &\times \left( (2p+1) +  \frac{4p}{\mLt r}+\frac{2p}{\mLt^2r^2}  \right)~.
  \end{split}
\label{Eq:IR_On_YS}
\end{equation}
\end{itemize}

In the heavy-field dominated regime, we perform consistency checks between different quantities computed by \phienics{}. In this regime, it is helpful to express the Laplacian $\lapRad^2\phi$ in terms of the equation of motion of the heavy field: $\lapRad^2\fldHvy = -\frac{\higgsNL}{6\kinmix}\fldHvy^3 + \frac{1}{\kinmix} \lapRad^2\fldHvy - \mHvy^2 \fldHvy$. We can see in Fig.~\ref{Fig:Eqn_terms_across_box} that, at $r \sim \rSrc$, the Laplacian $\lapRad^2\fldLt$ tracks the nonlinear term 
$\frac{\higgsNL\fldHvy^3}{6\kinmix}$, which is therefore expected to have a large contribution. Although the Laplacian 
$\lapRad^2\fldHvy/\kinmix$ is strongly subdominant in that regime, it has a large gradient, so we also expect it to have a large contribution on the operator $O_n$. The mass term instead does not show large gradients and is subdominant compared to the nonlinear term, so that we can expect it to generate the smallest correction. A consistency check can be therefore obtained by comparing the \phienics{} numerical result for $O_n$ against the increasingly accurate approximations $\lapRad^2\fldLt \approx -\frac{\higgsNL}{6\kinmix}\fldHvy^3$, $\lapRad^2\fldLt \approx -\frac{\higgsNL}{6\kinmix}\fldHvy^3 + \frac{1}{\kinmix} \lapRad^2 \fldHvy$ and $\lapRad^2\fldLt = -\frac{\higgsNL}{6\kinmix}\fldHvy^3 + \frac{1}{\kinmix} \lapRad^2\fldHvy - \mHvy^2\fldHvy$. This is obtained by evaluating \eqnref{Eq:Qn_expanded} after having set $Y$ to coincide with the desired function (\eg $Y=\frac{\higgsNL}{6 \kinmix} \fldHvy^3$), also using \phienics{}. The inset of Fig.~\ref{Fig:Operators} compares the \phienics{} postprocessing result for $\opCorr{p}$ against the leading contributions coming from the heavy-field dominated regime: the nonlinear term and the Laplacian $\lapRad^2 H$.
The mass term induces a further, very small correction that we do not show for simplicity.

\section{Summary and Outlook}
\label{sec_conclusion}

In this work we have introduced a new finite-element code, \phienics, which computes the behaviour of nonlinear scalar field theories around isolated, spherical compact objects. The motivation for introducing these fields comes from attempts to modify the theory of gravity, and understand the nature of dark energy, whilst still being compatible with local tests of gravity. In order to robustly test these theories against observations, precision predictions for their phenomenology are needed.  It is to facilitate this that we present the \phienics{}  code.

As scalar field theories with screening are, by construction, nonlinear theories, numerical simulation can be challenging. Additional difficulties come from the large hierarchies of scale in the problems we wish to solve and the presence of higher-derivative operators in the equations of motion. In this work, we have demonstrated the power of the finite element method to address these problems. We show explicitly how the code works for two very different nonlinear scalar field theories: one with multiple fields, and the other with a single field, but relevant higher order derivatives.  We have also shown that \phienics{} can solve for the behaviour of the fields around different source density profiles. This includes profiles that approximate a step function, and profiles where the density does not scale monotonically with distance. 

For both example theories, we find that we can accurately simulate the behaviour of the scalar field across all dynamical regimes, including deep inside the source object, regions where nonlinear and higher derivative terms dominate, as well as the Yukawa suppression far from the source due to the Compton wavelength of the scalar field. Where analytic predictions are available, we confirm that \phienics{} reproduces the expected scalar field profile.  We also found cases where, despite the lack of a full analytic solution for the profile, we were able to analytically estimate the scaling behaviour of the solution across the various regimes of behaviour, using our numerical results as a guide.  We again found that \phienics{} agrees with  all expectations.  An illustration of this excellent agreement is shown in Fig.~\ref{Fig:Operators}. We have also been able to identify new behaviours, such as oscillations in the scalar field profile at the surface of the source due to a rapid change in the source density.

This paper lays the ground work for future applications of the finite element method to more complex geometries, which will allow us to explore the behaviour of screened scalar fields around asymmetrical objects.  Currently \phienics{} assumes the sources are static and spherically symmetric. In future work we will aim to relax these two assumptions, allowing for arbitrary spatial source shapes and time-dependence, with the corresponding aspherical and time-dependent field profiles. This will allow us to study  the full spectrum of behaviour for screened scalar theories, providing new opportunities to detect or constrain these interesting cosmological theories.

\phienics{} is publicly available from \url{https://github.com/scaramouche-00/phienics} and can be easily adapted for other theories, by modifying the relevant module as described in \Appref{Sec:Code_structure}.

\acknowledgments
We would like to thank Peter Millington for very helpful discussions during the course of this work. We particularly thank Ben Coltman, Tobias Wilson and Antonio Padilla for their collaboration on the related project of Ref.~\cite{physics_paper}.

We acknowledge use of the \fenics{} library \cite{Old_FEniCS_book, FEniCS_citations, New_FEniCS_book}, available from \url{https://fenicsproject.org/}, in the version 2017.2.0 for the \texttt{Python 2} package and 2019.1.0 (\ie the latest at the time of writing) for the \texttt{Python 3} package.

CB, BE and DS were supported, in part, by a Research Leadership Award from the Leverhulme Trust. CB was also supported by a Royal Society University Research Fellowship. DS was also supported by the European Research Council under the European Union's Horizon 2020 research and innovation programme (grant agreement No. 646702 ``CosTesGrav''). JB is supported by the Simons Foundation Origins of the Universe program (Modern Inflationary Cosmology collaboration). JB also thanks the Kavli Institute for Theoretical Physics for its hospitality while a portion of this project was completed.

\appendix
\appendixpage
\addappheadtotoc
\section{Dimensionless equations}\label{Rescalings}

In this appendix we present the dimensionless variables and corresponding equations of motion used in our code.  Throughout, we work in units with $\hbar = c = 1$ and assume three spatial dimensions, so that time and space have units of inverse mass, and fields have units of mass.  To fix notation, we denote dimensionless units by an overbar $\nounits{\cdot}$.  We introduce a length scale $\rscl$ to define the dimensionless radius $\nounits{r} = \rscl^{-1} r$, and mass scales $\fldscl_i$ to define dimensionless fields $\nounits{\pi} = \fldscl_{\fldOne}^{-1}\fldOne$, $\nounits{\fldHvy} = \fldscl_{\fldHvy}^{-1}\fldHvy$, and $\nounits{\fldLt} = \fldscl_\fldLt^{-1}\fldLt$.  It is further convenient to express the source energy density $\rho = \frac{\mSrc}{\rSrc^3}f\left(\frac{r}{\rSrc}\right)$ in terms of a dimensionless $\mathcal{O}(1)$ function $f(x)$ which vanishes for $x\gtrsim 1$. For the profiles used in this paper, the function $f(x)$ is shown in Fig.~\ref{Fig:source_profiles}. 
For generality, we keep these choices arbitrary for the moment, with the specific choices made in \phienics\ indicated below.

In dimensionless units, the equations of motion are:
\begin{enumerate}
\item Single-field theory:
\begin{equation} \label{Eq:IR_general_rescaling}
  \nounits{\lapRad}^2\nounits{\pi} - \left(\mOne\rscl \right)^2\nounits{\pi} - \epsilon\left(\fldscl_\fldOne\rscl\right)^{n-1}\left(\rscl\UVScl\right)^{1-3n}\nounits{\lapRad}^2\left[\left(\nounits{\lapRad}^2\nounits{\pi}\right)^n \right] = \frac{\mSrc}{\mpl}\left(\frac{\rscl}{\rSrc}\right)^3\frac{1}{\fldscl_\fldOne\rscl}f\left(\frac{\rscl}{\rSrc}\nounits{r}\right) \, .
\end{equation}

The single-field theory displays highly nonlinear behaviour: as a guide to obtaining optimum choice of units, it is then helpful to consider the deep nonlinear regime of Eq.~\eqref{Eq:NL_initial_guess}, which gives the initial guess for the Newton algorithm. In dimensionless units, it is:
\begin{equation}
- \epsilon\left(\fldscl_\fldOne\rscl\right)^{n-1}\left(\rscl\UVScl\right)^{1-3n}\nounits{\lapRad}^2\left[\left(\nounits{\lapRad}^2\nounits{\pi}\right)^n \right] = \frac{\mSrc}{\mpl}\left(\frac{\rscl}{\rSrc}\right)^3\frac{1}{\fldscl_\fldOne\rscl}f\left(\frac{\rscl}{\rSrc}\nounits{r}\right)~.
\end{equation}
By setting $\rscl=\rSrc$ and $\fldscl_\fldOne\rscl = (\rscl\UVScl)^{(3n-1)/n} (\mSrc/\mpl)^{1/n}$, one obtains the Poisson equation $- \epsilon\nounits{\lapRad}^2\left[\left(\nounits{\lapRad}^2\nounits{\pi}\right)^n \right] = f\left(\nounits{r}\right)$, which is particularly suited to solve for $\nounits{W}\equiv\left(\nounits{\lapRad}^2\nounits{\pi}\right)^n$.

However, to better match between the nonlinear and linear regimes of the single-field theory, it is helpful to include information on the Galileon mass $\mOne$ in the rescaling $\fldscl_{\fldOne}$. This is the default choice in \phienics{}, where we set:
\begin{equation} \label{Eq:def_IR_Mf}
 \fldscl_\fldOne = \left(\rscl\UVScl\right)^{(3n-1)/n} (\mSrc/\mpl)^{1/n} \mOne \, ,
\end{equation}
together with $\rscl = \rSrc$. 
With this choice of units, Eq.~\eqref{Eq:IR_general_rescaling} can be now expressed as:
\begin{equation} \label{Eq:IR_NL_rescaling}
  \nounits{\lapRad}^2\nounits{\fldOne} - \left( \mOne \rSrc \right)^2 \nounits{\fldOne} - \tilde{\paramNL}\nounits{M}\nounits{\lapRad}^2\left[\left(\nounits{\lapRad}^2\nounits{\fldOne}\right)^n \right] = \nounits{M}f\left(\nounits{r}\right) \, ,
\end{equation}
where we have defined $\tilde{\epsilon} = \epsilon(\mOne\rSrc)^n$ and $\nounits{M} \equiv \left( \frac{\mSrc}{\mpl} \right)^{1-\frac{1}{n}} \left( \rSrc \UVScl \right)^{-3+\frac{1}{n}}(\mOne\rSrc)^{-1}$.

For a given source profile $f$, the solution is determined (up to global scalings) by the two dimensionless parameters $\nounits{m}_\fldOne\equiv\mOne\rSrc$ and $\nounits{\paramNL}\equiv\paramNL\left(\frac{\mSrc}{\mpl}\right)^{n-1}\left(\rSrc\UVScl\right)^{1-3n}$, which we see by choosing $\rscl = \rSrc$ and $\fldscl_\fldOne\rscl = \mSrc/\mpl$, resulting in
\begin{equation}
  \nounits{\lapRad}^2\nounits{\pi} - {\nounits{m}_\fldOne}^2\nounits{\pi} - \nounits{\paramNL}\nounits{\lapRad}^2\left[\left(\nounits{\lapRad}^2\nounits{\pi}\right)^n \right] = f\left(\nounits{r}\right) \, .
\end{equation}
However, this form of the equation is normally numerically less-suited to the computation of the solution, as $\nounits{\paramNL}$ can become very large.

\item Two-field theory:
\begin{subequations}
\begin{empheq}[left=\empheqlbrace]{align}
  &\nounits{\lapRad}^2\nounits{\fldLt} - \left(\mLt\rscl\right)^2\nounits{\fldLt} - \kinmix\left(\frac{\fldscl_\fldHvy}{\fldscl_\fldLt}\right)\nounits{\lapRad}^2\nounits{\fldHvy} = \frac{\mSrc}{\mpl}\left(\frac{\rscl}{\rSrc}\right)^3\frac{1}{\fldscl_\fldLt\rscl}f\left(\frac{\kappa}{\rSrc}\nounits{r}\right)~,\\
  &\nounits{\lapRad}^2\nounits{\fldHvy} - \left(\mHvy\rscl\right)^2\nounits{\fldHvy} - \kinmix\left(\frac{\fldscl_\fldLt}{\fldscl_\fldHvy}\right)\nounits{\lapRad}^2\nounits{\fldLt} - \frac{\lambda}{3!}\left(\fldscl_\fldHvy\rscl\right)^2\nounits{\fldHvy}^3 = 0 \, .
\end{empheq}
\end{subequations}
To normalise to the source profile, it is convenient to choose $\rscl=\rSrc$, $\fldscl_\fldLt\rscl = \frac{\mSrc}{\mpl}$, and $\fldscl_\fldHvy = \lvert \alpha \rvert \fldscl_\fldLt$,\footnote{The choice $\fldscl_\fldHvy = \lvert \alpha \rvert \fldscl_\fldLt$ ensures the dominant term in the second order equations of motion have similar scales, which makes residuals easier to interpret.} resulting in
\begin{subequations}
\begin{empheq}[left=\empheqlbrace]{align}
  &\nounits{\lapRad}^2\nounits{\fldLt} - \left(\mLt\rSrc\right)^2\nounits{\fldLt} - \mathrm{sgn}(\kinmix)\kinmix^2\nounits{\lapRad}^2\nounits{\fldHvy} = f\left(\nounits{r}\right)\\
  & \nounits{\lapRad}^2\nounits{\fldHvy} - \left(\mHvy\rSrc\right)^2\nounits{\fldHvy} - \mathrm{sgn}(\kinmix)\nounits{\lapRad}^2\nounits{\fldLt} - \frac{\lambda}{3!}\kinmix^2\left(\frac{\mSrc}{\mpl}\right)^2\nounits{\fldHvy}^3 = 0 \, .
\end{empheq}
\end{subequations}
This time, for a given source profile $f$, the solution is determined (up to global scalings) by four dimensionless parameters: $\mLt\rSrc$, $\mHvy\rSrc$, $\kinmix$, and $\frac{\lambda}{3!}\kinmix^2\frac{\mSrc^2}{\mpl^2}$.

\end{enumerate}

For both theories, \phienics{}'s default choices are a helpful guide to obtaining optimal units that allow for convergence in a physical model of interest. However, for every specific model, setting the quantities $\fldscl_{\fldOne}, \fldscl_{\fldHvy}, \fldscl_\fldLt$ `by hand' may give the best results.

Before being solved, the rescaled equations of motion are expressed in the weak form detailed in Sec.~\ref{Sec:weak_form} and \ref{Sec:Examples_weak_form}.

\section{Numerical Convergence Tests}
\label{Sec:Tests}

In this appendix, we demonstrate the numerical convergence properties of our solver.
In particular, we provide independent tests of both our iterative Newtonian solver and the convergence of the solution with varying spatial discretisations.
For concreteness, we focus on the UV complete model \ref{Ex:Test_model}. In the following, we will solve the field profiles around a smoothed top-hat source profile of width $w=0.02 \,\rSrc$.

\subsection{Convergence of Newton iterations}

\begin{figure}[h!]
\centering
\includegraphics[width=0.9\textwidth]{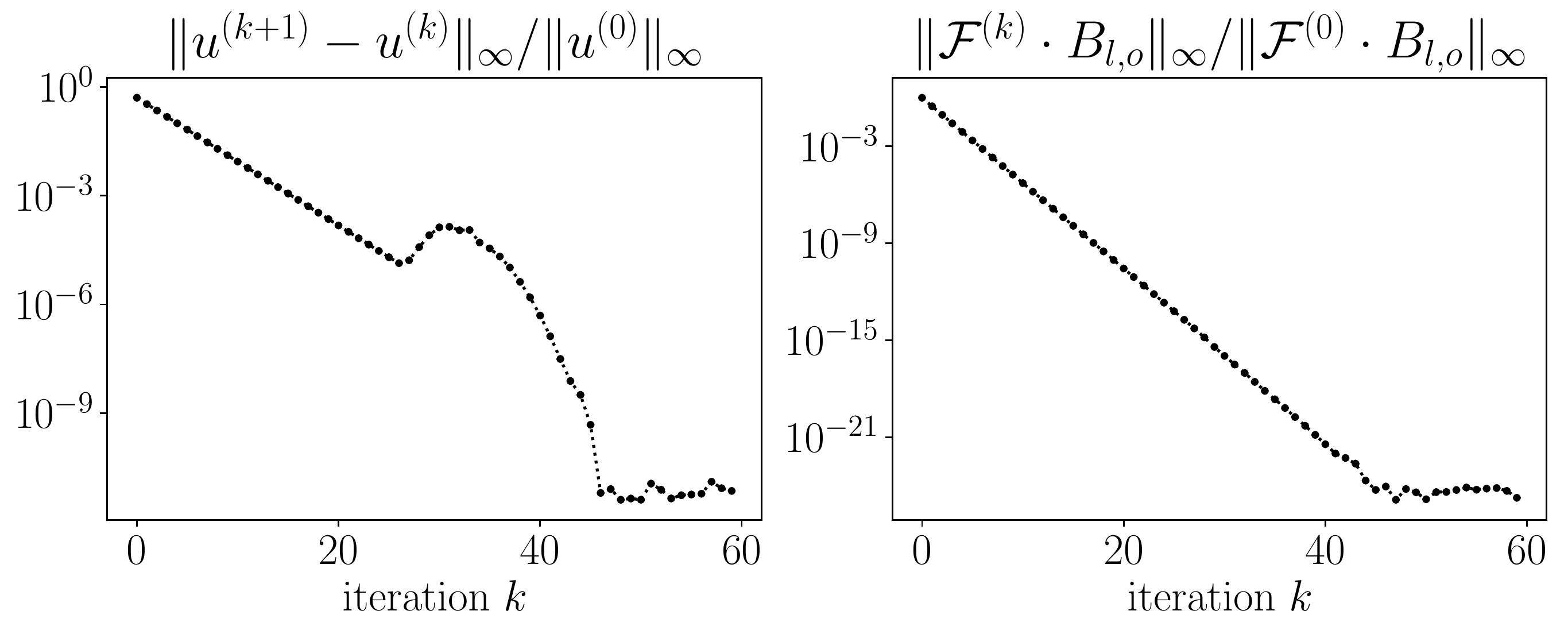}
\caption{The error in the solution as a function of iteration of the Newton solver, expressed as change in the solution (\emph{left panel}) and residuals of the equation (\emph{right panel}), for model \ref{Ex:Test_model}. The norms used for this test are defined in Eqns.~\eqref{eqn:projection_norm}-\eqref{eqn:collocation_norm}. This plot has been obtained with the following settings: \meshExp{} mesh with $250$ elements, $k=25, a=0.5, b=6, r_{\rm max}=10^9 \rSrc$; degree of interpolating polynomials: $5$; $\fldscl_{\fldLt}=10^{13} \mpl$, $\fldscl_{\fldHvy}=10^{12} \mpl$.}\label{Fig:Convergence}
\end{figure}

We start with convergence of our Newton iterations. In Fig.~\ref{Fig:Convergence}, we show the behaviour of the error in the solution at every iteration: we quantify the error both in terms of the change in the solution (left panel, also see Eq.~\ref{Eq:Change_criterion}) and residual (right panel and Eq.~\ref{Eq:Residual_criterion}). In this run, tolerances have been adjusted to exceedingly small values so that the full behaviour of the error could be observed. The norms in Fig.~\ref{Fig:Convergence} are as defined in Eqns.~\eqref{eqn:projection_norm}-\eqref{eqn:collocation_norm}.

As expected, the residual decreases with each iteration (in this case exponentially), eventually reaching a plateau when the scheme reaches its accuracy limits. 
Interestingly, in this case we find an intermediate regime where the required field perturbation temporarily increases with iteration, while the residual always decreases. %
We typically find a larger number of iterations is required as $\higgsNL$ is increased. This is easy to understand when considering that, for the two-field model, we take the $\higgsNL=0$ solution as our initial guess.  As a result, the ``distance'' between the initial guess and the converged solution for $\higgsNL \neq 0$ increases as the nonlinear parameter $\higgsNL$ is turned on.  
For the single-field model, on the other hand, our initial guess is instead based on assuming the nonlinear term dominates, and as a result it typically requires fewer iterations as the parameter $\paramNL$ is increased.

\subsection{Spatial Discretisation Convergence}

We shall now present a test that the obtained solution is a good approximation of the true continuous solution. If that is the case, the properties of the solution should be independent of the particular chosen discretisation. In Fig.~\ref{Fig:MI_profiles}, we show the field profiles as obtained for two meshes characterised by the same number of elements ($500$) but with considerable difference in their distributions:

{\begin{itemize}
      \item \meshExp{} mesh, $k=15, a=5\times10^{-2}, b=3\times10^{-2}, r_{\rm max}=10^9 \, \rSrc$~,
      \item \meshPow{} mesh, $k=40, \gamma=20, r_{\rm max}=10^9 \, \rSrc$~.
\end{itemize}
In both cases, we use interpolating polynomials of degree $5$.

Under every plot in Fig.~\ref{Fig:MI_profiles}, we show the relative difference between the two obtained solutions, which can be taken to be a measure of the error in the computation of the true, unknown continuous profile. We can see that, for $r \lesssim 10^4 \, \rSrc$, we have a precision of $10^{-9}$ or better in the characterisation of $\phi$'s profile, and one of $10^{-5}$ or better in the characterisation of $H$'s profile. As the radius grows larger, the relative difference grows: this is due to the \meshExp{} mesh increasing its sizing exponentially at large radii, which is less suited to describe exponential decay compared to the \meshPow{}, which increases element sizes polynomially. It should be noted, however, that Yukawa suppression, unlike other features in the field profiles, is completely captured by linear physics: as such, it is less of a focus for numerical codes, like \phienics{}, tailored to the description of nonlinear physics inaccessible to analytical understanding. In particular, the interior solution is insensitive to any numerical errors appearing in the Yukawa-suppressed tail. 
Finally, note that the large increase in the relative error of $H$ around $r \sim 2-3 \, \rSrc$ is due to the field going to $0$, and does not reflect a correspondingly large increase in the absolute difference between the two computed solutions.

\begin{figure}[h!]
\centering
\includegraphics[width=\textwidth]{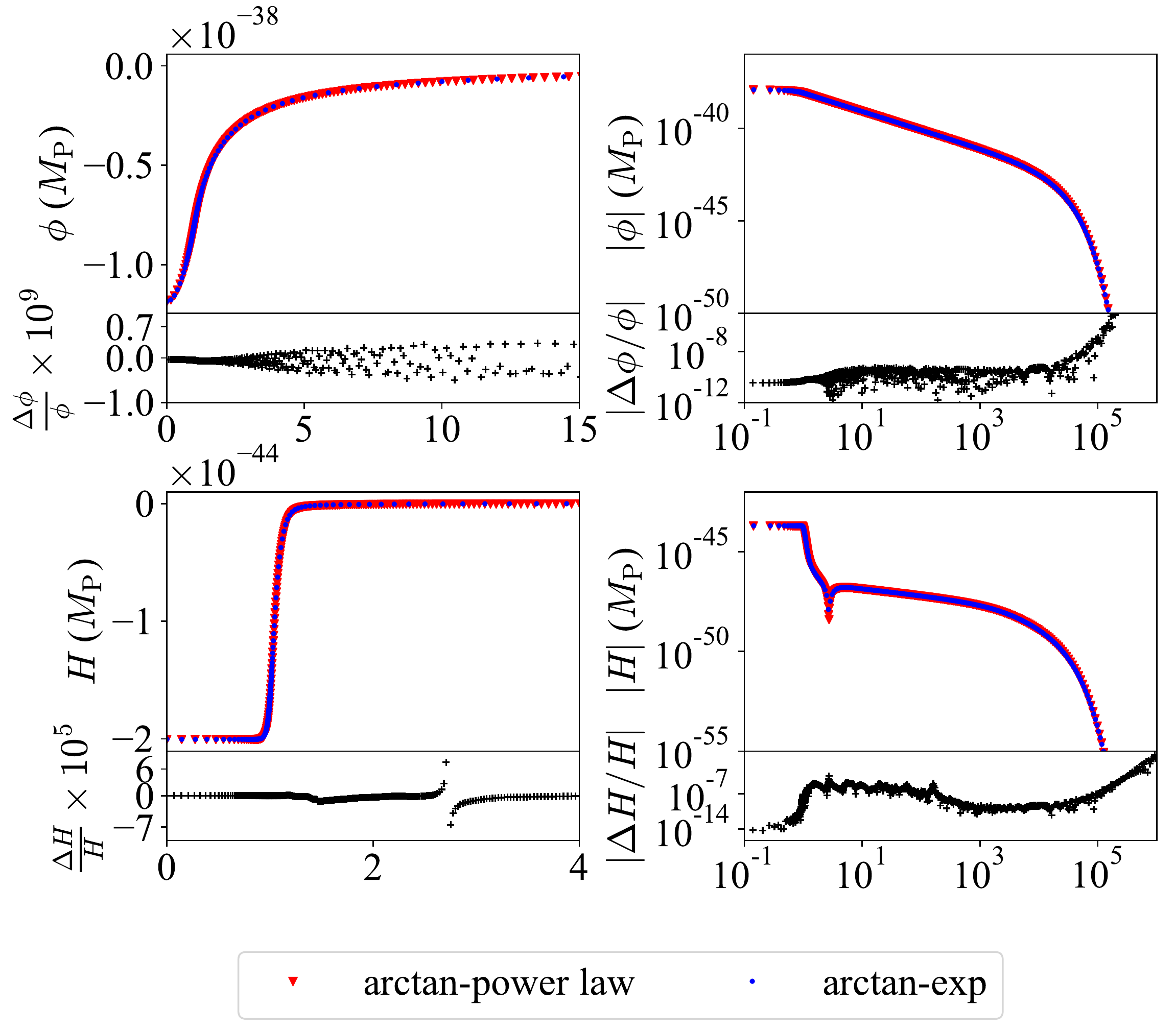}
\caption{\phienics{} computation of the field profiles for model \ref{Ex:Test_model}, as obtained with an \meshExp{} mesh (\emph{blue circles}) with $500$ elements and $k=15, a=5\times10^{-2}, b=3\times10^{-2}, r_{\rm max}=10^9 \, \rSrc$, and an \meshPow{} mesh (\emph{red triangles}) with the same number of elements and $k=40, \gamma=20, r_{\rm max}=10^9 \, \rSrc$. In both cases, we use order $5$ interpolating polynomials. Underneath the profiles, the relative difference in the solutions are shown. Please refer to the the text for discussion of the features of these plots.} 
\label{Fig:MI_profiles}
\end{figure}

\begin{figure}
\centering
\includegraphics[width=\textwidth]{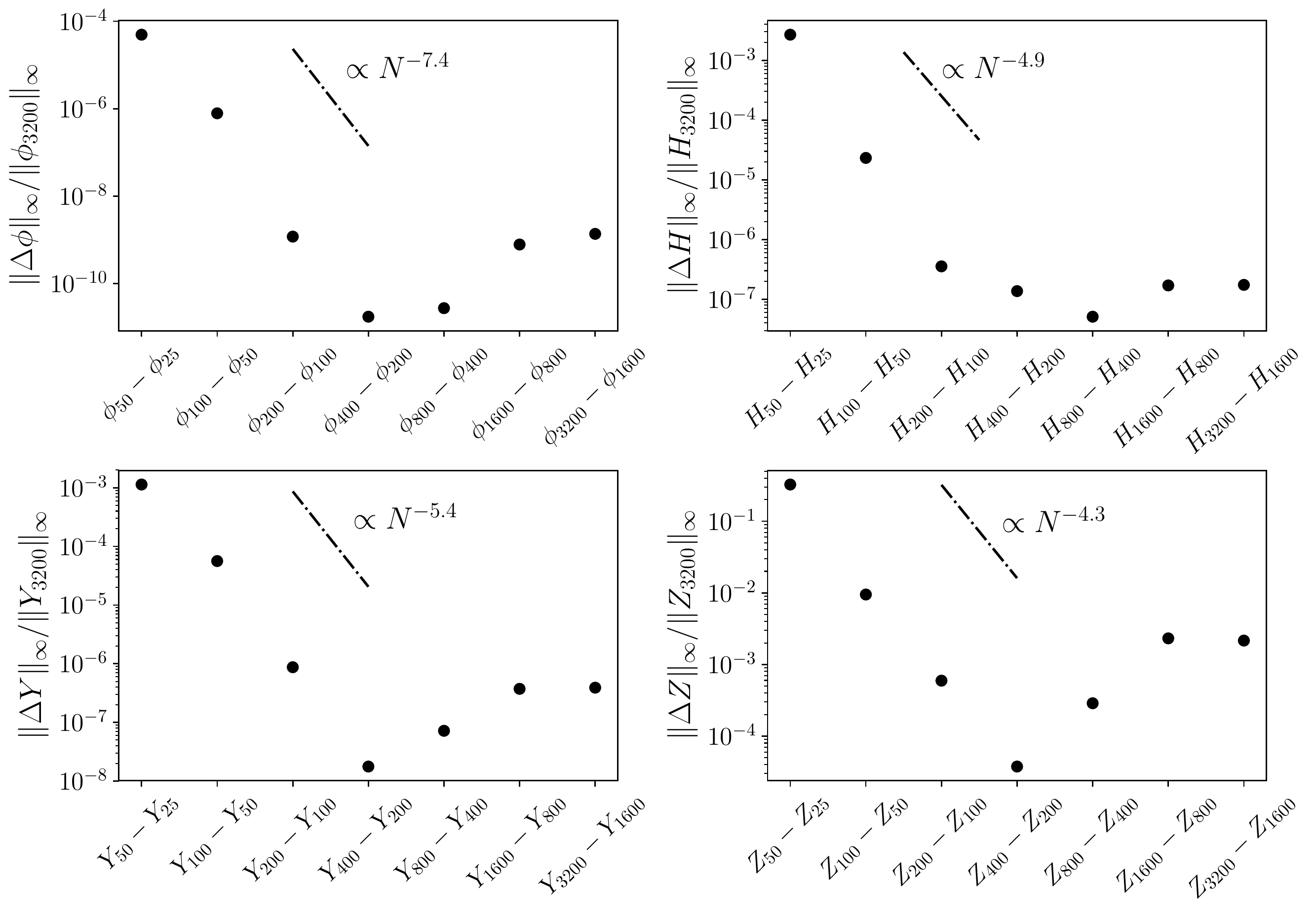}
\caption{Convergence of our discrete solution to the continuum limit.  We consider a family of solutions evaluated on an \meshExp{} mesh ($k=7$, $a=5\times 10^{-2}$, and $b = 5\times 10^{-2}$), with each solution indexed by the number of elements varying by factors of $2$ from $25$ to $3200$.  To compare solutions, we define an interpolation grid comprised of the union of all mesh vertices from the meshes of different resolutions. We then compute the maximal pointwise difference on this interpolation grid (denoted here by $\lVert\cdot\rVert_{\infty}$) between solutions on grids of neighbouring resolutions.  This provides an estimate of the difference between the continuum and discrete solution. As the number of elements increases, the error in the solution decreases as a power of $N$, until it reaches the final accuracy in the scheme and no further improvement is obtained. 
For this model and choice of mesh, we find that the error saturates with around $N=200$ elements, corresponding to roughly $1000$ degrees of freedom.}
\label{Fig:giant_mesh_interp}
\end{figure}

We now consider the spatial convergence properties of our solution with varying resolution (\ie number of elements).
As we refine our spatial discretisation, we expect our discrete solutions to approach the continuum limit.
We can determine the rate at which we approach the continuum by comparing solutions as the resolution is varied.
This standard convergence testing is illustrated in \figref{Fig:giant_mesh_interp} for a collection of \meshExp{} meshes with mapping parameters $k=7$, $a=5 \times 10^{-2}$, and $b=5 \times 10^{-2}$. We consider a series of eight of these meshes, each constructed from twice as many elements as the previous one: \ie $25$, $50$, $100$, $200$, $400$, $800$, $1600$, and $3200$ elements. For each mesh, $r_{\rm max}$ is fixed to $r_{\rm max}=10^9 \, \rSrc$, and we employ interpolating polynomials of order $5$. 
Due to the nonlinear nature of the transformation \eqnref{Eq:ATExpMesh}, the sequence of meshes are non-nested.
To facilitate a direct comparison, we thus introduce a separate ``global'' mesh consisting of the union of the element vertices from each of the individual meshes.  We then interpolate each of the individual solutions onto this global mesh for comparison.
For notational convenience, we denote the resulting interpolated profile arising from a mesh with $N$ elements by $\fldLt_N$ (and similarly for $\fldHvy$, $Z$, and $W$). 
\Figref{Fig:giant_mesh_interp} shows the maximal pointwise difference between solutions for neighbouring resolutions. For this test we will calculate the norm $\lVert \cdot \rVert_{\infty}$ on the ``global'' mesh vertices only, without including intermediate nodes.
As expected, the maximal pointwise difference decreases with increasing resolution, until it saturates due to roundoff errors inherent in finite precision arithmetic. 
Continuing to enhance the resolution beyond this point is not only computationally wasteful, but can also be detrimental as the additional modes will simply be populated with random numerical noise.
This is especially harmful when computing the $Q_n$ operators in post-processing, since derivatives enhance these noise-dominated short-scale modes. 
In Fig.~\ref{Fig:giant_mesh_interp}, we can see that the relative error for the auxiliary variable $Z$ is significantly larger than for the other fields.  Examining Fig.~\ref{Fig:Eqn_terms_across_box}, we see this occurs because the magnitude of $Z$ is considerably smaller than the other variables, so that the precision in its characterisation will be limited by the roundoff errors in the remaining variables.

 \begin{figure}[h!]
 \centering
 \includegraphics[width=0.7\textwidth]{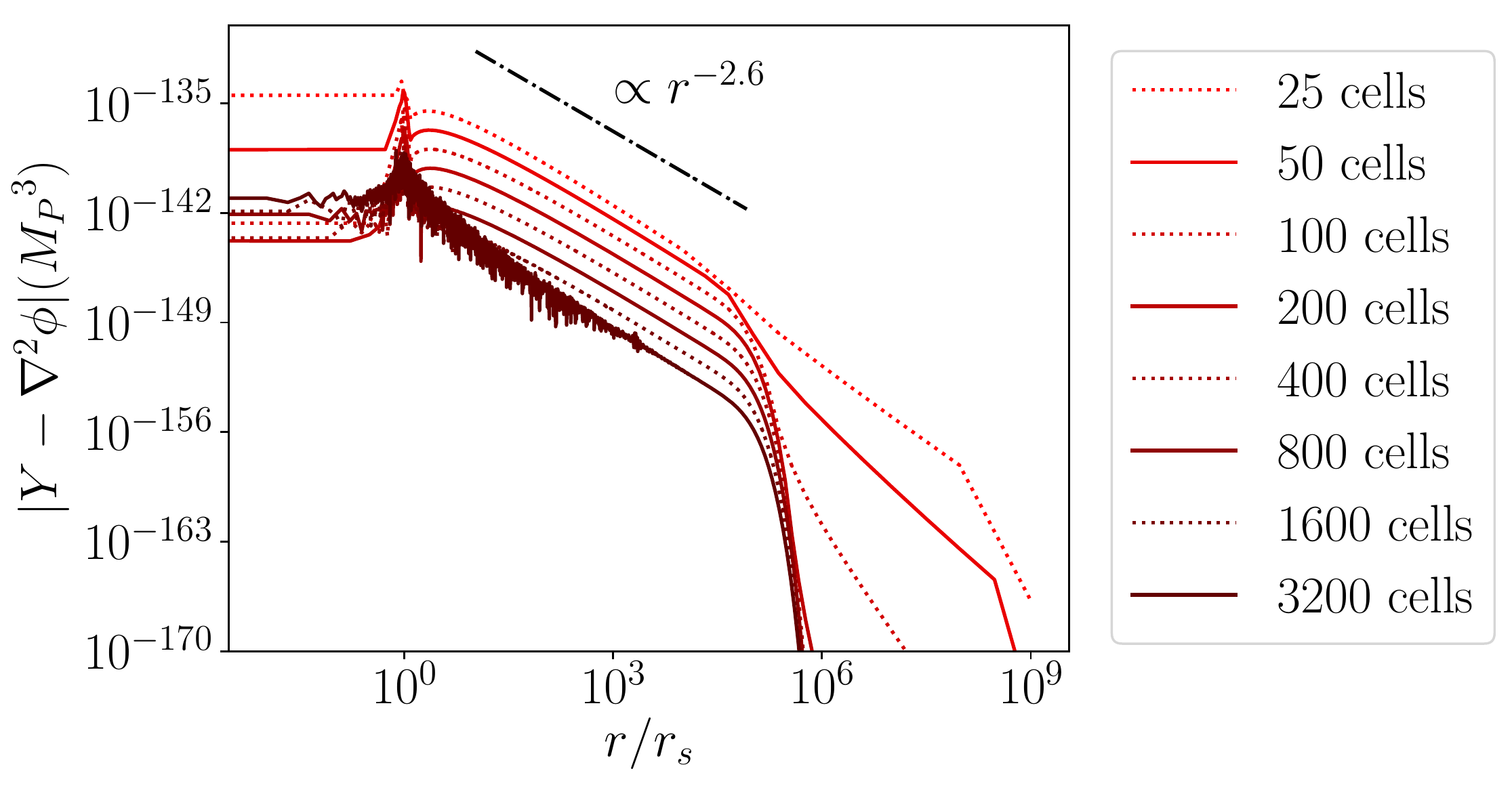}
 \caption{This figure shows the strong residual of the consistency relation $Y - \lapRad^2\fldLt$, for model \ref{Ex:Test_model} and increasing numbers of elements. In this plot, we consider the same family of \meshExp{} meshes as in Fig.~\ref{Fig:giant_mesh_interp}: mapping parameters $k=7$, $a=5 \times 10^{-2}$, $b=5 \times 10^{-2}$, $r_{\rm max} = 10^9 \, \rSrc$ and resolution increasing from $25$ to $3200$ elements doubling each time. Similarly to Fig.~\ref{Fig:giant_mesh_interp}, we can see that a higher number of elements initially decreases the error at mesh vertices, until no additional improvement is obtained. Once the element number become excessive and an overly high number of modes is used to described the target function, the effects of round-off errors dominate.}
 \label{Fig:F3_residuals}
 \end{figure}

A similar test of accuracy can be made by studying the strong residuals of the system of equations in Eq.~\eqref{Eq:UV_strong_system} for varying numbers of mesh elements. We make this comparison in Fig.~\ref{Fig:F3_residuals}, for the same meshes considered in the previous test and in Fig.~\ref{Fig:giant_mesh_interp}. In particular, we focus on the consistency relation $Y - \lapRad^2\fldLt$ (third equation), which is the dominant residual for the example model considered --- this suggests the largest source of error comes from inaccuracies in the discretisation of the product rule, an effect we will further investigate below. To make this comparison, we compute (in post-processing) the Laplacian $\lapRad^2\fldLt$ as the sum of derivatives $\left( \frac{\partial^2}{\partial r^2} + \frac{2}{r}\frac{\partial}{\partial r}\right) \fldLt$, projecting this expression onto a discontinuous function space of Lagrange elements of degree $5$ (\ie the same degree as used when solving the system of equations). We can again see that increasing the number of elements initially improves the error in the solution, until an excessive number of them just enhances the effects of numerical noise without further improvement. Between $r\sim10^1$ and $r \sim 10^5$ the residuals decrease polynomially with $r$; their magnitude further decreases with the number of elements as $\sim N^{-3.7}$.

\begin{figure}[h!]
\centering
\includegraphics[width=\textwidth]{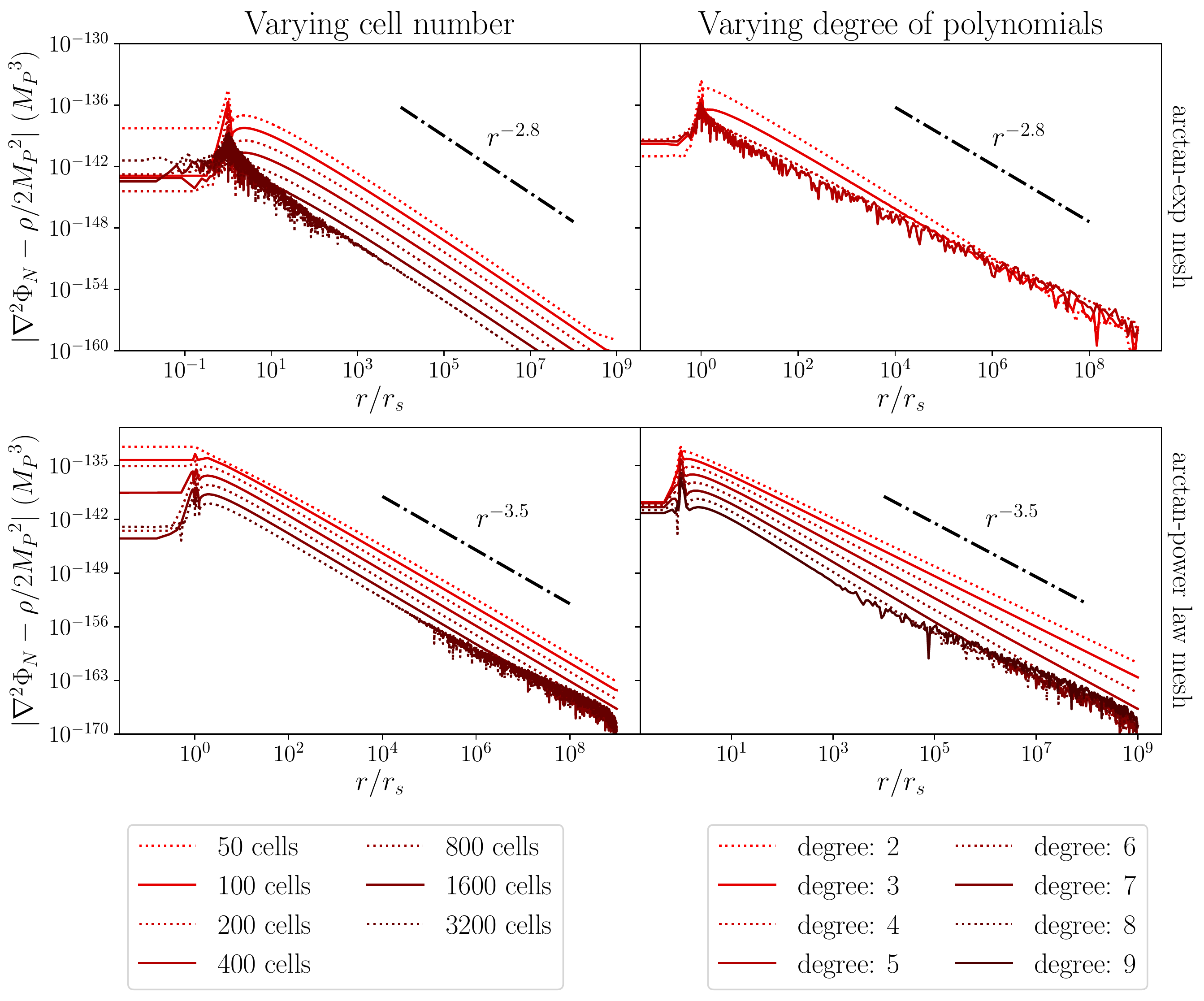}
\caption{Residuals of the Poisson equation, for the same source profile as in model \ref{Ex:Test_model}. For the Poisson equation, the residuals are ascribable to the discretisation of the Laplacian alone: as such, they are a useful measure of the breakdown of the product rule in the discretum. In the left panels, we show the residuals for varying element numbers, from $50$ to $3200$ doubling each time, whilst keeping the order of interpolating polynomials fixed to $5$. In the right panels, we instead vary the order of interpolating polynomials from $2$ to $9$, whilst keeping the number of elements fixed to $400$. We repeat the experiment for both of the meshes described in Sec.~\ref{Sec:Meshes}: the results for the \meshExp{} meshes are shown in the top panels, and in the bottom panels for the \meshPow{} meshes. The transformation parameters for the family of \meshExp{} meshes are $k=7$, $a=5 \times 10^{-2}$, $b=5 \times 10^{-2}$, and for the \meshPow{} meshes $k=15, \gamma=6.5$. In both cases, we use $r_{\rm max} = 10^9 \, \rSrc$. The residuals of the Poisson equation improve with increasing number of elements and degree of interpolating polynomials, until numerical noise resulting from round-off errors dominates, similarly to what is shown in Figs.~\ref{Fig:giant_mesh_interp}-\ref{Fig:F3_residuals}. }
\label{Fig:Poisson_residuals}
\end{figure}
 
As discussed above, the dominant source of error in our numerical solution to model \ref{Ex:Test_model} appears to stem from the breakdown of the product rule upon discretisation.  When computing the weak form of the equations of motion, the Laplacian is expressed as a product of gradients via integration by parts (see Sec.~\ref{Sec:IR_theory_weak_form}-\ref{Sec:UV_theory_weak_form}). For continuous functions and the given boundary conditions, the equality $\int \lapRad^2 f v r^2 dr = \int \lapRad f \cdot \lapRad v r^2 dr$ is exact. However, for discretised functions, it is approximate: the result is an extra, artificial, boundary term.\footnote{Note that it is, in principle, possible to discretise the domain in such a way that the product rule is preserved in the discretum.}

In order to study this effect more closely, we now examine the residuals for numerical solutions to the Poisson equation $\lapRad^2\eta=\frac{1}{2}\frac{\rho}{{M_p}^2}$, for the same source profile as in model~\ref{Ex:Test_model}. 
For the Poisson equation, the source term $\rho/w \mpl^2$ can be evaluated within machine precision at every vertex, so that the residuals are ascribable to the Laplacian alone: this setup is then particularly convenient to explore the breakdown of the product rule in the discretum.
We solve the Poisson equation for a collection of seven \meshExp{} meshes with the same transformation parameters as in Figs.~\ref{Fig:giant_mesh_interp}-\ref{Fig:F3_residuals} (i.e. $k=7$, $a=5 \times 10^{-2}$, $b=5 \times 10^{-2}$ and $r_{\rm max} = 10^9 \, \rSrc$), as well as a collection of \meshPow{} meshes with transformation parameters $k=15, \gamma=6.5, r_{\rm max} = 10^9 \, \rSrc$. For both meshes, we use a resolution of $50, 100, 200, 400, 800, 1600$, and $3200$ elements.
To explore the effect of element degree on the accuracy of the product rule, we also vary the degree of interpolating polynomials from $2$ to $9$. The results are shown in Fig~.\ref{Fig:Poisson_residuals}. We can see (left panels) that the error in the product rule decreases as the number of elements increases (as approximately $N^{-3.9}$ for the \meshExp{} meshes and $N^{-3.5}$ for the \meshPow{} meshes) until the effects of numerical noise are felt again. Similarly, the residuals decrease with increasing order of interpolating polynomials (right panels), until the number of nodes similarly becomes too large and round-off errors dominate.

\section{Code structure} \label{Sec:Code_structure}

In this appendix, we briefly describe \phienics{}'s structure, and outline the steps that prospective users will need to take if they wish to adapt the code to other screening models. Further details on the listed code elements may be found in \phienics{}'s documentation; 
additionally, the Jupyter notebooks provided with the code (\texttt{IR\_main.ipynb} and \texttt{UV\_main.ipynb}) show a suggested workflow that generalises to any static and spherically symmetric screening problem.

\subsection{Discretisation}

Any prospective user of \phienics{} should start by discretising the domain. For the example theories considered in this paper, the \texttt{ArcTanExpMesh} and \texttt{ArcTanPowerLawMesh} meshes presented in Sec.~\ref{Sec:Meshes} were implemented with the provided mesh classes of the same name. We expect them to be useful for other problems of screening and not to be specific of our models.
However, prospective users may wish to implement different transformations based on specific knowledge of their problem and thus generate different meshes. 
\phienics{}'s \texttt{ArcTanExpMesh} and \texttt{ArcTanPowerLawMesh} were implemented as subclasses of a \texttt{Mesh} base class to allow this while retaining other useful functionalities, including linear refinement, point declustering, and checks that mesh vertices are not closer than machine precision allows or that the derivative of the transformation is negative (for declustering and $k_{\rm rm}$ negative). These general functionalities, useful for any mesh, can therefore be performed without any specific knowledge of the nonlinear transformation used.

To add a new mesh to \phienics{}, we encourage users to implement a new \texttt{Mesh} subclass modelled on the \texttt{ArcTanExpMesh} and \texttt{ArcTanPowerLawMesh} subclasses provided. 
Any new subclass should contain a definition of the transformation to be applied, together with its first and second derivatives, plus either an exact or approximate inverse transformation at small and large radii.

Further finite-element settings are given in the \texttt{fem} module: they are expected to work `as is' for generic models of screening. In particular, they define continuous and discontinuous Lagrange elements to represent functions as described in Sec.~\ref{Sec:FEM}.

\subsection{Source}

The \texttt{source} modules contain definitions of static and spherically symmetric source profiles of interest for general models of screening. They are: smoothed top-hat (\texttt{TopHatSource}, Eq.~\ref{Eq:top_hat_source}), step function (\texttt{StepSource}), truncated cosine (\texttt{CosSource}, Eq.~\ref{Eq:cos_source}), Gaussian (\texttt{GaussianSource}), and `Gaussian cake' (\texttt{GCakeSource}, Eq.~\ref{Eq:GCake_source}). Any user wishing to implement a different source profile can add a corresponding subclass of \texttt{Source} following the examples provided.

\subsection{Solver}

\phienics{} contains a base solver class (\texttt{Solver}) that takes care of functionalities that are general to any nonlinear solver, and not restricted to specific models of screening. These functionalities include: advancing the Newton iterations, computing the error at every iteration (given a weak residual form and/or the change in the solution at that iteration) and implementing the stopping criteria.

Similarly to the way in which the mesh is built, model-specific tasks for the single- and two-field theories are implemented in the \texttt{Solver} subclasses \texttt{IRSolver} (\texttt{IR} module) and \texttt{UVSolver} (\texttt{UV} module). Users wishing to apply \phienics{} to different theories are encouraged to write a model-specific subclass of \texttt{Solver} following the examples provided, and use the base-class functionalities for non-model specific tasks. Information to be supplied in the custom subclass includes:

\begin{itemize}
\item a linear solver for the Newton iteration (Eq.~\ref{eqn:newton-frechet}), expressed in weak form. For our example theories, the analytic formulae for this step have been written down explicitly in Eq.~\eqref{IR_NL_weak_form} and Eq.~\eqref{Eq:UV_Newton_iteration}. Note that the weak form must correctly implement the Neumann boundary conditions;

\item the Dirichlet boundary conditions for the theory;

\item the weak and strong form of the residuals; 

\item where the equation of motion is expanded into a system of equations (or is natively a system of equations), the definition of the composite function space for the vector equation. The \texttt{IRSolver} and \texttt{UVSolver} subclasses perform this task and can be taken as examples.

\end{itemize}

\bibliographystyle{unsrt}
\bibliography{references}

\begin{thebibliography}{10}

\bibitem{Aghanim:2018eyx}
N.~Aghanim et~al.
\newblock {Planck 2018 results. VI. Cosmological parameters}.
\newblock 2018.

\bibitem{Bernal:2016gxb}
Jose~Luis Bernal, Licia Verde, and Adam~G. Riess.
\newblock {The trouble with $H_0$}.
\newblock {\em JCAP}, 1610(10):019, 2016.

\bibitem{Freedman:2017yms}
Wendy~L. Freedman.
\newblock {Cosmology at a Crossroads}.
\newblock {\em Nat. Astron.}, 1:0121, 2017.

\bibitem{Riess:2019cxk}
Adam~G. Riess, Stefano Casertano, Wenlong Yuan, Lucas~M. Macri, and Dan
  Scolnic.
\newblock {Large Magellanic Cloud Cepheid Standards Provide a 1\% Foundation
  for the Determination of the Hubble Constant and Stronger Evidence for
  Physics beyond $\Lambda$CDM}.
\newblock {\em Astrophys. J.}, 876(1):85, 2019.

\bibitem{Wong:2019kwg}
Kenneth~C. Wong et~al.
\newblock {H0LiCOW XIII. A 2.4\% measurement of $H_{0}$ from lensed quasars:
  $5.3\sigma$ tension between early and late-Universe probes}.
\newblock 2019.

\bibitem{Will:2014bqa}
Clifford~M. Will.
\newblock {Was Einstein Right? A Centenary Assessment}.
\newblock 2014.

\bibitem{Joyce:2014kja}
Austin Joyce, Bhuvnesh Jain, Justin Khoury, and Mark Trodden.
\newblock {Beyond the Cosmological Standard Model}.
\newblock {\em Phys. Rept.}, 568:1--98, 2015.

\bibitem{Ishak:2018his}
Mustapha Ishak.
\newblock {Testing General Relativity in Cosmology}.
\newblock {\em Living Rev. Rel.}, 22(1):1, 2019.

\bibitem{Sotiriou:2008rp}
Thomas~P. Sotiriou and Valerio Faraoni.
\newblock {f(R) Theories Of Gravity}.
\newblock {\em Rev. Mod. Phys.}, 82:451--497, 2010.

\bibitem{Hinterbichler:2011tt}
Kurt Hinterbichler.
\newblock {Theoretical Aspects of Massive Gravity}.
\newblock {\em Rev. Mod. Phys.}, 84:671--710, 2012.

\bibitem{deRham:2010kj}
Claudia de~Rham, Gregory Gabadadze, and Andrew~J. Tolley.
\newblock {Resummation of Massive Gravity}.
\newblock {\em Phys. Rev. Lett.}, 106:231101, 2011.

\bibitem{Horndeski:1974wa}
Gregory~Walter Horndeski.
\newblock {Second-order scalar-tensor field equations in a four-dimensional
  space}.
\newblock {\em Int. J. Theor. Phys.}, 10:363--384, 1974.

\bibitem{Deffayet:2009mn}
C.~Deffayet, S.~Deser, and G.~Esposito-Farese.
\newblock {Generalized Galileons: All scalar models whose curved background
  extensions maintain second-order field equations and stress-tensors}.
\newblock {\em Phys. Rev.}, D80:064015, 2009.

\bibitem{Desmond:2018sdy}
Harry Desmond, Pedro~G. Ferreira, Guilhem Lavaux, and Jens Jasche.
\newblock {Fifth force constraints from the separation of galaxy mass
  components}.
\newblock {\em Phys. Rev.}, D98(6):064015, 2018.

\bibitem{Desmond:2018kdn}
Harry Desmond, Pedro~G. Ferreira, Guilhem Lavaux, and Jens Jasche.
\newblock {Fifth force constraints from galaxy warps}.
\newblock {\em Phys. Rev.}, D98(8):083010, 2018.

\bibitem{Bull:2015stt}
Philip Bull et~al.
\newblock {Beyond $\Lambda$CDM: Problems, solutions, and the road ahead}.
\newblock {\em Phys. Dark Univ.}, 12:56--99, 2016.

\bibitem{Koyama:2015vza}
Kazuya Koyama.
\newblock {Cosmological Tests of Modified Gravity}.
\newblock {\em Rept. Prog. Phys.}, 79(4):046902, 2016.

\bibitem{Slosar:2019flp}
Anže Slosar et~al.
\newblock {Dark Energy and Modified Gravity}.
\newblock 2019.

\bibitem{Nicolis:2008in}
Alberto Nicolis, Riccardo Rattazzi, and Enrico Trincherini.
\newblock {The Galileon as a local modification of gravity}.
\newblock {\em Phys. Rev.}, D79:064036, 2009.

\bibitem{Gleyzes:2014dya}
Jérôme Gleyzes, David Langlois, Federico Piazza, and Filippo Vernizzi.
\newblock {Healthy theories beyond Horndeski}.
\newblock {\em Phys. Rev. Lett.}, 114(21):211101, 2015.

\bibitem{Gleyzes:2014qga}
Jérôme Gleyzes, David Langlois, Federico Piazza, and Filippo Vernizzi.
\newblock {Exploring gravitational theories beyond Horndeski}.
\newblock {\em JCAP}, 1502:018, 2015.

\bibitem{Langlois:2015cwa}
David Langlois and Karim Noui.
\newblock {Degenerate higher derivative theories beyond Horndeski: evading the
  Ostrogradski instability}.
\newblock {\em JCAP}, 1602(02):034, 2016.

\bibitem{Langlois:2015skt}
David Langlois and Karim Noui.
\newblock {Hamiltonian analysis of higher derivative scalar-tensor theories}.
\newblock {\em JCAP}, 1607(07):016, 2016.

\bibitem{Achour:2016rkg}
Jibril Ben~Achour, David Langlois, and Karim Noui.
\newblock {Degenerate higher order scalar-tensor theories beyond Horndeski and
  disformal transformations}.
\newblock {\em Phys. Rev.}, D93(12):124005, 2016.

\bibitem{Crisostomi:2016czh}
Marco Crisostomi, Kazuya Koyama, and Gianmassimo Tasinato.
\newblock {Extended Scalar-Tensor Theories of Gravity}.
\newblock {\em JCAP}, 1604(04):044, 2016.

\bibitem{deRham:2017imi}
Claudia de~Rham, Scott Melville, Andrew~J. Tolley, and Shuang-Yong Zhou.
\newblock {Massive Galileon Positivity Bounds}.
\newblock {\em JHEP}, 09:072, 2017.

\bibitem{Ogawa:2018srw}
Hiromu Ogawa, Takashi Hiramatsu, and Tsutomu Kobayashi.
\newblock {Anti-screening of the Galileon force around a disk center hole}.
\newblock {\em Mod. Phys. Lett.}, A34(02):1950013, 2018.

\bibitem{Hiramatsu:2012xj}
Takashi Hiramatsu, Wayne Hu, Kazuya Koyama, and Fabian Schmidt.
\newblock {Equivalence Principle Violation in Vainshtein Screened Two-Body
  Systems}.
\newblock {\em Phys. Rev.}, D87(6):063525, 2013.

\bibitem{Kuntz2019}
Adrien {Kuntz}.
\newblock {Two-body potential of Vainshtein screened theories}.
\newblock {\em Physical Review D}, 100(2):024024, July 2019.

\bibitem{White2020}
Nicholas~C. {White}, Sandra~M. {Troian}, Jeffrey~B. {Jewell}, Curt~J. {Cutler},
  Sheng-wey {Chiow}, and Nan {Yu}.
\newblock {Robust numerical computation of the 3D scalar potential field of the
  cubic Galileon gravity model at solar system scales}.
\newblock {\em Physical Review D}, 102(2):024033, July 2020.

\bibitem{Dar:2018dra}
Furqan Dar, Claudia De~Rham, J.~Tate Deskins, John~T. Giblin, and Andrew~J.
  Tolley.
\newblock {Scalar Gravitational Radiation from Binaries: Vainshtein Mechanism
  in Time-dependent Systems}.
\newblock {\em Class. Quant. Grav.}, 36(2):025008, 2019.

\bibitem{Frolov_2017}
Andrei~V. Frolov, José~T. Gálvez~Ghersi, and Alex Zucca.
\newblock Unscreening scalarons with a black hole.
\newblock {\em Physical Review D}, 95(10), May 2017.

\bibitem{Upadhye_2006}
Amol Upadhye, Steven~S. Gubser, and Justin Khoury.
\newblock Unveiling chameleon fields in tests of the gravitational
  inverse-square law.
\newblock {\em Physical Review D}, 74(10), Nov 2006.

\bibitem{Elder_2016}
Benjamin Elder, Justin Khoury, Philipp Haslinger, Matt Jaffe, Holger Müller,
  and Paul Hamilton.
\newblock Chameleon dark energy and atom interferometry.
\newblock {\em Physical Review D}, 94(4), Aug 2016.

\bibitem{Jaffe_2017}
Matt Jaffe, Philipp Haslinger, Victoria Xu, Paul Hamilton, Amol Upadhye,
  Benjamin Elder, Justin Khoury, and Holger Müller.
\newblock Testing sub-gravitational forces on atoms from a miniature in-vacuum
  source mass.
\newblock {\em Nature Physics}, 13(10):938–942, Jul 2017.

\bibitem{Brax:2018zfb}
Philippe Brax, Anne-Christine Davis, Benjamin Elder, and Leong~Khim Wong.
\newblock {Constraining screened fifth forces with the electron magnetic
  moment}.
\newblock {\em Phys. Rev. D}, 97(8):084050, 2018.

\bibitem{Elder_2020}
Benjamin Elder, Valeri Vardanyan, Yashar Akrami, Philippe Brax, Anne-Christine
  Davis, and Ricardo~S. Decca.
\newblock Classical symmetron force in casimir experiments.
\newblock {\em Physical Review D}, 101(6), Mar 2020.

\bibitem{Burrage:2017shh}
Clare Burrage, Edmund~J. Copeland, Adam Moss, and James~A. Stevenson.
\newblock {The shape dependence of chameleon screening}.
\newblock {\em JCAP}, 1801(01):056, 2018.

\bibitem{Barreira:2015xvp}
Alexandre Barreira, Sownak Bose, and Baojiu Li.
\newblock {Speeding up N-body simulations of modified gravity: Vainshtein
  screening models}.
\newblock {\em JCAP}, 1512(12):059, 2015.

\bibitem{Winther:2014cia}
Hans~A. Winther and Pedro~G. Ferreira.
\newblock {Fast route to nonlinear clustering statistics in modified gravity
  theories}.
\newblock {\em Phys. Rev.}, D91(12):123507, 2015.

\bibitem{Bose:2016wms}
Sownak Bose, Baojiu Li, Alexandre Barreira, Jian-hua He, Wojciech~A. Hellwing,
  Kazuya Koyama, Claudio Llinares, and Gong-Bo Zhao.
\newblock {Speeding up $N$-body simulations of modified gravity: Chameleon
  screening models}.
\newblock {\em JCAP}, 1702(02):050, 2017.

\bibitem{Llinares:2013jza}
Claudio Llinares, David~F. Mota, and Hans~A. Winther.
\newblock {ISIS: a new N-body cosmological code with scalar fields based on
  RAMSES. Code presentation and application to the shapes of clusters}.
\newblock {\em Astron. Astrophys.}, 562:A78, 2014.

\bibitem{Hagala:2015paa}
R.~Hagala, C.~Llinares, and D.~F. Mota.
\newblock {Cosmological simulations with disformally coupled symmetron fields}.
\newblock {\em Astron. Astrophys.}, 585:A37, 2016.

\bibitem{Winther:2015wla}
Hans~A. Winther et~al.
\newblock {Modified Gravity N-body Code Comparison Project}.
\newblock {\em Mon. Not. Roy. Astron. Soc.}, 454(4):4208--4234, 2015.

\bibitem{Barreira:2013eea}
Alexandre Barreira, Baojiu Li, Wojciech~A. Hellwing, Carlton~M. Baugh, and
  Silvia Pascoli.
\newblock {Nonlinear structure formation in the Cubic Galileon gravity model}.
\newblock {\em JCAP}, 1310:027, 2013.

\bibitem{Li:2013nua}
Baojiu Li, Gong-Bo Zhao, and Kazuya Koyama.
\newblock {Exploring Vainshtein mechanism on adaptively refined meshes}.
\newblock {\em JCAP}, 1305:023, 2013.

\bibitem{Li:2013tda}
Baojiu Li, Alexandre Barreira, Carlton~M. Baugh, Wojciech~A. Hellwing, Kazuya
  Koyama, Silvia Pascoli, and Gong-Bo Zhao.
\newblock {Simulating the quartic Galileon gravity model on adaptively refined
  meshes}.
\newblock {\em JCAP}, 1311:012, 2013.

\bibitem{Dossett:2011tn}
Jason~N. Dossett, Mustapha Ishak, and Jacob Moldenhauer.
\newblock {Testing General Relativity at Cosmological Scales: Implementation
  and Parameter Correlations}.
\newblock {\em Phys. Rev.}, D84:123001, 2011.

\bibitem{Hojjati:2011ix}
Alireza Hojjati, Levon Pogosian, and Gong-Bo Zhao.
\newblock {Testing gravity with CAMB and CosmoMC}.
\newblock {\em JCAP}, 1108:005, 2011.

\bibitem{Zucca:2019xhg}
Alex Zucca, Levon Pogosian, Alessandra Silvestri, and Gong-Bo Zhao.
\newblock {MGCAMB with massive neutrinos and dynamical dark energy}.
\newblock {\em JCAP}, 1905(05):001, 2019.

\bibitem{Hu:2013twa}
Bin Hu, Marco Raveri, Noemi Frusciante, and Alessandra Silvestri.
\newblock {Effective Field Theory of Cosmic Acceleration: an implementation in
  CAMB}.
\newblock {\em Phys. Rev.}, D89(10):103530, 2014.

\bibitem{Hu:2014oga}
Bin Hu, Marco Raveri, Noemi Frusciante, and Alessandra Silvestri.
\newblock {EFTCAMB/EFTCosmoMC: Numerical Notes v3.0}.
\newblock 2014.

\bibitem{Zumalacarregui:2016pph}
Miguel Zumalacárregui, Emilio Bellini, Ignacy Sawicki, Julien Lesgourgues, and
  Pedro~G. Ferreira.
\newblock {hi\_class: Horndeski in the Cosmic Linear Anisotropy Solving
  System}.
\newblock {\em JCAP}, 1708(08):019, 2017.

\bibitem{Bellini:2019syt}
Emilio {Bellini}, Ignacy {Sawicki}, and Miguel {Zumalac{\'a}rregui}.
\newblock {hi\_class: Background Evolution, Initial Conditions and
  Approximation Schemes}.
\newblock {\em JCAP}, 2002(02):008, 2020.

\bibitem{Bellini:2017avd}
E.~Bellini et~al.
\newblock {Comparison of Einstein-Boltzmann solvers for testing general
  relativity}.
\newblock {\em Phys. Rev.}, D97(2):023520, 2018.

\bibitem{Arnold:2019zup}
Christian Arnold and Baojiu Li.
\newblock {Simulating galaxy formation in f(R) modified gravity: Matter, halo,
  and galaxy-statistics}.
\newblock 2019.

\bibitem{Llinares:2018maz}
Claudio Llinares.
\newblock {Simulation techniques for modified gravity}.
\newblock {\em Int. J. Mod. Phys.}, D27(15):1848003, 2018.

\bibitem{physics_paper}
Clare {Burrage}, Ben {Coltman}, Antonio {Padilla}, Daniela {Saadeh}, and Toby
  {Wilson}.
\newblock {Massive Galileons and Vainshtein screening}.
\newblock {\em JCAP}, 2021(2):050, February 2021.

\bibitem{boyd01}
John~P. Boyd.
\newblock {\em {Chebyshev} and {Fourier} Spectral Methods}.
\newblock Dover Books on Mathematics. Dover Publications, Mineola, NY, second
  edition, 2001.

\bibitem{Bond:2015zfa}
J.~Richard Bond, Jonathan Braden, and Laura Mersini-Houghton.
\newblock {Cosmic bubble and domain wall instabilities III: The role of
  oscillons in three-dimensional bubble collisions}.
\newblock {\em JCAP}, 09:004, 2015.

\bibitem{langtangen2019introduction}
H.P. Langtangen and K.A. Mardal.
\newblock {\em Introduction to Numerical Methods for Variational Problems}.
\newblock Texts in Computational Science and Engineering. Springer
  International Publishing, 2019.

\bibitem{FEniCS_citations}
Martin~S. Aln{\ae}s, Jan Blechta, Johan Hake, August Johansson, Benjamin
  Kehlet, Anders Logg, Chris Richardson, Johannes Ring, Marie~E. Rognes, and
  Garth~N. Wells.
\newblock The fenics project version 1.5.
\newblock {\em Archive of Numerical Software}, 3(100), 2015.

\bibitem{Old_FEniCS_book}
Anders Logg, Kent-Andre Mardal, Garth~N. Wells, et~al.
\newblock {\em Automated Solution of Differential Equations by the Finite
  Element Method}.
\newblock Springer, 2012.

\bibitem{MINPACK}
Jorge~J. {Moré}, Burton~S. {Garbow}, and Kenneth~E. {Hillstrom}.
\newblock {\em User Guide for MINPACK-1}.
\newblock Argonne National Laboratory and United States. Department of Energy.
  Office of Scientific and Technical Information, 1980.

\bibitem{Tobys_thesis}
Toby Wilson.
\newblock Scalar fields and gravity, July 2018.

\bibitem{New_FEniCS_book}
Hans~Petter Langtangen and Anders Logg.
\newblock {\em Solving PDEs in Python}.
\newblock Springer, 2017.

\end{thebibliography}

\end{document}